\documentclass[a4paper,showpacs,notitlepage,floatfix,nofootinbib,superscriptaddress,prd]{revtex4-2}
\usepackage{graphicx,color}
\usepackage{amsfonts}
\usepackage{amsmath}
\usepackage{hyperref}
\usepackage{float}
\usepackage{amssymb}
\usepackage{epstopdf}
\usepackage{oldgerm}

\begin{document}

\title{Exact solutions for the moments of the binary collision integral and
its relation to the relaxation-time approximation in leading-order
anisotropic fluid dynamics}
\author{Etele Moln\'ar}
\affiliation{Incubator of Scientiﬁc Excellence--Centre for Simulations of Superdense Fluids,\\
	University of Wroc{\l}aw, pl. M. Borna 9, PL-50204 Wroc{\l}aw, Poland} 
\affiliation{Institut f\"ur Theoretische Physik,
	Johann Wolfgang Goethe--Universit\"at,\\
	Max-von-Laue-Str.\ 1, D--60438 Frankfurt am Main, Germany}

\pacs{12.38.Mh, 24.10.Nz, 47.75.+f, 51.10.+y}

\begin{abstract}
We compute the moments of the nonlinear binary collision integral in the
ultrarelativistic hard-sphere approximation for an arbitrary anisotropic
distribution function in the local rest frame. 
This anisotropic distribution function has an angular asymmetry controlled 
by the parameter of anisotropy $\xi$, such that in the limit of a vanishing anisotropy
$\lim_{\xi \rightarrow 0} \hat{f}_{0 \mathbf{k}} = f_{0 \mathbf{k}}$, 
it approaches the spherically symmetric local equilibrium distribution function.
The corresponding moments of the binary collision integral are obtained in terms 
of quadratic products of different moments of the anisotropic distribution function 
and couple to a well-defined set of lower-order moments.
To illustrate these results we compare the moments of the binary
collision integral to the moments of the widely used relaxation-time
approximation of Anderson and Witting in the case of a spheroidal distribution function.
We found that in an expanding system the nonlinear Boltzmann collision term leads 
to twice slower equilibration than the relaxation-time approximation.
Furthermore, we also show that including two dynamical moments 
helps to resolve the ambiguity of which additional moment of the Boltzmann equation 
to choose to close the conservation laws. 
\end{abstract}

\maketitle


\section{Introduction}

\label{Introduction}

The earliest application of fluid dynamics in relativistic particle
collisions was formulated by Landau~\cite{Landau_hydro} in the early 1950's with the following
assumptions~\cite{Landau_hydro56}. The initial state of the system forms at
the instant of the collision when a large number of particles with mean free
paths smaller than the dimensions of the system is born in local
equilibrium, i.e., local isotropic momentum-space distribution $f_{0 \mathbf{k}}$, 
the J\"uttner distribution~\cite{Juttner,Juttner_quantum}. This system
then expands as a relativistic ideal fluid until the particle interactions
become gradually weaker and the mean free path becomes comparable to the
dimensions of the system. The final or freeze-out stage happens when the
fluid dynamical description is no longer appropriate and the particle nature
of the system becomes of primary importance.

Over the past decades this so-called hydrodynamical model has been constantly 
improved and now relativistic fluid dynamics became an indispensable tool in 
the description of the space-time evolution of the matter created in relativistic 
heavy-ion collisions, astrophysical phenomena as well as of the early 
universe~\cite{Rezzolla_book,Denicol_Rischke_book,Chabanov:2021dee,Heinz:2024jwu,Rocha:2023ilf}. 
Among the most important recent advancements in relativistic fluid dynamics
has been the treatment of irreversible non-equilibrium phenomena that includes 
the transport properties of matter, such as shear and bulk viscosities, charge
diffusivity, thermal and/or electric conductivity, and magnetization. 
These novel fluid dynamical theories extended beyond the constraints of ideal fluid
dynamics and hence the requirement that the fluid is in local thermodynamic
equilibrium but still assumes that the deviations from local equilibrium are
relatively small, i.e., $|\delta f_{\mathbf{k}}/f_{0\mathbf{k}}|\ll 1$. 
This means that the non-equilibrium corrections to ideal fluid dynamics may be
approximated order by order through the corresponding irreducible moments of 
$\delta f_{\mathbf{k}}$ representing the dissipative quantities. 
While the first-order or relativistic Navier-Stokes theory may be acausal and unstable 
\cite{Muller:1967zza,Israel:1979wp,Hiscock:1983zz,Hiscock:1985zz,Hiscock:1987zz,Pu:2009fj,Rocha:2023ilf}, the well-known and widely used equations of second-order dissipative
fluid dynamics of M\"uller~\cite{Muller:1967zza}, and of Israel and Stewart~\cite%
{Israel:1979wp}, relax on finite timescales to their corresponding
Navier-Stokes values. Henceforth under certain conditions second-order
theories avoid the problems of acausality and instability through hyperbolic
equations of motion for the dissipative fields.

Even so, in the case of a rapid longitudinal expansion as in the early
stages of heavy-ion collisions the local momentum anisotropies can be very large;
hence, the series expansion of the distribution function near equilibrium is
no longer an appropriate approximation. 
To overcome these limitations of traditional fluid dynamical theories and to 
better account for large deviations from local equilibrium, in the
late 1980's Barz, K\"ampfer \textit{et al.}~\cite{Barz:1987pq,Kampfer:1990qg} 
proposed an energy-momentum tensor which incorporated the momentum anisotropy 
in terms of a space-like four-vector $l^\mu$, introducing an anisotropic decomposition 
of the isotropic pressure.
As matter of fact, anisotropic matter distributions have been
studied in general relativity already in the early 1970's in the seminal
work by Bowers and Liang~\cite{Bowers:1974tgi}. On the other hand, it was also known 
that the strongly directed nature of various dynamical processes
in heavy-ion collisions such as the hadronization and the freeze-out in
hydrodynamical models, also favors an anisotropic distribution function over
a locally isotropic distribution function \cite{Anderlik:1998cb,Anderlik:1998et}.

In the early 2010's the framework of (leading-order) anisotropic fluid dynamics for
ultrarelativistic heavy-ion collisions was rediscovered by two different
groups: Florkowski and Ryblewski~\cite%
{Florkowski:2008ag,Florkowski:2010cf,Ryblewski:2010bs,Ryblewski:2011aq,Ryblewski:2012rr}
and Martinez and Strickland~\cite{Martinez:2009ry,Martinez:2010sc,Martinez:2010sd}.
This fluid dynamical framework extends beyond ideal fluid dynamics, i.e., the simplification that 
$f_{\mathbf{k}}\equiv f_{0\mathbf{k}}$ is a local isotropic distribution function, 
and instead it is based on a local anisotropic distribution function 
$f_{\mathbf{k}} \equiv \hat{f}_{0\mathbf{k}}$ which incorporates an anisotropy parameter $\xi$ and 
a new spacelike four-vector $l^\mu$ in the direction of the momentum-space anisotropy. 
Therefore the corresponding moment equations derived from the relativistic Boltzmann equation 
for such an anisotropic distribution function form the framework of leading-order 
anisotropic fluid dynamics. 
This means that leading-order anisotropic fluid dynamics describes a specific 
class of dissipative fluids with one additional (when compared to ideal fluids)
independent variable specified by $\hat{f}_{0\mathbf{k}}$. 
An explicit connection to second-order dissipative fluid dynamics is obtained through 
a series expansion of the fluid-dynamical quantities for small $\xi$, or equivalently
to the method of moments applied to $\hat{f}_{0\mathbf{k}}(\xi) = 
f_{0\mathbf{k}} + \delta f_{\mathbf{k}}(\xi)$, see for example Refs.~\cite{Florkowski:2013lya,Tinti:2015xwa,Alqahtani:2017mhy}.
Furthermore, this leading-order framework can also be improved further by generalizing the
expansion of the distribution function as $f_{\mathbf{k}} \equiv \hat{f}_{0%
\mathbf{k}} + \delta \hat{f}_{\mathbf{k}}$, and hence similar to the method
formulated in second-order dissipative fluid dynamics it leads to a more
complete theory of anisotropic dissipative fluid dynamics~\cite%
{Bazow:2013ifa,Alqahtani:2017mhy,Molnar:2016vvu}. 
Furthermore, taking into account the
coupling to the external electromagnetic field naturally extends the
equations of leading-order anisotropic fluid dynamics to electrically
conducting fluids and leads to a novel theory of dissipative and resistive
anisotropic magnetohydrodynamics~\cite{Molnar:2024fgy}.

In the derivation of fluid dynamical theories from the relativistic Boltzmann equation, 
the moments of the collision term reveal the timescales over 
which the matter approaches local equilibrium. 
However, unlike in traditional dissipative fluid dynamical theories,
so far leading-order anisotropic fluid dynamical theories employed the moments 
of the simplified relaxation-time approximation (RTA) instead of the moments 
of the binary collision integral.
In this paper we go beyond this approximation and
present a simple and at the same time general projection method to calculate
the moments of the nonlinear binary collision term. 
As expected the corresponding moments of the binary collision integral are found 
to be quadratic products of anisotropic thermodynamic integrals which can be computed
exactly and in some cases analytically in terms of various moments of the
underlying anisotropic distribution function. 
Therefore these new results also extend and generalize the well-known 
Bobylev-Krook-Wu (BKW)~\cite{Bobylev:1976,Krook_Wu:1976,Krook_Wu:1977} 
model based on the homogeneous and isotropic
solutions of the non-relativistic Boltzmann equation to anisotropic distribution functions.
The relativistic BKW model, the Bazow-Denicol-Heinz-Martinez-Noronha (BDHMN)~\cite{Bazow:2015dha,Bazow:2016oky}
also obtained exact solutions to the relativistic Boltzmann equation through an infinite set of
nonlinear ordinary differential equations for the moments of an isotropic single-particle 
distribution function and the corresponding moments of the nonlinear binary collision integral 
computed for an isotropic distribution function.

The equations of leading-order fluid dynamics derived from the relativistic Boltzmann equation
correspond to an infinite hierarchy of coupled moment equations, where the moments of the 
binary collision term now couple to a well-defined set of lower-order moments in the hierarchy. 
These nonlinear couplings between moments of different order represent the nonlinear 
dependence on the distribution function of the binary collision integral. 
To be practical one must truncate and close the hierarchy of moment
equations of leading-order anisotropic fluid dynamics. 
But even after a well-defined truncation there remains an ambiguity, which
higher-order moment to use for closure. In this paper we also go beyond
a single dynamical moment to close the conservation equations~\cite{Molnar:2016gwq,Niemi:2017stb}, 
and show that including two dynamical moments 
helps to resolve this ambiguity and improves the solutions to the fluid dynamical equations and 
the approximation to the anisotropic distribution function.

As an important application in heavy-ion collision we have used the well-known 
spheroidal distribution function of Romatschke and Strickland (RS)~\cite%
{Romatschke:2003ms} to explicitly study the moments of the binary collision
integral and the fluid dynamical evolution of the system. 
We found that different anisotropic moments approach equilibrium on different
timescales, alike in second-order fluid dynamical theories. 
Furthermore, the widely used relativistic relaxation-time approximation of 
Anderson and Witting (AW)~\cite{Anderson:1974nyl} for the collision integral is 
also included for comparison. 
We found that the collision term in the RTA drives the system 
about two times faster toward equilibrium than the binary collision
integral irrespective of the choice of the dynamical moment(s). 
This resulting time difference may have important implications in the 
understanding and the description of the momentum-space and pressure anisotropies 
of the quark gluon plasma (QGP) which relaxes toward an 
isotropic distribution on timescales of a few fm/c after the initial impact of collision.
Finally, it is shown that the rescaling of the relaxation time of the RTA, based 
on the asymptotic values of the ratio of the binary collision integral and 
the RTA collision integral, captures reasonably well the characteristics of 
the binary collision integral. Therefore this mapping defines the 
"proper" relaxation-time parameters $\tau_{ij}(\tau_R)$ of the nonlinear binary collision integral 
of higher-order moments, facilitating direct comparisons between the two collision terms.

This paper is organized as follows. A brief introduction to leading-order
anisotropic fluid dynamics is given in Sec.~\ref{sec:aniso_hydro}. In Sec.~%
\ref{sect:Binary_collision_integral} we define the moments of the binary
collision integral in the general case of an arbitrary anisotropic
distribution function. Then, in Sec.\ref{sec:Loss_Terms} and Sec.\ref%
{sec:Gain_Terms} we compute the moments of the corresponding loss and gain
terms for an arbitrary anisotropic distribution function. These results are
then applied in Sec.~\ref{sec:RS_binary_vs_RTA} to the spheroidal distribution 
function introduced by Romatschke and Strickland. The evolution
of matter in $(0+1)$-dimensional boost invariant expansion is studied using
the moments of the binary collision integral as well as using its well-known
relativistic approximation, the relaxation-time approximation of Anderson and
Witting, in Sec.~\ref{sec:Results_BJ} and in Sec.~\ref{sec:Results_BJ_two}.
The differences between the collision terms are analyzed, and the
rescaling of the relaxation times to reduce these differences is
presented in Sec.~\ref{sec:RTA_matching}.
We conclude this work in Sec.~\ref{sec:Conclusions}. 
Technical details and additional computations are relegated to the appendixes.

\subsection{Notation and definitions}
\label{sec:definitions}

In this paper we work in natural units by setting $\hbar =c=k_{B}=1$, and
in flat space-time with metric tensor $g^{\mu \nu }=\text{diag}(1,-1,-1,-1)$. 
The contravariant fluid four-velocity $u^{\mu }\left( t,\mathbf{x}\right)
=\gamma \left( 1,\mathbf{v}\right)$ with $\gamma =(1-\mathbf{v}^{2})^{-1/2}$
has timelike normalization $u^{\mu }u_{\mu }\equiv c^{2}=1$. 
The unit four-vector in the direction of the momentum anisotropy 
$l^{\mu }\left(t,\mathbf{x}\right) $ is chosen orthogonal to the fluid four-velocity; 
i.e., $u^{\mu}l_{\mu }=0$, and has spacelike normalization $l^{\mu }l_{\mu } \equiv -l^2=-1$. 
For example, in the context of heavy-ion collision choosing $l^\mu$ in the direction of 
the beam axis sets $l^{\mu }=\gamma _{z}\left( v_{z},0,0,1\right)$ with 
$\gamma _{z}=(1-v_{z}^{2})^{-1/2}$. 
In the local rest (LR) frame of the fluid $u_{LR}^{\mu }=(1,0,0,0)$ and $l_{LR}^{\mu }=(0,0,0,1)$.

The symmetric rank-two projection operator orthogonal to $u^{\mu }$, i.e., $%
\Delta ^{\mu \nu }u_{\nu }=0$, is defined as in Refs.~\cite%
{deGroot_book,Cercignani_book,Denicol_Rischke_book},
\begin{equation}
\Delta^{\mu \nu } \equiv g^{\mu \nu }-u^{\mu }u^{\nu }\, ,
\label{Delta_munu}
\end{equation}%
while the symmetric rank-two projection operator that is orthogonal to both $u^{\mu }$
and $l^{\mu }$, i.e., $\Xi^{\mu \nu }u_{\nu }=\Xi^{\mu \nu }l_{\nu}=0$, is
defined as in Refs.~\cite%
{Gedalin_1991,Gedalin_1995,Huang:2011dc,Denicol_Rischke_book} 
\begin{equation}
\Xi ^{\mu \nu }\equiv g^{\mu \nu }-u^{\mu }u^{\nu }+l^{\mu }l^{\nu }
=\Delta^{\mu \nu }+l^{\mu }l^{\nu }\, .  \label{Xi_munu}
\end{equation}

The particle four-momentum $k^{\mu }=\left( k^{0},\mathbf{k}\right)$ is normalized 
to the rest mass squared of the particle, $k^{\mu }k_{\mu }=m_{0}^{2}$. 
The energy $E_{\mathbf{k}u}$ and the momentum in the direction of the
anisotropy $E_{\mathbf{k}l}$ are 
\begin{equation}
E_{\mathbf{k}u}\equiv k^{\mu }u_{\mu }\, ,\quad E_{\mathbf{k}l}\equiv -k^{\mu}l_{\mu }\, .
\end{equation}%
The four-momentum orthogonal to the four-velocity $u^{\mu }$ is denoted by 
$k^{\left\langle \mu \right\rangle }\equiv \Delta ^{\mu \nu }k_{\nu }$, while
the four-momentum orthogonal to both $u^{\mu }$ and $l^{\mu }$ is denoted by 
$k^{\left\{ \mu \right\} }\equiv \Xi ^{\mu \nu }k_{\nu }$, such that 
$k^{\left\langle \mu \right\rangle }=E_{\mathbf{k}l}l^{\mu }+k^{\left\{ \mu\right\} }$ 
and thus 
\begin{equation}
k^{\mu }\equiv E_{\mathbf{k}u}u^{\mu }+k^{\left\langle \mu \right\rangle
}=E_{\mathbf{k}u}u^{\mu }+E_{\mathbf{k}l}l^{\mu }+k^{\left\{ \mu \right\}} \, .
\label{k_mu_decomposition}
\end{equation}

The total momentum in binary collision is denoted by 
\begin{equation}
P_{T}^{\mu }\equiv k^{\mu }+k^{\prime \mu }=p^{\mu }+p^{\prime \mu }\, ,
\label{P_T}
\end{equation}%
and it is normalized to the square of the center of mass energy, 
$s\equiv P_{T}^{\mu }P_{T,\mu}$. 
The symmetric rank-two projection operator orthogonal to the total momentum,
i.e., $\Delta _{T}^{\mu \nu }P_{T,\nu }=0$, is defined similarly 
to Eq.~(\ref{Delta_munu}) as 
\begin{equation}
\Delta _{T}^{\mu \nu }\equiv g^{\mu \nu }-\frac{P_{T}^{\mu }P_{T}^{\nu }}{s}\,,  
\label{Delta_munu_T}
\end{equation}%
and therefore the particle four-momentum can be decomposed as 
$k^{\mu}=P_{T}^{\mu }(P_{T}^{\nu }p_{\nu })/s +\Delta _{T}^{\mu\nu}k_{\nu }$.

The local distribution function of an anisotropic state as a function of 
$\hat{\alpha}$, $\hat{\beta}_{u}$, and $\hat{\beta}_{l}$, as well as of 
$E_{\mathbf{k}u}$ and $E_{\mathbf{k}l}$, is denoted by 
$\hat{f}_{0\mathbf{k}}\left( \hat{\alpha},\hat{\beta}_{u}E_{\mathbf{k}u},
\hat{\beta}_{l}E_{\mathbf{k}l}\right) $. 
Such anisotropic distribution functions distinguish particle momenta parallel 
and perpendicular to the direction of the anisotropy, and hence are commonly interpreted by two 
different ``temperatures''. 
Here $\hat{\beta}^{-1}_l$ is the inverse temperature in the direction of the anisotropy, 
and $\hat{\beta}^{-1}_u$ is in the direction perpendicular to the anisotropy, 
while $\hat{\alpha}$ is related to the chemical potential. 
In the limit of vanishing anisotropy parameter $\hat{\beta}_l \rightarrow 0$, 
the anisotropic distribution converges to the distribution function in local equilibrium, 
\begin{equation}
\lim_{\hat{\beta}_{l}\rightarrow 0}\hat{f}_{0\mathbf{k}}\left( \hat{\alpha},
\hat{\beta}_{u}E_{\mathbf{k}u},\hat{\beta}_{l}E_{\mathbf{k}l}\right) 
=f_{0\mathbf{k}}\left( \hat{\alpha},\hat{\beta}_{u}E_{\mathbf{k}u}\right) \, ,
\label{hat_f->f_0}
\end{equation}%
where the J\"{u}ttner distribution \cite{Juttner,Juttner_quantum} defines
local thermodynamic equilibrium,
\begin{equation}
f_{0\mathbf{k}} \left(\alpha, \beta E_{\mathbf{k}u} \right) 
\equiv \left[\exp \left( \beta E_{\mathbf{k}u}-\alpha \right) +a\right] ^{-1}\, .
\label{f_0k}
\end{equation}%
Here $\alpha =\mu \beta $, such that $\mu $ is the chemical potential and 
$\beta =1/T$ is the inverse temperature, while $a=\pm 1$ for fermions/bosons
and $a=0$ for classical Boltzmann particles. The equilibrium
distribution function also defines~\cite{Van:2013sma} the inverse temperature four-vector 
$\beta^\mu \equiv \beta u^{\mu}= u^\mu/T$ normalized to 
$\beta^\mu \beta_\mu \equiv\beta^2 = 1/T^2$ 
and so $\beta^\mu k_\mu \equiv \beta E_{\mathbf{k}u}$. 

Note that in principle there are infinitely many different distribution functions
which are anisotropic in momentum-space according to Eq.~(\ref{hat_f->f_0}).
A straightforward  approach is to stretch or squeeze an isotropic distribution function 
in some direction or along a particular axis. 
Thereby along the direction of the momentum anisotropy we may introduce a 
``longitudinal temperature'', $T_l$, and consequently the inverse temperature 
four-vector now is
$\beta^\mu \equiv \hat{\beta}_u u^\mu+ \hat{\beta}_{l} l^\mu $, and it is normalized to 
$\beta^\mu \beta_\mu = 1/T_u^2 - 1/T_l^2 >0$.
Here $\hat{\beta}_u \equiv \beta$ is the reciprocal of the J\"uttner temperature, 
and  $\hat{\beta}_l \equiv 1/T_l = \beta \xi$ is the inverse temperature along the anisotropy. 
The temperature ratio, $\xi = \hat{\beta}_{l}/\hat{\beta}_{u} = T_u/T_l$, 
defines the so-called anisotropy parameter that quantifies the strength of the anisotropy.
This straightforward  anisotropic decomposition of $\beta^\mu$  defines an anisotropic J\"uttner distribution function, $\hat{f}_{0\mathbf{k}} = \left[\exp \left( \hat{\beta}_{u}E_{\mathbf{k}u} 
- \hat{\beta}_{l}E_{\mathbf{k}l} -\hat{\alpha} \right) +a\right] ^{-1}$, which in the limit of 
$\hat{\beta}_l \rightarrow 0$ leads to the local equilibrium distribution
function~(\ref{f_0k}).

Another simple anisotropic distribution function can be constructed by modifying
the argument of the equilibrium distribution function $\beta^\mu k_\mu
\rightarrow \sqrt{k^\mu \Omega_{\mu \nu} k^\mu}$, where $\Omega^{\mu \nu } =
u^{\mu }u^{\nu }+\xi l^{\mu }l^{\nu }$; hence $\hat{\beta}_u = \beta $ and $%
\hat{\beta}_l = \beta \sqrt\xi$. This then yields the well-known spheroidal or
RS distribution function~\cite{Romatschke:2003ms},
which will be used and discussed in this paper. Other anisotropic
distribution functions are well-known in plasma physics, where the momentum
anisotropy is induced by the presence of an external magnetic field, e.g.,
the bi-Maxwellian distribution function, etc. 
Note that the main results in this paper, i.e., Sec.~\ref{sect:Binary_collision_integral},
are valid for the arbitrary anisotropic 
distribution function satisfying Eq.~(\ref{hat_f->f_0}), and we only explicitly 
use the RS distribution function starting from Sec.~\ref{sec:Applications}.

The moments of tensor rank $n$ of the anisotropic distribution function 
$\hat{f}_{0\mathbf{k}}$ are defined as 
\begin{align}
\hat{\mathcal{I}}_{ij}^{\mu _{1}\cdots \mu _{n}} & \equiv \int dK E_{\mathbf{k}u}^{i}
E_{\mathbf{k}l}^{j} k^{\mu _{1}}\cdots k^{\mu _{n}}\hat{f}_{0\mathbf{k}}
\notag \\
& =\sum_{q=0}^{\lfloor n/2\rfloor }\sum_{r=0}^{n-2q}\left( -1\right)
^{q}b_{nrq}\hat{I}_{i+j+n,j+r,q}\Xi ^{\left( \mu _{1}\mu _{2}\right. }\cdots
\Xi ^{\mu _{2q-1}\mu _{2q}}l^{\mu _{2q+1}}\cdots l^{\mu _{2q+r}} 
u^{\mu_{2q+r+1}}\cdots u^{\left. \mu _{n}\right) }\, ,  \label{I_ij_tens}
\end{align}
where $i\geq-1$ and $j\geq0$ define the power of $E_{\mathbf{k}u}$ and $E_{\mathbf{k}l}$, 
while $dK=gd^{3}\mathbf{k}/\left[ \left( 2\pi \right) ^{3}k^{0}\right] $ is
the invariant measure and $g$ is the degeneracy of the state. 
Here $\lfloor n/2\rfloor $ denotes the greatest integer less than equal to $n/2$,
i.e., the floor function, and the round brackets around indices denote the symmetrization 
of indices, i.e., $T^{\left( \mu _{1}\cdots \mu _{n}\right) } 
= \frac{1}{n!}\sum T^{\mu_{1}\mu _{2}\cdots \mu _{n}}$.

The anisotropic thermodynamic integrals $\hat{I}_{nrq}$ are defined as in
Refs.~\cite{Molnar:2016vvu,Molnar:2016gwq} 
\begin{equation}
\hat{I}_{nrq}\left( \hat{\alpha},\hat{\beta}_{u},\hat{\beta}_{l}\right) 
\equiv \frac{\left( -1\right) ^{q}}{\left( 2q\right) !!}\int dK 
E_{\mathbf{k}u}^{n-r-2q} E_{\mathbf{k}l}^{r} 
\left( \Xi ^{\mu \nu }k_{\mu }k_{\nu }\right)^{q}\hat{f}_{0\mathbf{k}}\, ,  
\label{I_nrq}
\end{equation}%
where $\left( 2q\right) !!=2^{q}q!$ is the double factorial of even numbers,
while the $b_{nrq}$ coefficient counts the distinct terms of the symmetrized
product $\Xi ^{\left( \mu _{1}\mu _{2}\right. }\cdots l^{\mu _{2q+r}}\cdots
u^{\left. \mu _{n}\right)}$, 
\begin{equation}
b_{nrq}\equiv \frac{n!}{2^{q}q!r!\left( n-r-2q\right) !}=\frac{n!\left(
2q-1\right) !!}{\left( 2q\right) !r!\left( n-r-2q\right) !}\, ,  
\label{b_nrq}
\end{equation}
where the double factorial of odd numbers is 
$\left( 2q-1\right) !!=\left(2q\right) !/\left( 2^{q}q!\right)$.

The equilibrium moments of tensor rank $n$ and of power $r$ in energy 
$E_{\mathbf{k}u}$ are defined as 
\begin{align}
\mathcal{I}_{i}^{\mu _{1}\cdots \mu _{n}}& \equiv \int dK 
E_{\mathbf{k}u}^{i} k^{\mu _{1}}\cdots k^{\mu _{n}} f_{0\mathbf{k}}  \notag \\
&= \sum_{q=0}^{\lfloor n/2\rfloor }\left( -1\right)^{q} b_{nq} 
I_{i+n,q}\Delta ^{\left( \mu _{1}\mu _{2}\right. }\cdots 
\Delta^{\mu _{2q-1}\mu _{2q}}u^{\mu _{2q+1}}\cdots u^{\left. \mu _{n}\right) }\, ,
\label{I_i_tens}
\end{align}%
where $i\geq -1$ and the equilibrium thermodynamic integrals are defined as 
\begin{equation}
I_{nq}\left( \alpha ,\beta \right) \equiv \frac{\left( -1\right) ^{q}}{\left( 2q+1\right) !!}
\int dKE_{\mathbf{k}u}^{n-2q}\left( \Delta ^{\mu \nu}k_{\mu }k_{\nu }\right)^{q}f_{0\mathbf{k}}\, ,  \label{I_nq}
\end{equation}%
while the number of distinct terms in the symmetrized tensor product~(\ref{I_i_tens}) is
given by Eq.~(\ref{b_nrq}) for $r=0$, 
\begin{equation}
b_{nq}\equiv b_{n0q}=\frac{n!}{2^{q}q!\left( n-2q\right) !}\, .  \label{b_nq}
\end{equation}%
The anisotropic thermodynamic integrals in equilibrium $I_{nrq}\left( \alpha,\beta \right)$ 
are defined using Eq.~(\ref{I_nrq}) with $\hat{f}_{0\mathbf{k}}\rightarrow f_{0\mathbf{k}}$, 
and hence from Eq.~(\ref{hat_f->f_0}) it also follows that 
$I_{nrq}=\lim_{\hat{\beta}_{l}\rightarrow 0}\hat{I}_{nrq}$. 
Furthermore, replacing Eq.~(\ref{Xi_munu}) into Eq.~(\ref{I_nq}) we find
the following relation between the equilibrium thermodynamic integrals, 
$I_{nq} =\frac{1}{\left( 2q+1\right) !!}\sum_{r=0}^{q} \frac{q!}{r!} 2^{q-r} I_{n,2r,q-r}$.

Note that the local-equilibrium distribution function is isotropic in
momentum space, and thus the orthogonal projections of the equilibrium moments
vanish for any tensor rank larger than one; i.e., for $n\geq 1$ we have 
$\mathcal{I}_{r}^{\left\langle \mu_{1}\right\rangle \cdots \left\langle
\mu_{n}\right\rangle }=0$. Similarly, the orthogonal projections with respect 
to both $u^{\mu }$ and $l^{\mu }$ of the anisotropic moments also vanish 
$\hat{\mathcal{I}}_{ij}^{\left\{ \mu_{1}\right\} \cdots \left\{\mu_{n}\right\} }=0$.

\section{The conservation equations of leading-order anisotropic fluid dynamics}
\label{sec:aniso_hydro}

The relativistic Boltzmann equation for the space-time evolution of the
single-particle distribution function of classical indistinguishable
particles corresponding to $f_{\mathbf{k}}\equiv \hat{f}_{0\mathbf{k}}
\left( \hat{\alpha},\hat{\beta}_{u}E_{\mathbf{k}u},\hat{\beta}_{l}E_{\mathbf{k}l}\right)$ 
reads, 
\begin{equation}
k^{\mu }\partial _{\mu }\hat{f}_{0\mathbf{k}}= C\!\left[ \hat{f}_{0}\right] \, ,  
\label{BTE_aniso}
\end{equation}%
where $C\left[ \hat{f}_{0}\right] $ is the Boltzmann collision term specified in Sec.~\ref{sect:Binary_collision_integral}.

Relativistic fluid dynamics is an effective theory for the long-wavelength and low-frequency dynamics of macroscopic systems. The fluid-dynamical
equations of motion are derived from the Boltzmann equation
by integrating over the microscopic equation~(\ref{BTE_aniso}) with respect to the particle momentum. 
This leads to an infinite hierarchy of moment equations being equivalent to the
Boltzmann equation.
The lower-order moments of the distribution function, i.e., lower power of $E_{\mathbf{k}u}$ and $E_{\mathbf{k}l}$, describe lower frequency or longer wavelength dynamics while higher-order moments  capture the higher frequency or shorter wavelength dynamics.
For an anisotropic distribution function we obtain
\begin{equation}
\partial _{\nu }\hat{\mathcal{I}}_{00}^{\mu _{1}\cdots \mu _{n}\nu }=\hat{C}%
_{00}^{\mu _{1}\cdots \mu _{n}}\, ,  \label{Hierarchy_moments}
\end{equation}%
where the rank $n$ moment of the collision integral is defined as 
\begin{equation}
\hat{C}_{ij}^{\mu _{1}\cdots \mu _{n}}\equiv \int dKE_{\mathbf{k}u}^{i} 
E_{\mathbf{k}l}^{j} k^{\mu _{1}}\cdots k^{\mu _{n}}C\!\left[ \hat{f}_{0}\right]\, .  
\label{hat_coll_int}
\end{equation}%
The general equations of motion which follow from Eq.~(\ref{Hierarchy_moments}) 
in terms of scalar moments $\hat{\mathcal{I}}_{ij}$ were derived in Refs.~\cite{Molnar:2016gwq,Molnar:2024fgy} and for reasons of completeness are presented 
in Appendix~\ref{app:equations_of_motion}.

The particle four-current conservation, and the energy-momentum conservation follow from Eq.~(\ref{Hierarchy_moments}),
\begin{align}
\partial _{\mu }\hat{N}^{\mu }&\equiv \partial_{\mu }\hat{\mathcal{I}}^{\mu }_{00} 
= \hat{C}_{00}=0\, , \\
\partial _{\mu }\hat{T}^{\mu \nu } &\equiv \partial_{\mu}\hat{\mathcal{I}}^{\mu \nu}_{00}  
= \hat{C}_{00}^{\nu }=0\, ,
\end{align}%
where the conservation of particle number or charge, the conservation
of energy, the conservation of momentum in the direction of the momentum
anisotropy as well as transverse to it leads to 
\begin{equation}
\hat{C}_{00}=0\, ,\quad \hat{C}_{10}=0\, ,\quad \hat{C}_{01}=0\, ,\quad 
\hat{C}_{00}^{\{\mu \}}=0\, .  \label{C_0_C_1_C_0_mu_aniso}
\end{equation}%
The tensor decompositions~(\ref{I_ij_tens}) of the primary or the lowest-order moments of the anisotropic
distribution function in leading-order anisotropic fluid dynamics give
\begin{align}
\hat{N}^{\mu } &\equiv \hat{\mathcal{I}}_{00}^{\mu }=\hat{n}u^{\mu } 
+\hat{n}_{l}l^{\mu }\, ,  \label{hat_N_0_mu} \\
\hat{T}^{\mu \nu } &\equiv \hat{\mathcal{I}}_{00}^{\mu \nu } 
=\hat{e}u^{\mu}u^{\nu }+2\hat{M}u^{\left( \mu \right. }l^{\left. \nu \right) } 
+\hat{P}_{l}l^{\mu }l^{\nu }-\hat{P}_{\perp }\Xi ^{\mu \nu }\, ,
\label{hat_T_0_munu}
\end{align}%
where the particle density and energy density are $\hat{n}$ and $\hat{e}$,
respectively. The particle diffusion current along the $l^{\mu }$ direction
is denoted by $\hat{n}_{l}$, while the energy-momentum diffusion current is
denoted by $\hat{M}$. The pressure in the direction of the anisotropy is $\hat{P}_{l}$, 
while the pressure in the direction transverse to it is denoted by $\hat{P}_{\perp }$. 
These quantities are expressed in terms of different tensor 
projections or equivalently through Eq.~(\ref{I_ij_tens}) as 
\begin{align}
\hat{n} &\equiv \hat{N}^{\mu }u_{\mu }=\hat{\mathcal{I}}_{10}=\hat{I}_{100}\, ,  
\label{hat_n} \\
\hat{n}_{l} &\equiv -\hat{N}^{\mu }l_{\mu }=\hat{\mathcal{I}}_{01}=\hat{I}_{110}\, ,  
\label{hat_n_l} \\
\hat{e} &\equiv \hat{T}^{\mu \nu }u_{\mu }u_{\nu }=\hat{\mathcal{I}}_{20}=\hat{I}_{200}\, ,  
\label{hat_e} \\
\hat{M} &\equiv -\hat{T}^{\mu \nu }u_{\mu }l_{\nu }=\hat{\mathcal{I}}_{11}=\hat{I}_{210}\, ,  
\label{hat_M} \\
\hat{P}_{l} &\equiv \hat{T}^{\mu \nu }l_{\mu }l_{\nu }=\hat{\mathcal{I}}_{02}=\hat{I}_{220}\, ,  
\label{hat_P_l} \\
\hat{P}_{\perp } &\equiv -\frac{1}{2}\hat{T}^{\mu \nu }\Xi _{\mu \nu }
=-\frac{1}{2}\left( m_{0}^{2}\hat{\mathcal{I}}_{00}-\hat{\mathcal{I}}_{20}
+\hat{\mathcal{I}}_{02}\right) =\hat{I}_{201}\, ,  
\label{hat_P_T}
\end{align}%
while the isotropic pressure is $\hat{P}\equiv -\frac{1}{3}\hat{T}^{\mu \nu}
\Delta _{\mu \nu }=\frac{1}{3}\left( \hat{P}_{l}+2\hat{P}_{\perp }\right) $. 
The fluid dynamical four-velocity and the local rest frame is usually
chosen according to the definition of Landau and Lifshitz~\cite{Landau_book}; i.e., 
$u^{\mu }=\hat{T}^{\mu \nu }u_{\nu }/(u_{\alpha }\hat{T}^{\alpha \beta}u_{\beta })$ 
which leads to a vanishing energy diffusion current; i.e., $\hat{M}=0$.

The particle four-current and the energy-momentum tensor in local
thermodynamic equilibrium are 
\begin{align}
N_{0}^{\mu } &\equiv \mathcal{I}_{0}^{\mu }=n_{0}u^{\mu }\,,  
\label{N_0_mu} \\
T_{0}^{\mu \nu } &\equiv \mathcal{I}_{0}^{\mu \nu }=e_{0}u^{\mu }u^{\nu}-P_{0}\Delta^{\mu \nu }\, ,  \label{T_0_munu}
\end{align}%
where the particle density, the energy density, and the pressure in local
thermal equilibrium are obtained either by different projections of the
conserved quantities or through Eq.~(\ref{I_i_tens}) leading to
\begin{align}
n_{0} &\equiv N_{0}^{\mu }u_{\mu }=\mathcal{I}_{1}=I_{10}\,,  \label{n_0} \\
e_{0} &\equiv T_{0}^{\mu \nu }u_{\mu }u_{\nu }=\mathcal{I}_{2}=I_{20}\,,
\label{e_0} \\
P_{0} &\equiv -\frac{1}{3}T_{0}^{\mu \nu }\Delta _{\mu \nu }=-\frac{1}{3}%
\left( m_{0}^{2}\mathcal{I}_{0}-\mathcal{I}_{2}\right) =I_{21}\,.
\label{P_0}
\end{align}%
The pressure $P_{0}\equiv P_{0}\left(e_{0},n_{0}\right)$ is specified by an
equation of state (EoS) and hence the temperature $T(e_{0},n_{0})$ and
chemical potential $\mu (e_{0},n_{0})$ are also determined. 
In an arbitrary anisotropic state, $\hat{\alpha}$ and $\hat{\beta}_{u}$, are obtained from
the Landau matching conditions $\left( \hat{N}^{\mu }-N_{0}^{\mu }\right)
u_{\mu }\equiv \hat{n}-n_{0}=0$ and $\left( \hat{T}^{\mu \nu }-T_{0}^{\mu\nu }\right) 
u_{\mu }u_{\nu }\equiv \hat{e}-e_{0}=0$ leading to 
$\mu (\hat{\alpha},\hat{\beta}_{u},\hat{\beta}_{l})$ and 
$T(\hat{\alpha},\hat{\beta}_{u},\hat{\beta}_{l})$, while the new parameter 
$\hat{\beta}_{l}$ must be determined from an additional equation of motion.
Note that the conservation equations, $\partial _{\mu }\hat{N}^{\mu }=0$
and $\partial _{\mu }\hat{T}^{\mu \nu }=0$, 
provide only five constraints for the six independent variables, and hence
we need an additional equation of motion to determine the remaining variable. 
Naturally, this can be provided by choosing other equation(s) from
the infinite hierarchy of moment equations of the Boltzmann equation~(\ref{Hierarchy_moments}). 
This is studied and discussed in detail in Sec.~\ref{sec:Results_BJ} and Sec.~\ref{sec:Results_BJ_two}.

\section{The moments of the binary collision integral}
\label{sect:Binary_collision_integral}

The collision term of the relativistic Boltzmann equation~(\ref{BTE_aniso})
in the case of binary elastic collisions is defined as~\cite%
{deGroot_book,Cercignani_book,Denicol_Rischke_book} 
\begin{equation}
C\!\left[ \hat{f}_{0}\right] \equiv \frac{1}{2}\int dK^{\prime }dP dP^{\prime}
W_{\mathbf{kk}^{\prime }\rightarrow \mathbf{pp}^{\prime }} 
\left( \hat{f}_{0\mathbf{p}}\hat{f}_{0\mathbf{p}^{\prime }} 
-\hat{f}_{0\mathbf{k}}\hat{f}_{0\mathbf{k}^{\prime }}\right) \, ,  
\label{binary_collision_integral}
\end{equation}%
where the $1/2$ factor removes the double counting due to the
indistinguishability of identical particles. 
The invariant transition rate satisfies detailed balance, 
$W_{\mathbf{kk}^{\prime }\rightarrow \mathbf{pp}^{\prime }}
=W_{\mathbf{pp}^{\prime }\rightarrow \mathbf{kk}^{\prime }}$, and
it is also symmetric with respect to the exchange of particle four-momentum
in binary collisions, $W_{\mathbf{kk}^{\prime }\rightarrow \mathbf{pp}^{\prime }}
=W_{\mathbf{k}^{\prime }\mathbf{k}\rightarrow \mathbf{pp}^{\prime}}
=W_{\mathbf{kk}^{\prime }\rightarrow \mathbf{p}^{\prime }\mathbf{p}}$. 
The invariant transition rate for elastic binary collisions is defined as 
\begin{equation}
g^{2}W_{\mathbf{kk}^{\prime }\rightarrow \mathbf{pp}^{\prime }}\equiv s(2\pi
)^{6}\frac{d\sigma (s,\Omega )}{d\Omega }\delta (k^{\mu }+k^{\prime \mu
}-p^{\mu }-p^{\prime \mu })\, ,  
\label{W_kk_pp}
\end{equation}%
where the conservation of the energy and momentum in binary collision is
enforced by the Dirac delta function. 
The transition rate depends on the total center-of-momentum (CM) energy squared 
$s\equiv (k^{\mu }+k^{\prime\mu })^{2}=(p^{\mu }+p^{\prime \mu })^{2}$, 
while the total cross section integrated over the solid angle $\Omega $ 
is defined as 
\begin{equation}
\sigma _{T}(s)\equiv \frac{2\pi }{2}\int_{0}^{\pi } d\theta _{s}\sin \theta_{s} 
\frac{d\sigma (s,\Omega )}{d\Omega }\,,
\end{equation}%
where $\theta _{s} \equiv \arccos \frac{(k^{\mu }-k^{\prime \mu })(p^{\mu
}-p^{\prime \mu })}{(k^{\mu }-k^{\prime \mu })^{2}}$ is the scattering
angle. From now on the so-called hard-sphere approximation will be used
assuming that the total cross section is isotropic and independent of
the total CM energy, 
\begin{equation}
\sigma _{T}\equiv 2\pi \frac{d\sigma (s,\Omega )}{d\Omega }=\frac{1}{\hat{n}%
\lambda _{\mathrm{mfp}}}\,,  \label{sigma_T}
\end{equation}%
where $\hat{n}$ is the particle density and $\lambda _{\mathrm{mfp}}$ is the
mean free path between collisions.

The moments~(\ref{hat_coll_int}) of tensor rank $n$ of the binary collision integral~(\ref{binary_collision_integral}) naturally 
separates into gain and loss parts, 
\begin{equation}
\hat{C}_{ij}^{\mu _{1}\cdots \mu _{n}}\equiv \int dKE_{\mathbf{k}u}^{i}
E_{\mathbf{k }l}^{j}k^{\mu _{1}}\cdots k^{\mu _{n}}C\!\left[ \hat{f}_{0}\right]
=\hat{G}_{ij}^{\mu _{1}\cdots \mu _{n}}-\hat{L}_{ij}^{\mu _{1}\cdots\mu_{n}} \, ,  
\label{C_ij_general}
\end{equation}%
where using Eq.~(\ref{binary_collision_integral}) and the symmetries of the
transition rate the corresponding gain and loss moments of tensor rank $n$ are defined as 
\begin{align}
\hat{G}_{ij}^{\mu _{1}\cdots \mu _{n}} 
&\equiv \frac{1}{2} \int_{K\!K^{\prime}} \!\int_{P\! P^{\prime }} 
W_{\mathbf{kk}^{\prime }\rightarrow \mathbf{pp}^{\prime }} 
\hat{f}_{0\mathbf{k}}\hat{f}_{0\mathbf{k}^{\prime }}E_{\mathbf{p}u}^{i} 
E_{\mathbf{p}l}^{j}p^{\mu _{1}}\cdots p^{\mu _{n}} \, ,  
\label{Gain_ij} \\
\hat{L}_{ij}^{\mu _{1}\cdots \mu _{n}} 
&\equiv \frac{1}{2}\int_{K\!K^{\prime}} \!\int_{P\! P^{\prime }} 
W_{\mathbf{kk}^{\prime }\rightarrow \mathbf{pp}^{\prime }} 
\hat{f}_{0\mathbf{k}}\hat{f}_{0\mathbf{k}^{\prime }}
E_{\mathbf{k}u}^{i}E_{\mathbf{k}l}^{j}k^{\mu _{1}}\cdots k^{\mu _{n}} \, ,
\label{Loss_ij}
\end{align}%
where we introduced the shorthand notation $\int dKdK^{\prime} =\int_{K\!K^{\prime }}$ 
and $\int dP dP^{\prime} = \int_{P\!P^{\prime }}$.

To evaluate these 12-dimensional integrals, first we need to
compute the following simpler 6-dimensional integrals over $P$ and $P^{\prime }$:
\begin{eqnarray}
\mathcal{P}_{ij}^{\mu _{1}\cdots \mu _{n}} &\equiv &\frac{1}{2} \int_{P\!
P^{\prime }} W_{\mathbf{kk}\prime \rightarrow \mathbf{pp}\prime }E_{\mathbf{p%
}u}^{i}E_{\mathbf{p}l}^{j}p^{\mu _{1}}\cdots p^{\mu _{n}}  \notag \\
&=&\left( -1\right) ^{j}u_{\mu _{1}}\cdots u_{\mu _{i}}l_{\mu _{i+1}}\cdots
l_{\mu _{i+j}}\Theta ^{\mu _{1}\cdots \mu _{n+i+j}}\, ,  \label{P_ij}
\end{eqnarray}%
which in turn require the calculation of the following unprojected auxiliary integrals 
(see also Refs.~\cite{Molnar:2013lta,Wagner:2023joq}):
\begin{align}
\Theta ^{\mu _{1}\cdots \mu _{n}} &\equiv 
\frac{1}{2}\int_{P\! P^{\prime }} W_{\mathbf{kk}\prime \rightarrow \mathbf{pp}\prime }
p^{\mu _{1}}\cdots p^{\mu _{n}} \notag \\
&= \sum_{q=0}^{\lfloor n/2\rfloor }\left( -1\right)^{q} b_{nq}\mathcal{B}_{nq} 
\Delta _{T}^{\left( \mu_{1}\mu_{2}\right. }\cdots \Delta_{T}^{\mu_{2q-1}\mu_{2q}} 
P_{T}^{\mu _{2q+1}}\cdots P_{T}^{\left. \mu_{n}\right) }\, .  
\label{Theta_mu1_mun}
\end{align}%
Here the $b_{nq}$ coefficient was defined in Eq.~(\ref{b_nq}), while the 
$\mathcal{B}_{nq}$ coefficient is defined similar to Eq.~(\ref{I_nq})
(see Appendix~\ref{app:CM_frame} for more details), and it reads
\begin{eqnarray}
\mathcal{B}_{nq} &\equiv &\frac{\left( -1\right) ^{q}}{2\left( 2q+1\right) !!}
\int_{P\! P^{\prime }} W_{\mathbf{kk}\prime \rightarrow \mathbf{pp}\prime}
\left( \frac{P_{T}^{\mu }p_{\mu }}{s}\right) ^{n-2q} 
\left( \Delta _{T}^{\mu\nu }p_{\mu }p_{\nu }\right) ^{q}  \notag \\
&=&\frac{\sigma _{T}}{2^{n+1}\left( 2q+1\right) !!}\sqrt{s} 
\left( \sqrt{s-4m_{0}^{2}}\right) ^{\left( 2q+1\right) }\, ,  
\label{B_nq}
\end{eqnarray}%
which in the ultrarelativistic limit when $m_0\rightarrow 0$ and hence when 
$s\equiv 2k^{\mu }k_{\mu }^{\prime }$ leads to
\begin{equation}
\lim_{m_{0}\rightarrow 0}\mathcal{B}_{nq}=\frac{\sigma _{T}}{2^{n+1}\left(
2q+1\right) !!}\left( \sqrt{s}\right) ^{2+2q}\, .  
\label{B_nq_massless}
\end{equation}
Note that the $\mathcal{P}_{ij}^{\mu_{1}\cdots \mu_{n}}$ integrals are built
contracting the $\Theta ^{\mu_{1}\cdots \mu_{n}}$ tensors by the
four-vectors $u^\mu$ and/or $l^\mu$. Therefore integrals containing negative
powers of energy, e.g., $E^{-1}_{\mathbf{p}u}$, are not possible to obtain through 
these tensor projections.

\subsection{The loss terms}
\label{sec:Loss_Terms}

Using the results of the previous section, the loss term from Eq.~(\ref{Loss_ij}) 
is easily computed in the massless limit, noting that for $i=j=0$ 
the $P$ and $P^{\prime }$ integral from Eq.~(\ref{P_ij}) leads
to $\mathcal{P}_{00}\equiv \Theta =\sigma _{T}s/2 \equiv \sigma_{T} 
k^{\lambda}k_{\lambda }^{\prime }$, and hence the remaining $K$ and $K^{\prime }$ 
integrals evaluate to
\begin{align}
\hat{L}_{ij}^{\mu _{1}\cdots \mu _{n}} &\equiv \int_{K\!K^{\prime }\!}
\hat{f}_{0\mathbf{k}}\hat{f}_{0\mathbf{k}^{\prime }}E_{\mathbf{k}u}^{i} E_{\mathbf{k}l}^{j} 
k^{\mu _{1}}\cdots k^{\mu _{n}}\mathcal{P}_{00}  \notag \\
&=\sigma _{T}\int dK\hat{f}_{0\mathbf{k}}E_{\mathbf{k}u}^{i}E_{\mathbf{k}l}^{j} 
k^{\mu _{1}}\cdots k^{\mu _{n}}k^{\lambda } \! \int dK^{\prime } 
\hat{f}_{0\mathbf{k}^{\prime }}k_{\lambda }^{\prime }  \notag \\
&= \sigma _{T}\hat{\mathcal{I}}_{ij}^{\mu _{1}\cdots \mu _{n}\lambda }
\hat{\mathcal{I}}_{00,\lambda }\, ,
\end{align}%
where in the last step we used the definition of the moments of the
anisotropic distribution function from Eq.~(\ref{I_ij_tens}). 
Now using the decomposition of the particle four-current from Eq.~(\ref{hat_N_0_mu}) 
we obtain (see Appendix~\ref{app:loss_terms} for the details of the derivation)
\begin{align}
\hat{L}_{ij}^{\mu _{1}\cdots \mu _{n}} 
&=\sigma _{T}\sum_{q=0}^{\lfloor n/2\rfloor }\sum_{r=0}^{n-2q}\left( -1\right) ^{q} b_{nrq} 
\left( \hat{I}_{i+j+n+1,j+r,q}\hat{I}_{100}-\hat{I}_{i+j+n+1,j+r+1,q}\hat{I}_{110}\right) 
\notag \\
&\times \Xi ^{\left( \mu _{1}\mu _{2}\right. }\cdots \Xi ^{\mu _{2q-1} \mu_{2q}} 
l^{\mu _{2q+1}}\cdots l^{\mu _{2q+r}}u^{\mu _{2q+r+1}}\cdots u^{\left. \mu _{n}\right) }\, .  
\label{L_ij_final}
\end{align}%
Note that in leading-order anisotropic fluid dynamics we only need to 
compute the scalar and vector moments of the collision integral.
However, we can easily show that the orthogonal projection of the rank-one moment loss 
term vanishes equivalently 
$\Xi _{\alpha }^{\mu }\hat{L}_{ij}^{\alpha } \equiv \hat{L}_{ij}^{\{ \mu \}} =0$.
Recalling that the vector moment of the loss term,
$\hat{L}_{ij}^{\alpha }= (\cdots)u^{\alpha} + (\cdots) l^{\alpha}$, is proportional to both 
$u^{\alpha}$ and $l^{\alpha}$; consequently, when contracted by the
elementary projection operator $\Xi _{\alpha }^{\mu }$ which
is orthogonal to both $u^{\alpha}$ and $l^{\alpha}$, leads to zero.
Furthermore, the rank-one gain term $\hat{G}_{ij}^{\alpha }$ shares the same tensor structure 
as the loss term; therefore, in summary, the orthogonal projection of the rank-one moment of 
the collision integral vanishes in leading-order anisotropic fluid dynamics; i.e., 
$\Xi_{\alpha }^{\mu }\hat{C}_{ij}^{\alpha} \equiv \hat{C}_{ij}^{\{ \mu \}} = 0$.

Thus, for the purposes of this paper dealing with leading-order anisotropic
fluid dynamics we only need to compute the scalar moments of the collision
integral, i.e., when $n=r=q=0$, and hence all scalar loss terms read 
\begin{align}
\hat{L}_{ij} &\equiv \frac{1}{2}\int_{K\!K^{\prime }\!}\int_{P\!P^{\prime}\!} 
W_{\mathbf{kk}^{\prime }\rightarrow \mathbf{pp}^{\prime }} \hat{f}_{0\mathbf{k}}
\hat{f}_{0\mathbf{k}^{\prime }}E_{\mathbf{k}u}^{i} E_{\mathbf{k}l}^{j}  \notag \\
&= \sigma _{T}\left( \hat{I}_{i+j+1,j,0}\hat{n}-\hat{I}_{i+j+1,j+1,0}\hat{n}_{l}\right) \, .
\end{align}%
We are only interested in the first few entries of the collision matrix
which we will compute here explicitly. For the values $i=0,1,2,3,4$ and $j=0$
the loss term leads to 
\begin{eqnarray}
\hat{L}_{00} &=&\sigma _{T}\left( \hat{n}^{2}-\hat{n}_{l}^{2}\right) \, ,
\label{L00} \\
\hat{L}_{10} &=&\sigma _{T}\left( \hat{e}\hat{n}-\hat{M}\hat{n}_{l}\right)\, ,  
\label{L10} \\
\hat{L}_{20} &=&\sigma _{T}\left( \hat{I}_{300}\hat{n}-\hat{I}_{310}\hat{n}_{l}\right) \, ,  
\label{L20} \\
\hat{L}_{30} &=&\sigma _{T}\left( \hat{I}_{400}\hat{n}-\hat{I}_{410}\hat{n}_{l}\right) \, ,  
\label{L30} \\
\hat{L}_{40} &=&\sigma _{T}\left( \hat{I}_{500}\hat{n}-\hat{I}_{510}\hat{n}_{l}\right) \, .  
\label{L40}
\end{eqnarray}%
Similarly for $i=0,1,2,3$ and $j=1$ we have%
\begin{eqnarray}
\hat{L}_{01} &=&\sigma _{T}\left( \hat{M}\hat{n}-\hat{P}_{l}\hat{n}_{l}\right) \, ,  
\label{L01} \\
\hat{L}_{11} &=&\sigma _{T}\left( \hat{I}_{310}\hat{n}-\hat{I}_{320}\hat{n}_{l}\right) \, ,  
\label{L11} \\
\hat{L}_{21} &=&\sigma _{T}\left( \hat{I}_{410}\hat{n}-\hat{I}_{420}\hat{n}_{l}\right) \, ,  
\label{L21} \\
\hat{L}_{31} &=&\sigma _{T}\left( \hat{I}_{510}\hat{n}-\hat{I}_{520}\hat{n}_{l}\right) \, ,  
\label{L31}
\end{eqnarray}%
and finally for $i=0,1,2$ and $j=2$ we have%
\begin{eqnarray}
\hat{L}_{02} &=&\sigma _{T}\left( \hat{I}_{320}\hat{n}-\hat{I}_{330}\hat{n}_{l}\right) \, ,  
\label{L02} \\
\hat{L}_{12} &=&\sigma _{T}\left( \hat{I}_{420}\hat{n}-\hat{I}_{430}\hat{n}_{l}\right) \, ,  
\label{L12} \\
\hat{L}_{22} &=&\sigma _{T}\left( \hat{I}_{520}\hat{n}-\hat{I}_{530}\hat{n}_{l}\right) \, .  
\label{L22}
\end{eqnarray}%
Note that the loss terms $\hat{L}_{00}$ and $\hat{L}_{10}$ correspond to
the moments of the particle number and energy conservation equations, while $\hat{L}_{01}$
to the moment of the momentum conservation in the direction of the anisotropy.

\subsection{The gain terms}
\label{sec:Gain_Terms}

The gain terms are substantially more difficult to calculate than the loss
terms due to the complicated structure of the $P$ and $P^{\prime }$
integrals; i.e., now we need to compute $\mathcal{P}_{ij}$ for $i>0$ and $j \geq 0$, 
while we only need $\mathcal{P}_{00}$ for the loss terms. 
For the purposes of this paper dealing with
leading-order anisotropic distribution functions we only need to evaluate
the scalar gain term $\mathcal{P}_{ij}$; hence, now Eq.~(\ref{Gain_ij}) reads
\begin{equation}
\hat{G}_{ij}\equiv \int_{K\!K^{\prime }\!}\hat{f}_{0\mathbf{k}} 
\hat{f}_{0\mathbf{k}^{\prime }}\mathcal{P}_{ij} 
=\left( -1\right)^{j} \int_{K\!K^{\prime }\!}\hat{f}_{0\mathbf{k}} 
\hat{f}_{0\mathbf{k}^{\prime}}
\Theta ^{\mu _{1}\cdots \mu _{i+j}}u_{\mu _{1}}\cdots u_{\mu _{i}}
l_{\mu_{i+1}}\cdots l_{\mu _{i+j}}\, ,  
\label{hat_G_ij}
\end{equation}%
where the unprojected auxiliary integrals were defined in Eq.~(\ref{Theta_mu1_mun}). 
Note that although it is possible to compute the gain terms to arbitrary tensor rank, 
for practical purposes here we only compute a small subset of scalar gain terms explicitly.
These follow from various projections with up to tensor rank four of the auxiliary
integrals $\Theta^{\mu _{1}\mu _{2}\mu _{3}\mu _{4}}$ by the four-vectors $u^\mu$ and/or $l^\mu$.

We start by computing the simplest scalar gain term which only contains
projections in the direction of the fluid four-velocity and no projections in the
direction of the anisotropy.
Therefore, substituting $i=n$ and $j=0$ into Eq.~(\ref{hat_G_ij}) (see Appendix~\ref{app:Gn0} for the derivation), we obtain
\begin{align}
\hat{G}_{n0} &\equiv \int_{K\!K^{\prime }\!}\hat{f}_{0\mathbf{k}} 
\hat{f}_{0\mathbf{k}^{\prime }} \mathcal{P}_{n0}
=\int_{K\!K^{\prime }}\hat{f}_{0\mathbf{k}}\hat{f}_{0\mathbf{k}^{\prime }}
\Theta ^{\mu _{1}\cdots \mu _{n}}u_{\mu_{1}}\cdots u_{\mu _{n}}  \notag \\
&= \int_{K\!K^{\prime }\!}\hat{f}_{0\mathbf{k}}\hat{f}_{0\mathbf{k}^{\prime}}
\sum_{q=0}^{\lfloor n/2\rfloor }\left( -1\right) ^{q}b_{nq}\mathcal{B}_{nq} 
\left(P_{T}^{\mu }u_{\mu }\right)^{n-2q}\left(\Delta _{T}^{\mu \nu}u_{\mu }u_{\nu }\right)^{q}\, .  \label{Gn0}
\end{align}%
Now using this result in the massless limit when $s\equiv 2k^{\mu}k_{\mu}^{\prime }$ 
together with the definitions,
\begin{align}
P_{T}^{\mu }u_{\mu } &= E_{\mathbf{k}u}+E_{\mathbf{k}^{\prime }u}\, , 
\label{PT_u} \\
\Delta _{T}^{\mu \nu }u_{\mu }u_{\nu } &= 1-s^{-1}\left( E_{\mathbf{k}u} 
+E_{\mathbf{k}^{\prime }u}\right) ^{2}\, ,  
\label{DeltaT_uu}
\end{align}%
we get
\begin{equation}
\hat{G}_{n0} =\frac{\sigma _{T}}{2^{n}} 
\int_{K\!K^{\prime }\!}\hat{f}_{0\mathbf{k}} \hat{f}_{0\mathbf{k}^{\prime }} 
k^{\mu}k_{\mu }^{\prime}
\sum_{q=0}^{\lfloor n/2\rfloor }\frac{\left( -1\right)^{q}b_{nq}} {\left(2q+1\right) !!} 
\left( E_{\mathbf{k}u}+E_{\mathbf{k}^{\prime }u}\right)^{n-2q} \left( 2k^{\nu }k_{\nu }^{\prime } 
-\left( E_{\mathbf{k}u}+E_{\mathbf{k}^{\prime }u}\right)^{2}\right)^{q}\, .
\end{equation}

Thus, for $n=0$ and $n=1$, also see Appendix~\ref{app:Gn0} for more details, we obtain
\begin{align}
\hat{G}_{00}& =\sigma _{T}\left( \hat{n}^{2}-\hat{n}_{l}^{2}\right) \, ,
\label{G00} \\
\hat{G}_{10}& =\sigma _{T}\left( \hat{e}\hat{n}-\hat{M}\hat{n}_{l}\right) \, .  
\label{G10}
\end{align}%
These gain terms precisely correspond to loss terms in Eqs.~(\ref{L00}) and~(\ref{L10}) 
as they should according to the conservation of particle number and energy in binary collisions. 
The gain terms for $n=2$ and $n=3$ are 
\begin{align}
\hat{G}_{20}& =\frac{\sigma _{T}}{6}\left( 4\hat{I}_{300}\hat{n}-4\hat{I}%
_{310}\hat{n}_{l}+3\hat{e}^{2}\right) -\frac{\sigma _{T}}{6}\left( 2\hat{M}%
^{2}+\hat{P}_{l}^{2}+2\hat{P}_{\perp }^{2}\right) \, ,  \label{G20} \\
\hat{G}_{30}& =\frac{\sigma _{T}}{2}\left( \hat{I}_{400}\hat{n}-\hat{I}_{410}%
\hat{n}_{l}+2\hat{I}_{300}\hat{e}\right) -\frac{\sigma _{T}}{2}\left( \hat{I}%
_{310}\hat{M}+\hat{I}_{320}\hat{P}_{l}+2\hat{I}_{301}\hat{P}_{\perp}\right)
\, ,  \label{G30}
\end{align}%
and finally for $n=4$ we have
\begin{align}
\hat{G}_{40}& =\frac{\sigma _{T}}{5}\left( 2\hat{I}_{500}\hat{n}-2\hat{I}%
_{510}\hat{n}_{l}+5\hat{I}_{400}\hat{e}-2\hat{I}_{410}\hat{M}-3\hat{I}_{420}%
\hat{P}_{l}-6\hat{I}_{401}\hat{P}_{\perp }\right)  \notag \\
& +\frac{\sigma _{T}}{20}\left( 13\hat{I}_{300}^{2}-3\hat{I}_{310}^{2}-9\hat{%
I}_{320}^{2}-\hat{I}_{330}^{2}-18\hat{I}_{301}^{2}-6\hat{I}%
_{311}^{2}\right)\, .  \label{G40}
\end{align}

The next set of gain terms from Eq.~(\ref{hat_G_ij}) is defined for $i=n$ and $j=1$,
and thus contains an additional projection with respect to $l^{\mu}$.
This additional projection along the direction of the anisotropy will give birth to new scalar products
when compared to the previous case in Eq.~(\ref{Gn0}) (see Appendix~\ref%
{app:Gn1} for the derivation),
\begin{align}
\hat{G}_{n1}& \equiv \int_{K\!K^{\prime }\!}\hat{f}_{0\mathbf{k}}\hat{f}_{0%
\mathbf{k}^{\prime }}\mathcal{P}_{n1}=-\int_{K\!K^{\prime }\!}\hat{f}_{0%
\mathbf{k}}\hat{f}_{0\mathbf{k}^{\prime }}\Theta ^{\mu _{1}\cdots \mu
_{n+1}}u_{\mu _{1}}\cdots u_{\mu _{n}}l_{\mu _{n+1}}  \notag \\
& =-\int_{K\!K^{\prime }\!}\hat{f}_{0\mathbf{k}}\hat{f}_{0\mathbf{k}^{\prime
}}\left( l_{\mu _{n+1}}P_{T}^{\mu _{n+1}}\right) \sum_{q=0}^{\lfloor
n/2\rfloor }\left( -1\right) ^{q}b_{nq}\mathcal{B}_{n+1,q}\left( P_{T}^{\mu
}u_{\mu }\right) ^{n-2q}\left( \Delta _{T}^{\mu \nu }u_{\mu }u_{\nu }\right)
^{q}  \notag \\
& -\int_{K\!K^{\prime }\!}\hat{f}_{0\mathbf{k}}\hat{f}_{0\mathbf{k}^{\prime
}}\left( l_{\mu _{n+1}}\Delta _{T}^{\mu _{n+1}\mu _{n}}u_{\mu _{n}}\right)
\sum_{q=1}^{\lfloor \left( n+1\right) /2\rfloor }\left( -1\right) ^{q}\left(
b_{n+1,q}-b_{nq}\right) \mathcal{B}_{n+1,q}\left( P_{T}^{\mu }u_{\mu
}\right) ^{n+1-2q}\left( \Delta _{T}^{\mu \nu }u_{\mu }u_{\nu }\right)
^{q-1}\, ,  \label{Gn1}
\end{align}%
where the new invariant scalars are
\begin{align}
P_{T}^{\mu }l_{\mu } &= -\left( E_{\mathbf{k}l}+E_{\mathbf{k}^{\prime}l}\right) \, ,  
\label{PT_l} \\
\Delta _{T}^{\mu \nu }u_{\mu }l_{\nu } &= s^{-1}\left( E_{\mathbf{k}l}
+E_{\mathbf{k}^{\prime }l}\right) \left( E_{\mathbf{k}u}+E_{\mathbf{k}^{\prime}u}\right) \, .  \label{DeltaT_ul}
\end{align}

Now using this latter result for $n=0$ we obtain
\begin{equation}
\hat{G}_{01}=\sigma _{T}\left( \hat{M}\hat{n}-\hat{P}_{l}\hat{n}%
_{l}\right)\, ,  \label{G01}
\end{equation}%
which is precisely the same result as previously obtained for the loss term
in Eq.~(\ref{L01}), respecting the conservation of momentum in the
direction of the anisotropy. 
Similarly, using Eq.~(\ref{Gn1}) for $n=1$ and $n=2$ we get
\begin{align}
\hat{G}_{11}& =\frac{2\sigma _{T}}{3}\left( \hat{I}_{310}\hat{n}-\hat{I}%
_{320}\hat{n}_{l}+\hat{M}\hat{e}-\hat{M}\hat{P}_{l}\right) \, ,  \label{G11}
\\
\hat{G}_{21}& =\frac{\sigma _{T}}{6}\left( 3\hat{I}_{410}\hat{n}-3\hat{I}%
_{420}\hat{n}_{l}+5\hat{I}_{310}\hat{e}-4\hat{I}_{320}\hat{M}\right) +\frac{%
\sigma _{T}}{6}\left( 3\hat{I}_{300}\hat{M}-3\hat{I}_{310}\hat{P}_{l}-\hat{I}%
_{330}\hat{P}_{l}-2\hat{I}_{311}\hat{P}_{\perp }\right) \, ,  \label{G21}
\end{align}%
while the result for $n=3$ is%
\begin{align}
\hat{G}_{31}& =\frac{2\sigma _{T}}{5}\left( \hat{I}_{510}\hat{n}-\hat{I}%
_{520}\hat{n}_{l}+\hat{I}_{400}\hat{M}-\hat{I}_{410}\hat{P}_{l}\right) +%
\frac{3\sigma _{T}}{10}\left( 3\hat{I}_{410}\hat{e}-2\hat{I}_{420}\hat{M}-%
\hat{I}_{430}\hat{P}_{l}-2\hat{I}_{411}\hat{P}_{\perp }\right)  \notag \\
& +\frac{3\sigma _{T}}{10}\left( 3\hat{I}_{300}\hat{I}_{310}-2\hat{I}_{310}%
\hat{I}_{320}-\hat{I}_{320}\hat{I}_{330}-2\hat{I}_{301}\hat{I}%
_{311}\right)\, .  \label{G31}
\end{align}

Finally, the last gain term which we explicitly calculate follows from 
Eq.~(\ref{hat_G_ij}) for $i=n$ and $j=2$,
therefore generalizing our previous results by including two projections in
the direction of the anisotropy (see Appendix~\ref{app:Gn2} for details
of the derivation of this expression),
\begin{align}
\hat{G}_{n2}& \equiv \int_{K\!K^{\prime }\!}\hat{f}_{0\mathbf{k}}\hat{f}_{0%
\mathbf{k}^{\prime }}\mathcal{P}_{n2}=\int_{K\!K^{\prime }\!}\hat{f}_{0%
\mathbf{k}}\hat{f}_{0\mathbf{k}^{\prime }}\Theta ^{\mu _{1}\cdots \mu
_{n+2}}u_{\mu _{1}}\cdots u_{\mu _{n}}l_{\mu _{n+1}}l_{\mu _{n+2}}  \notag \\
& =\int_{K\!K^{\prime }\!}\hat{f}_{0\mathbf{k}}\hat{f}_{0\mathbf{k}^{\prime
}}\left( l_{\mu _{n+2}}P_{T}^{\mu _{n+2}}\right) \left( l_{\mu
_{n+1}}P_{T}^{\mu _{n+1}}\right) \sum_{q=0}^{\lfloor n/2\rfloor }\left(
-1\right) ^{q}b_{nq}\mathcal{B}_{n+2,q}\left( P_{T}^{\mu }u_{\mu }\right)
^{n-2q}\left( \Delta _{T}^{\mu \nu }u_{\mu }u_{\nu }\right) ^{q}  \notag \\
& +\int_{K\!K^{\prime }\!}\hat{f}_{0\mathbf{k}}\hat{f}_{0\mathbf{k}^{\prime
}}\left( l_{\mu _{n+2}}P_{T}^{\mu _{n+2}}\right) \left( l_{\mu _{n+1}}\Delta
_{T}^{\mu _{n+1}\mu _{n}}u_{\mu _{n}}\right)  \notag \\
& \times \sum_{q=1}^{\lfloor \left( n+2\right) /2\rfloor }\left( -1\right)
^{q}2\left( b_{n+1,q}-b_{nq}\right) \mathcal{B}_{n+2,q}\left( P_{T}^{\mu
}u_{\mu }\right) ^{n+1-2q}\left( \Delta _{T}^{\mu \nu }u_{\mu }u_{\nu
}\right) ^{q-1}  \notag \\
& +\int_{K\!K^{\prime }\!}\hat{f}_{0\mathbf{k}}\hat{f}_{0\mathbf{k}^{\prime
}}\left( l_{\mu _{n+2}}\Delta _{T}^{\mu _{n+2}\mu _{n+1}}l_{\mu
_{n+1}}\right) \sum_{q=1}^{\lfloor \left( n+2\right) /2\rfloor }\left(
-1\right) ^{q}b_{n,q-1}\mathcal{B}_{n+2,q}\left( P_{T}^{\mu }u_{\mu }\right)
^{n+2-2q}\left( \Delta _{T}^{\mu \nu }u_{\mu }u_{\nu }\right) ^{q-1}  \notag
\\
& +\int_{K\!K^{\prime }\!}\hat{f}_{0\mathbf{k}}\hat{f}_{0\mathbf{k}^{\prime
}}\left( l_{\mu _{n+2}}\Delta _{T}^{\mu _{n+2}\mu _{n}}u_{\mu _{n}}\right)
\left( l_{\mu _{n+1}}\Delta _{T}^{\mu _{n+1}\mu _{n-1}}u_{\mu _{n-1}}\right)
\notag \\
& \times \sum_{q=2}^{\lfloor \left( n+2\right) /2\rfloor }\left( -1\right)
^{q}2\left( q-1\right) b_{n,q-1}\mathcal{B}_{n+2,q}\left( P_{T}^{\mu }u_{\mu
}\right) ^{n+2-2q}\left( \Delta _{T}^{\mu \nu }u_{\mu }u_{\nu }\right)
^{q-2}\, ,  \label{Gn2}
\end{align}%
where the new invariant scalar is
\begin{equation}
\Delta _{T}^{\mu \nu }l_{\mu }l_{\nu }=-1-s^{-1}\left( E_{\mathbf{k}l}+E_{%
\mathbf{k}^{\prime }l}\right) ^{2}\ .  \label{DeltaT_ll}
\end{equation}

Using this result for $n=0$ and $n=1$, we obtain
\begin{align}
\hat{G}_{02}& =\frac{\sigma _{T}}{6}\left( 4\hat{I}_{320}\hat{n}-4\hat{I}%
_{330}\hat{n}_{l}+\hat{e}^{2}\right) +\frac{\sigma _{T}}{6}\left( 2\hat{M}%
^{2}-3\hat{P}_{l}^{2}+2\hat{P}_{\perp }^{2}\right) \ ,  \label{G02} \\
\hat{G}_{12}& =\frac{\sigma _{T}}{6}\left( 3\hat{I}_{420}\hat{n}-3\hat{I}%
_{430}\hat{n}_{l}+4\hat{I}_{310}\hat{M}-5\hat{I}_{320}\hat{P}_{l}\right) +%
\frac{\sigma _{T}}{6}\left( 3\hat{I}_{320}\hat{e}-3\hat{I}_{330}\hat{M}+\hat{%
I}_{300}\hat{e}+2\hat{I}_{301}\hat{P}_{\perp }\right) \, ,  \label{G12}
\end{align}%
and finally the result for $n=2$ reads
\begin{align}
\hat{G}_{22}& =\frac{\sigma _{T}}{10}\left( 4\hat{I}_{520}\hat{n}-4\hat{I}%
_{530}\hat{n}_{l}+7\hat{I}_{420}\hat{e}-7\hat{I}_{420}\hat{P}_{l}+4\hat{I}%
_{320}\hat{I}_{300}-4\hat{I}_{330}\hat{I}_{310}\right)  \notag \\
& +\frac{\sigma _{T}}{10}\left( 6\hat{I}_{410}\hat{M}-6\hat{I}_{430}\hat{M}+%
\hat{I}_{400}\hat{e}-\hat{I}_{440}\hat{P}_{l}+2\hat{I}_{401}\hat{P}_{\perp
}-2\hat{I}_{421}\hat{P}_{\perp }\right)  \notag \\
& +\frac{\sigma _{T}}{60}\left( 5\hat{I}_{300}^{2}-5\hat{I}_{330}^{2}+33\hat{%
I}_{310}^{2}-33\hat{I}_{320}^{2}+6\hat{I}_{301}^{2}-6\hat{I}%
_{311}^{2}\right) \, .  \label{G22}
\end{align}%
Note that the corresponding moments of the collision integral 
$\hat{C}_{ij}=\hat{G}_{ij}-\hat{L}_{ij}$ are computed and listed 
in Appendix~\ref{C_ij_binary}.

These results show that all loss and all gain terms and hence all the
moments of the binary collision integral in leading-order anisotropic fluid
dynamics are expressed as quadratic products of anisotropic
thermodynamic integrals $\hat{I}_{nrq}$ defined in Eq.~(\ref{I_nrq}). 
These nonlinear couplings between moments of different orders represent the nonlinear 
dependence on the distribution function of the binary collision integral. 
It is important to observe that the number of nonlinear terms increases with the
power of energy $E_{\mathbf{k}u}$ and the power of momentum in the direction
of the anisotropy $E_{\mathbf{k}l}$, corresponding to specific projections
of higher rank tensor moments. 
This also means that at any given order the collision term couples only to 
a well-defined set of lower-order moments in the hierarchy.

The method presented here is not only applicable but also essential
to the calculation of the arbitrary rank moment of the binary
collision integral $\hat{C}^{\mu _{1}\cdots \mu _{n}}_{ij}$ in more complete theories 
of anisotropic fluid dynamics~\cite{Molnar:2016vvu,Molnar:2024fgy}.
These also include $\delta \hat{f}_{\mathbf{k}}$ corrections to improve the leading-order anisotropic
framework further as $f_{\mathbf{k}} \equiv \hat{f}_{0\mathbf{k}} + \delta 
\hat{f}_{\mathbf{k}}$ and hence require the computation of $C\!\left[ \hat{f}_{0} + \delta 
\hat{f}\right]$ and corresponding irreducible moments of the collision integral. 
Nonetheless, for the sake of simplicity in this paper we only focus on leading-order 
anisotropic fluid dynamics.

\section{Applications}
\label{sec:Applications}

\subsection{The moments of the collision integral for the Romatschke-Strickland distribution function}
\label{sec:RS_binary_vs_RTA}

In this section we study the moments of collision integral using the
spheroidal distribution function of Ref.~\cite{Romatschke:2003ms}. The RS distribution function
represents the rescaling of the momentum in the direction of the anisotropy, 
\begin{equation}
\hat{f}_{RS}(\alpha _{RS},\beta _{RS},\xi )\equiv \exp 
\left( -\alpha_{RS}+\beta _{RS}\sqrt{E_{\mathbf{k}u}^{2}+\xi E_{\mathbf{k}l}^{2}}\right)\, ,  
\label{f_RS}
\end{equation}%
where $\xi $ denotes the anisotropy parameter. In the LR frame $u^\mu_{LR}=(1,0,0,0)$, and thus
$E_{\mathbf{k}u}=k^{0}$ and $E_{\mathbf{k}l}=k_{z}$; hence with respect to
the $z$ axis in momentum space $\hat{f}_{RS}$ is a prolate spheroid for 
$\xi<0$, while it is an oblate spheroid for $\xi >0$. The spherically symmetric
equilibrium distribution is obtained in the case $\xi=0$.

The anisotropic thermodynamic integrals corresponding to the RS distribution
function $\hat{I}_{nrq}^{RS}\left( \alpha _{RS},\beta _{RS},\xi \right) $
are given replacing $\hat{f}_{0\mathbf{k}}\rightarrow \hat{f}_{RS}$ in 
Eq.~(\ref{I_nrq}), 
\begin{equation}
\hat{I}_{nrq}^{RS}\equiv \frac{\left( -1\right) ^{q}}{\left( 2q\right) !!}
\int dKE_{\mathbf{k}u}^{n-r-2q}E_{\mathbf{k}l}^{r}
\left( \Xi ^{\mu \nu}k_{\mu }k_{\nu }\right) ^{q}\hat{f}_{RS}\, ,  
\label{I_nrq_RS}
\end{equation}%
which in the ultrarelativistic limit $m_{0}\rightarrow 0$ of a Boltzmann gas leads to
the following factorization~\cite{Martinez:2009ry}, 
\begin{equation}
\hat{I}_{nrq}^{RS}\left( \alpha _{RS},\beta _{RS},\xi \right) 
= I_{nq}\left(\alpha _{RS},\beta _{RS}\right) R_{nrq}\left( \xi \right) \, .
\label{R_nrq_RS}
\end{equation}%
All relevant thermodynamic integrals and ratios $R_{nrq}$ are explicitly
listed in Appendix~\ref{appendix_aniso_integrals}. Furthermore, for the RS
distribution function all ratios and thus all
anisotropic thermodynamic integrals vanish equivalently for all odd $r$, 
\begin{equation}
\hat{I}_{nrq}^{RS} \equiv R_{nrq}= 0\,,\quad \forall r\in \text{odd}\,;
\label{R_nrq_r_odd=0}
\end{equation}%
hence, due to this property the corresponding moments of the binary collision integral 
$\hat{C}_{ir}^{RS}=\hat{L}_{ir}^{RS}=\hat{G}_{ir}^{RS}=0$ also vanish equivalently 
for all odd $r$. 
Note that now the fluid dynamical flow velocity is also independent of
the choice of the LR frame, since not only the energy-momentum diffusion current, 
$\hat{M}\equiv \hat{I}_{210}^{RS}=0$, but also the particle diffusion current 
in the direction of the anisotropy vanishes $\hat{n}_{l}\equiv \hat{I}_{110}^{RS}=0$
equivalently. 
The former corresponds to Landau's definition of the LR frame~\cite{Landau_book}, while the 
latter corresponds to Eckart's definition of the LR frame~\cite{Eckart:1940te}.

The first and second moments of the RS distribution function are 
\begin{equation}
\hat{N}_{RS}^{\mu }=\hat{n}u^{\mu }\, , \quad \hat{T}_{RS}^{\mu \nu }=\hat{e}%
u^{\mu }u^{\nu }+\hat{P}_{l}l^{\mu }l^{\nu } - \hat{P}_{\perp }\Xi ^{\mu
\nu}\, ,  \label{N_mu_RS_and_T_munu_RS}
\end{equation}%
and with the help of Eq.~(\ref{R_nrq_RS}) the primary fluid dynamical quantities read 
\begin{eqnarray}
\hat{n} &\equiv &\hat{I}_{100}^{RS}=n_{0}\left( \alpha _{RS},\beta
_{RS}\right) R_{100}\left( \xi \right) \, ,  \label{n_hat_RS} \\
\hat{e} &\equiv &\hat{I}_{200}^{RS}=e_{0}\left( \alpha _{RS},\beta
_{RS}\right) R_{200}\left( \xi \right) \, ,  \label{e_hat_RS} \\
\hat{P}_{l} &\equiv &\hat{I}_{220}^{RS}=e_{0}\left( \alpha _{RS},\beta
_{RS}\right) R_{220}\left( \xi \right) \, ,  \label{Pl_hat_RS} \\
\hat{P}_{\perp } &\equiv &\hat{I}_{201}^{RS}=P_{0}\left( \alpha _{RS},\beta
_{RS}\right) R_{201}\left( \xi \right) \, .  \label{Pt_hat_RS}
\end{eqnarray}
For an ideal gas of massless particles, $P_0(n_0,e_0) \equiv e_0/3 = n_0 T$, 
where $\mu(n_0,e_0)$ is the chemical potential and $T(n_0,e_0)$ is the
temperature. Therefore the bulk viscous pressure, defined according to 
$\hat{\Pi}\equiv \hat{P}-P_{0}$, where $\hat{P} = (\hat{P}_{l} + 2\hat{P}_{\perp})/3$ 
is the isotropic pressure, also vanishes in the massless limit.

The parameters of the anisotropic distribution function, $\alpha _{RS}$, 
$\beta _{RS}$, and $\xi $, can be expressed in terms of a fictitious
equilibrium state specified by $\alpha=\mu/T$ and $\beta=1/T$, through the
so-called Landau matching conditions~\cite{Anderson:1974nyl}, 
\begin{equation}
\hat{n}\left( \alpha _{RS},\beta _{RS},\xi \right) = n_{0}\left(
\alpha,\beta \right) \, , \quad \hat{e}\left( \alpha _{RS},\beta
_{RS},\xi\right) = e_{0}\left( \alpha,\beta\right) \, ,  \label{matching_n_e}
\end{equation}%
and hence from Eqs.~(\ref{n_hat_RS}) and (\ref{e_hat_RS}), and the
corresponding fugacities, $\lambda\equiv \exp \left(\alpha \right)$ and 
$\lambda _{RS}=\exp \left( \mu _{RS}\beta_{RS}\right)$, we obtain 
\begin{equation}
\lambda =\lambda _{RS}\frac{\left[ R_{100}\left( \xi \right) \right] ^{4}}{%
\left[ R_{200}\left( \xi \right) \right] ^{3}}\, ,\quad \beta =\beta_{RS}%
\frac{R_{100}\left( \xi \right) }{R_{200}\left( \xi \right) }\, .
\label{Landau_matching_beta0}
\end{equation}%
Using these results together with Eq.~(\ref{R_nrq_RS}) leads to the following 
general relation between the anisotropic thermodynamic integrals of the RS 
distribution function and the equilibrium thermodynamic integrals~\cite{Molnar:2016gwq},
\begin{equation}
\hat{I}_{nrq}^{RS}\left( \alpha _{RS},\beta _{RS},\xi \right) =
I_{nq}\left(\alpha,\beta\right) R_{nrq}\left( \xi \right) 
\frac{\left[R_{200}\left(\xi\right)\right]^{1-n}}{\left[ R_{100}\left(\xi\right)\right]^{2-n}}\, .  \label{Matched_I_nrq_RS}
\end{equation}

Now taking into account the properties of the RS distribution function from
Eq.~(\ref{R_nrq_r_odd=0}) together with the matching condition 
$\hat{e}\left( \alpha_{RS},\beta_{RS},\xi \right)=e_{0}\left( \alpha,\beta \right)$, 
the moments of the binary collision integral from Appendix~\ref{C_ij_binary} lead to
\begin{align}
\hat{C}_{20}^{RS}& =-\frac{1}{\tau _{R}}\left( \frac{\hat{n}}{3n_{0}}\right) 
\hat{I}_{300}^{RS}+\frac{1}{\tau _{R}}\left[ \frac{1}{6n_{0}}\left(
3e_{0}^{2}-\hat{P}_{l}^{2}-2\hat{P}_{\perp }^{2}\right) \right] \, ,
\label{C_RS_20} \\
\hat{C}_{30}^{RS}& =-\frac{1}{\tau _{R}}\left( \frac{\hat{n}}{2n_{0}}\right) 
\hat{I}_{400}^{RS}+\frac{1}{\tau _{R}}\left[ \frac{1}{2n_{0}}\left( 2e_{0}%
\hat{I}_{300}^{RS}-\hat{P}_{l}\hat{I}_{320}^{RS}-2\hat{P}_{\perp }\hat{I}%
_{301}^{RS}\right) \right] \, ,  \label{C_RS_30}
\end{align}%
and%
\begin{align}
\hat{C}_{40}^{RS}& =-\frac{1}{\tau _{R}}\left( \frac{3\hat{n}}{5n_{0}}%
\right) \hat{I}_{500}^{RS}+\frac{1}{\tau _{R}}\left[ \frac{1}{5n_{0}}\left(
5e_{0}\hat{I}_{400}^{RS}-3\hat{P}_{l}\hat{I}_{420}^{RS}-6\hat{P}_{\perp }%
\hat{I}_{401}^{RS}\right) \right]  \notag \\
& +\frac{1}{\tau _{R}}\left[ \frac{1}{20n_{0}}\left( 13\left( \hat{I}%
_{300}^{RS}\right) ^{2}-9\left( \hat{I}_{320}^{RS}\right) ^{2}-18\left( \hat{%
I}_{301}^{RS}\right) ^{2}\right) \right] \, .  \label{C_RS_40}
\end{align}%
Similarly the other relevant moments of the binary collision integral are 
\begin{align}
\hat{C}_{02}^{RS}& =-\frac{1}{\tau _{R}}\left( \frac{\hat{n}}{3n_{0}}\right) 
\hat{I}_{320}^{RS}+\frac{1}{\tau _{R}}\left[ \frac{1}{6n_{0}}\left(
e_{0}^{2}-3\hat{P}_{l}^{2}+2\hat{P}_{\bot }^{2}\right) \right] \, ,
\label{C_RS_02} \\
\hat{C}_{12}^{RS}& =-\frac{1}{\tau _{R}}\left( \frac{\hat{n}}{2n_{0}}\right) 
\hat{I}_{420}^{RS}+\frac{1}{\tau _{R}}\left[ \frac{1}{6n_{0}}\left( e_{0}%
\hat{I}_{300}^{RS}+\left( 3e_{0}-5\hat{P}_{l}\right) \hat{I}_{320}^{RS}+2%
\hat{P}_{\perp }\hat{I}_{301}^{RS}\right) \right] \, ,  \label{C_RS_12}
\end{align}%
and finally%
\begin{align}
\hat{C}_{22}^{RS}& =-\frac{1}{\tau _{R}}\left( \frac{3\hat{n}}{5n_{0}}%
\right) \hat{I}_{520}^{RS}+\frac{1}{\tau _{R}}\left[ \frac{1}{10n_{0}}\left(
e_{0}\hat{I}_{400}^{RS}+\left( 7e_{0}-7\hat{P}_{l}\right) \hat{I}_{420}^{RS}-%
\hat{P}_{l}\hat{I}_{440}^{RS}+2\hat{P}_{\perp }\hat{I}_{401}^{RS}-2\hat{P}%
_{\perp }\hat{I}_{421}^{RS}\right) \right]  \notag \\
& +\frac{1}{\tau _{R}}\left[ \frac{1}{60n_{0}}\left( 5\left( \hat{I}%
_{300}^{RS}\right) ^{2}+24\hat{I}_{320}^{RS}\hat{I}_{300}^{RS}-33\left( \hat{%
I}_{320}^{RS}\right) ^{2}+6\left( \hat{I}_{301}^{RS}\right) ^{2}\right) %
\right] \, .  \label{C_RS_22}
\end{align}%
Note that here we have replaced the total cross section by 
$\sigma_{T}=1/(n_{0}\tau _{R})$, which introduces the mean free path or mean
free time between collisions based on the equilibrium particle density.
However according to Eq.~(\ref{sigma_T}) there is freedom to chose $%
\sigma_{T}=1/(\hat{n} \hat{\tau}_{R})$ based on the non-equilibrium particle
density with a different mean free time $\hat{\tau}_R$ between collisions. 
Nevertheless, in case the number of particles or charge(s) remain conserved, 
such as in the case of binary elastic collisions, then it is required that 
$\hat{n}\left(\alpha_{RS},\beta_{RS},\xi \right) = n_{0}\left( \alpha,\beta\right)$ 
and hence $\hat{\tau}_R=\tau_R$.

For the sake of comparison, here we recall the relativistic relaxation-time approximation 
model of Anderson and Witting (AW)~\cite{Anderson:1974nyl}, which is the relativistic 
generalization of the RTA model of Bhatnagar-Gross-Krook (BGK)~\cite{Bhatnagar:1954}. 
The RTA assumes that the nonlinear binary collision integral from Eq.~(\ref{BTE_aniso}) 
is approximated by 
\begin{equation}
\hat{C}\left[ \hat{f}_{0}\right] \approx \hat{C}_{AW}\left[\hat{f}_{0}\right] 
= -\frac{E_{\mathbf{k}u}}{\tau_{R}}\left(\hat{f}_{0\mathbf{k}}-f_{0\mathbf{k}}\right) \, ,  
\label{RTA}
\end{equation}%
where the relaxation time $\tau_{R}$ is an energy and momentum independent
parameter, and represents the timescale on which the anisotropic distribution 
function approaches the local equilibrium distribution function. 
This is a free parameter that is chosen either as a constant or according to 
some transport coefficient that relates to some intrinsic property of matter 
such as viscosity or diffusivity. For example in the 14-moment approximation of second-order 
fluid dynamics~\cite{Ambrus:2022vif}, the relaxation time of shear-viscosity and particle 
diffusivity are $\tau_\pi \equiv 5 \eta/(4 P_{0}) = (5/3)/(\sigma_{T} n_{0})$ and 
$\tau_V \equiv 12 T \kappa/P_0 = (9/4)/(\sigma_{T} n_{0}) $.
For reasons of simplicity and aiming for direct comparison to the binary collision integral, 
in the following the relaxation time is defined through the total cross section 
and the particle density, $\tau _{R}\equiv 1/(\sigma_{T} n_{0})$, and hence 
represents the mean free time between collisions.

Replacing the RS distribution function into Eq.~(\ref{RTA}), the
corresponding scalar moments of the collision integral~(\ref{C_ij_general})
in the RTA are given by the following simple formula, 
\begin{equation}
\hat{C}_{ij,AW}^{RS} = -\frac{1}{\tau _{R}}\hat{I}_{i+j+1,j,0}^{RS} +\frac{1%
}{\tau _{R}}I_{i+j+1,j,0}\, .  \label{Coll_Int_RTA}
\end{equation}%
Now comparing the moments of the nonlinear binary collision integral from
Eqs.~(\ref{C_RS_20})-(\ref{C_RS_40}) and Eqs.~(\ref{C_RS_02})-(\ref{C_RS_22}) to
the moments of the RTA approximation~(\ref{Coll_Int_RTA})
reveals that the latter omits an ever increasing number of
nonlinear couplings represented through the quadratic products of various 
anisotropic thermodynamic integrals.
Even though the very first terms, i.e., $- \hat{I}_{i+j+1,j,0}^{RS}/\tau_R$, 
on the right-hand sides appear in both cases, the numerical prefactors in the case 
of the binary collision integral are consistently smaller than 1. 
However, the major differences between the moments of the binary collision
integral and the moments in the RTA are explicitly given by all the remaining terms in square brackets 
in Eqs.~(\ref{C_RS_20})-(\ref{C_RS_40}) and Eqs.~(\ref{C_RS_02})-(\ref{C_RS_22}). 
In the relaxation-time approximation such terms are simply represented by an
equilibrium thermodynamic integral; see the second term in Eq.~(\ref{Coll_Int_RTA}). 
Hence due to these differences it should be expected that the  
approach to local equilibrium, i.e., $\xi \to 0$, will also happen on different timesscales
for the different approaches.
Since most terms in square brackets are non-equilibrium
quantities being some function of the anisotropy parameter we expect 
that the system approaches equilibrium slower than in the case of the RTA 
which directly and explicitly relaxes to local equilibrium on the shortest timescale. 
This observation is in agreement with earlier results based on the exact solutions of 
the relativistic Boltzmann equation for homogeneous and isotropic solutions~\cite{Bazow:2015dha,Bazow:2016oky}. 
Further detailed comparisons and discussions are presented in the next sections.

\subsection{(0+1)-dimensional boost-invariant expansion and a judicious choice
of moment}
\label{sec:Results_BJ}

We now study the direct influence of the collision term on the solution of
the fluid-dynamical equations of motion in the (0+1)-dimensional
boost-invariant expansion, known as Bjorken flow \cite{Bjorken:1982qr}. The
space-time coordinates $(t,z)$ are transformed to proper time $\tau =\sqrt{%
t^{2}-z^{2}}$, and space-time rapidity $\eta _{s}=\frac{1}{2}\ln \frac{t+z}{%
t-z}$, where the inverse transformations are $t=\tau \cosh \eta _{s}$, and $%
z=\tau \sinh \eta _{s}$. The longitudinal fluid velocity is defined as $v_{z}\equiv z/t=\tanh
\eta _{s}$, and hence now $u^{\mu }\equiv \left( \frac{t}{\tau },0,0,\frac{z}{%
\tau }\right) =\left( \cosh \eta _{s},0,0,\sinh \eta _{s}\right) $ and $%
l^{\mu }\equiv \left( \frac{z}{\tau },0,0,\frac{t}{\tau }\right) =\left(
\sinh \eta _{s},0,0,\cosh \eta _{s}\right) $. Furthermore, $D\equiv
u^{\mu }\partial _{\mu }=\frac{\partial }{\partial \tau }$, $D_{l}\equiv
-l^{\mu }\partial _{\mu }=-\frac{\partial }{\tau \partial \eta _{s}}$, and 
$Du^{\mu }=Dl^{\mu }=0$, $D_{l}u^{\mu }=-\frac{1}{\tau }l^{\mu }$, $%
D_{l}l^{\mu }=-\frac{1}{\tau }u^{\mu }$, while all thermodynamic quantities
and variables are independent of $\eta _{s}$ and only depend on the
proper time.

Applying these simplifications we obtain a hierarchy of coupled equations of motion 
represented by the following simple differential equation [see Eq.~(53) in 
Ref.~\cite{Molnar:2016gwq}]: 
\begin{equation}
\frac{\partial \hat{I}_{i+j,j,0}^{RS}}{\partial \tau } 
+\frac{1}{\tau } \left[\left( j+1\right) \hat{I}_{i+j,j,0}^{RS} 
+\left( i-1\right) \hat{I}_{i+j,j+2,0}^{RS}\right] 
= \hat{C}_{i-1,j}^{RS}\approx \hat{C}_{i-1,j,AW}^{RS}\, ,  
\label{Main_eq_motion}
\end{equation}%
where  $i,j\geq 0$, and on the rhs we use either the corresponding moments of the binary
collision integral $\hat{C}_{i-1,j}^{RS}$ or the relaxation-time
approximation to the collision integral denoted by $\hat{C}_{i-1,j,AW}^{RS}$.

The conservation of particle number and the conservation of energy are obtained
from Eq.~(\ref{Main_eq_motion}) for $i=1, j=0$ and $i=2,j=0$
respectively, 
\begin{align}
& \frac{\partial n_{0}\left( \alpha,\beta \right)}{\partial \tau} 
+\frac{1}{\tau }n_{0}\left( \alpha,\beta\right) = 0\, ,  
\label{BJ_n_cons} \\
& \frac{\partial e_{0}\left( \alpha,\beta \right)}{\partial \tau} 
+\frac{1}{\tau }\left[ e_{0}\left( \alpha,\beta \right) 
+\hat{P}_{l}\left( \alpha_{RS},\beta _{RS},\xi \right) \right] = 0\, ,  
\label{BJ_e_cons}
\end{align}
where according to Eqs.~(\ref{C_0_C_1_C_0_mu_aniso}) on the lhs we have 
$\hat{C}^{RS}_{00}\equiv \hat{C}^{RS}_{00,AW}= 0$ and 
$\hat{C}^{RS}_{10}\equiv \hat{C}^{RS}_{10,AW}= 0$.

The hierarchy of coupled moment equations~(\ref{Main_eq_motion}) provides
infinitely many possibilities to close the conservation equations 
and to determine the time evolution of the anisotropy parameter $\xi$. 
Here we will follow Ref.~\cite{Molnar:2016gwq} and restrict ourselves
to a few examples by choosing particular values for the indices $i$ and $j$ 
corresponding to the powers of $E_{\mathbf{k}u}$ and $E_{\mathbf{k}l}$ of the 
anisotropic integral $\hat{I}^{RS}_{i+j,j,0}$.
Therefore the corresponding anisotropic moment selected for closure is now treated 
dynamically and represented by the corresponding differential equation~(\ref{Main_eq_motion}).
This way we also obtain the proper time evolution of the anisotropic distribution function~(\ref{f_RS})
through its three parameters: $\alpha^{RS}_{ij}(\tau)$, $\beta^{RS}_{ij}(\tau)$, and $\xi_{ij}(\tau)$.
Using these solutions now all other (non-dynamical) moments in the hierarchy are 
subsequently obtained algebraically, i.e., using Eq.~(\ref{I_nrq_RS}) and leading to $\hat{I}^{RS}_{nrq}\left(\alpha^{RS}_{ij},\beta^{RS}_{ij},\xi_{ij}\right)$. 

Here we list the indices corresponding to the moments of the binary collision integral calculated 
in Eqs.~(\ref{C_RS_20})-(\ref{C_RS_40}).

(i) $i=3$, $j=0$: This choice is analogous to the one of Israel and Stewart 
\cite{Israel:1979wp} and follows from the projection of the rank $3$ tensor
moment equation $\hat{\mathcal{I}}_{00}^{\mu _{1}\mu_2 \mu _{3}}$ to close
the conservation equations, 
\begin{equation}
\frac{\partial \hat{I}_{300}^{RS}}{\partial \tau }+\frac{1}{\tau }
\left(\hat{I}_{300}^{RS}+2\hat{I}_{320}^{RS}\right) 
=\hat{C}_{20}^{RS}\approx \hat{C}_{20,AW}^{RS}\, ,  
\label{BJ_I300_relax}
\end{equation}

(ii) $i=4$, $j=0$: This choice is analogous to the previous but follows from a
rank $4$ tensor moment,
\begin{equation}
\frac{\partial \hat{I}_{400}^{RS}}{\partial \tau }+\frac{1}{\tau }
\left(\hat{I}_{400}^{RS}+3\hat{I}_{420}^{RS}\right) 
=\hat{C}_{30}^{RS}\approx \hat{C}_{30,AW}^{RS}\, ,  
\label{BJ_I400_relax}
\end{equation}

(iii) $i=5$, $j=0$: This choice is analogous to before and follows from
a rank $5$ tensor moment, 
\begin{equation}
\frac{\partial \hat{I}_{500}^{RS}}{\partial \tau }+\frac{1}{\tau }
\left(\hat{I}_{500}^{RS}+4\hat{I}_{520}^{RS}\right) 
=\hat{C}_{40}^{RS}\approx \hat{C}_{40,AW}^{RS}\, .  
\label{BJ_I500_relax}
\end{equation}

Now we fix $j=2$ and then choose $i$ to correspond to Eqs.~(\ref{C_RS_02})-(\ref{C_RS_22}):

(iv) $i=1$, $j=2$: This also results from a rank $3$ tensor moment but
through a different projection than in case (i), 
\begin{equation}
\frac{\partial \hat{I}_{320}^{RS}}{\partial \tau } +\frac{1}{\tau } \left(3\hat{I}_{320}^{RS} \right)
=\hat{C}_{02}^{RS}\approx \hat{C}_{02,AW}^{RS}\,,
\label{BJ_I320_relax}
\end{equation}

(v) $i=2$, $j=2$: This is similar to (ii) but through a different projection than in
case (ii), 
\begin{equation}
\frac{\partial \hat{I}_{420}^{RS}}{\partial \tau }+\frac{1}{\tau }
\left( 3\hat{I}_{420}^{RS}+\hat{I}_{440}^{RS}\right) 
=\hat{C}_{12}^{RS}\approx \hat{C}_{12,AW}^{RS}\, ,  
\label{BJ_I420_relax}
\end{equation}

(vi) $i=3$, $j=2$: This choice is similar to (iii) but projected differently
than in case (iii), 
\begin{equation}
\frac{\partial \hat{I}_{520}^{RS}}{\partial \tau }+\frac{1}{\tau }
\left( 3\hat{I}_{520}^{RS}+2\hat{I}_{540}^{RS}\right) 
=\hat{C}_{22}^{RS}\approx \hat{C}_{22,AW}^{RS}\, .  
\label{BJ_I520_relax}
\end{equation}
Note that the choices (i) and (iv) were already 
studied in Ref.~\cite{Molnar:2016gwq}, while the choice $i=0$, $j=2$, i.e., 
the equation for the longitudinal pressure 
$\hat{P}_{l}\left( \alpha _{RS},\beta_{RS},\xi \right) $, was also found 
to represent the best match~\cite{Tinti:2015xwa,Molnar:2016gwq,Niemi:2017stb} 
to the exact numerical solution of the Boltzmann equation in the RTA, 
\begin{equation}
\frac{\partial \hat{P}_{l}}{\partial \tau } 
+\frac{1}{\tau }\left( 3\hat{P}_{l}-\hat{I}_{240}^{RS}\right) 
=\hat{C}_{-12}^{RS}\approx -\frac{1}{\tau_{R}}\left( \hat{P}_{l}-P_{0}\right) \, .  
\label{BJ_Pl_relax}
\end{equation}%
This choice is also special because both $\hat{P}_{l}$ and $\hat{I}_{240}^{RS}$ 
expressed through Eq.~(\ref{Matched_I_nrq_RS}) 
remain formally unchanged and hence independent of whether we conserve the particle
number or not~\cite{Molnar:2016gwq,Ambrus:2023qcl}. 
Here we only make this choice in the RTA since the corresponding moments of the binary 
collision integral with negative powers of energy cannot be computed using the projection method.
Nevertheless, as shown in Ref.~\cite{Molnar:2016gwq} as well as in what follows 
one can reasonably well approximate the solutions of $\hat{I}_{220}$ by 
the solutions of higher moments, i.e., $\hat{I}_{i+2,2,0}$ for $i>0$.
Note that the small $\xi$ expansion (near equilibrium) of either of the moment equations (\ref{BJ_I300_relax})-(\ref{BJ_Pl_relax}) leads to an evolution equation for the shear-stress tensor component $\hat{\pi} \equiv 2(\hat{P}_\perp - \hat{P}_l)/3 = P_0- \hat{P}_l$ and corresponds to a 
so-called relaxation equation of second-order dissipative fluid dynamics
(see for example Refs.~\cite{Tinti:2015xwa,Molnar:2016gwq} for more details).

In what follows we will focus on the solutions to the conservation equations
(\ref{BJ_n_cons}) and (\ref{BJ_e_cons}) closed by one of the moment equations (\ref{BJ_I300_relax})-(\ref{BJ_I520_relax}) listed here. 
We will study the evolution of an ultrarelativistic anisotropic fluid 
either using the moments of the binary collision integral or
the moments in the relaxation-time approximation of the collision integral. 
The Landau matching conditions~(\ref{matching_n_e}) are used to infer the fugacity 
and temperature, while for the sake of comparison we also show the solutions 
of the ideal fluid dynamical equations, i.e., Eq.~(\ref{BJ_n_cons}) 
and Eq.~(\ref{BJ_e_cons}), where in the latter replacing 
$\hat{P}_l \rightarrow P_0(\alpha,\beta)$ leads to the ideal equation of motion, 
$\partial e_{0}\left( \alpha,\beta\right)/\partial \tau
+\left[ e_{0}\left( \alpha,\beta\right) + P_0\left( \alpha,\beta\right)\right]/\tau=0$.

\begin{figure*}[tbp]
\begin{tabular}{cc}
\includegraphics[width=.49\linewidth]{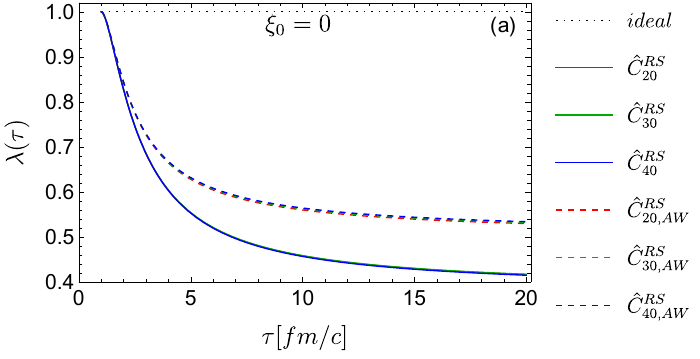} & %
\includegraphics[width=.49\linewidth]{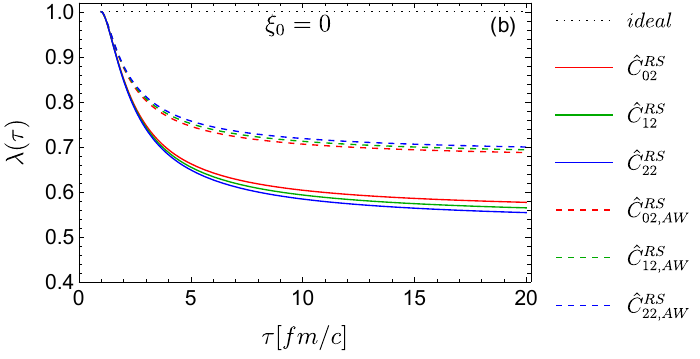} \\ 
\includegraphics[width=.49\linewidth]{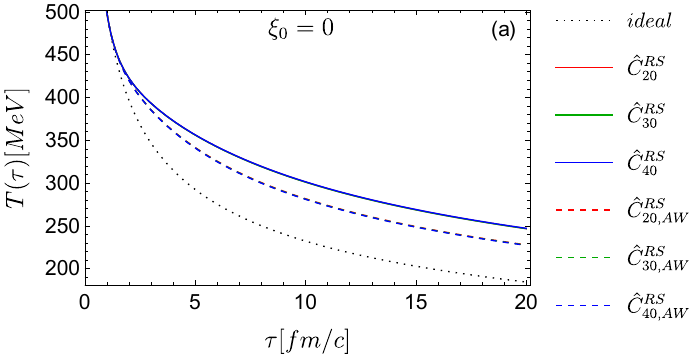} & %
\includegraphics[width=.49\linewidth]{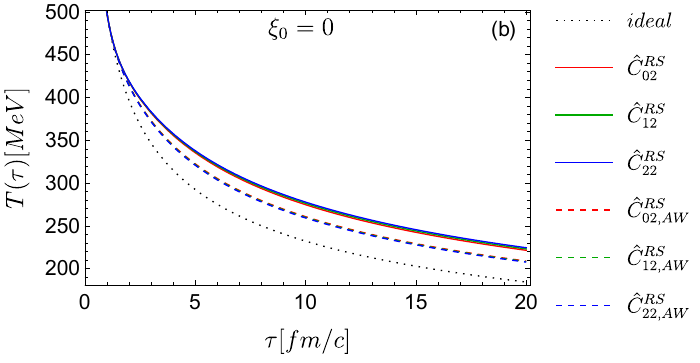} \\ 
\includegraphics[width=.49\linewidth]{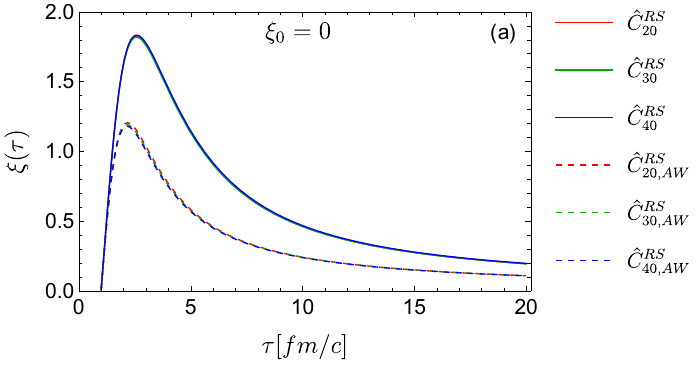} & %
\includegraphics[width=.49\linewidth]{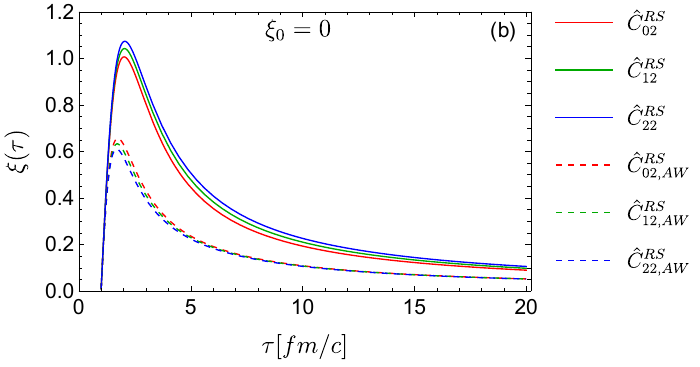} \\ 
\includegraphics[width=.49\linewidth]{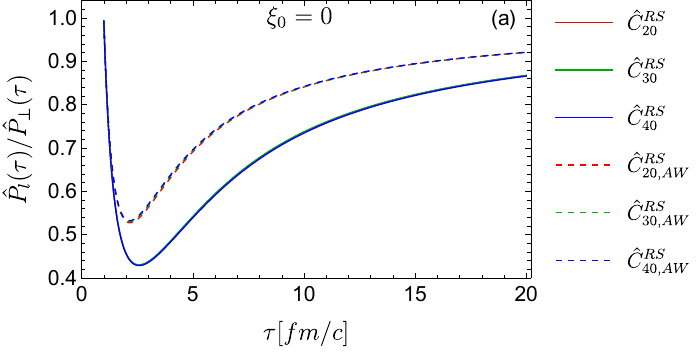} & %
\includegraphics[width=.49\linewidth]{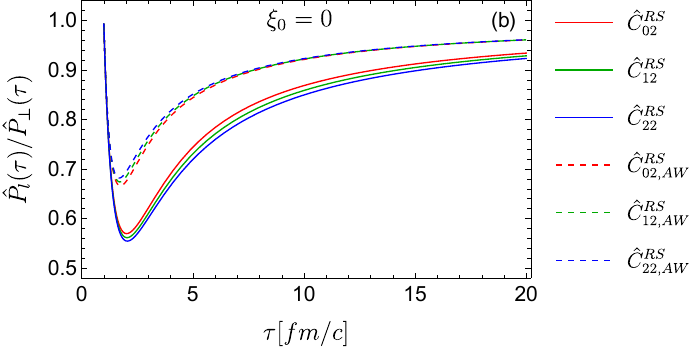}%
\end{tabular}%
\caption{From top to bottom as well as both left~(a) and right~(b) panels 
the initial anisotropy is $\xi_0 = 0$.
The evolution of the fugacity $\protect\lambda( \protect\tau)$, the
temperature $T(\protect\tau)$, the anisotropy parameter $\protect\xi(\protect%
\tau)$, and the ratio of the longitudinal pressure and the transverse pressure
 $\hat{P}_l(\protect\tau)/\hat{P}_{\perp}( \protect\tau)$, as a
function of proper time. The black dotted lines represent the time evolution
of the fugacity and temperature of an ideal fluid. The solid lines with red,
with green, and with blue, represent the moments of the binary collision
integral, while the dashed lines (with AW labels) of the same color
represent the RTA of the collision integral. All left~(a) panels show the
solutions of the conservation equations closed by the moment equations, (%
\protect\ref{BJ_I300_relax}) with red, (\protect\ref{BJ_I400_relax}) with
green, and (\protect\ref{BJ_I500_relax}) with blue. Similarly all right~(b)
panels show the solutions of the conservation equations closed by the moment
equations, (\protect\ref{BJ_I320_relax}) with red, (\protect\ref%
{BJ_I420_relax}) with green, and (\protect\ref{BJ_I520_relax}) with blue.
Note that all solutions presented in the left~(a) panels are very similar,
and they differ by less than the thickness of the line; hence, they overlap
and cover each other.}
\label{fig:set_xi_0=0}
\end{figure*}

\begin{figure*}[tbp]
\begin{tabular}{cc}
\includegraphics[width=.49\linewidth]{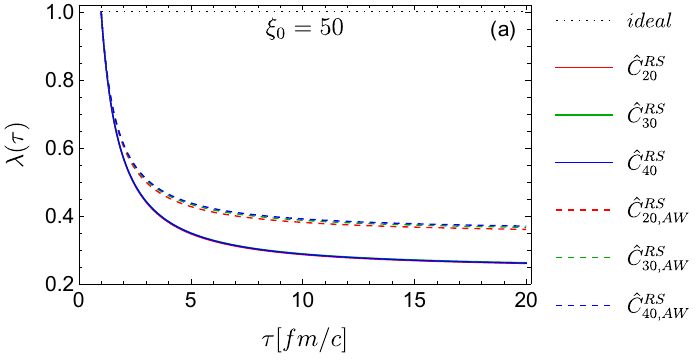} & %
\includegraphics[width=.49\linewidth]{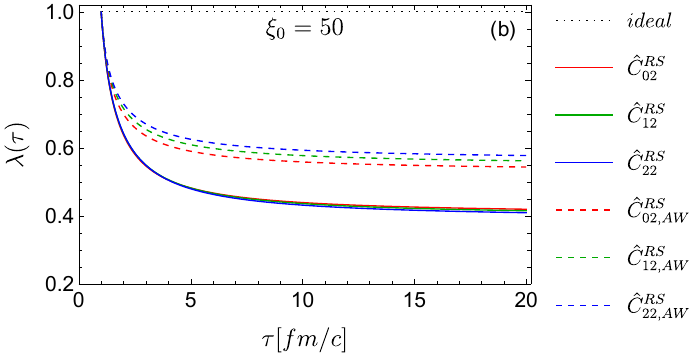} \\ 
\includegraphics[width=.49\linewidth]{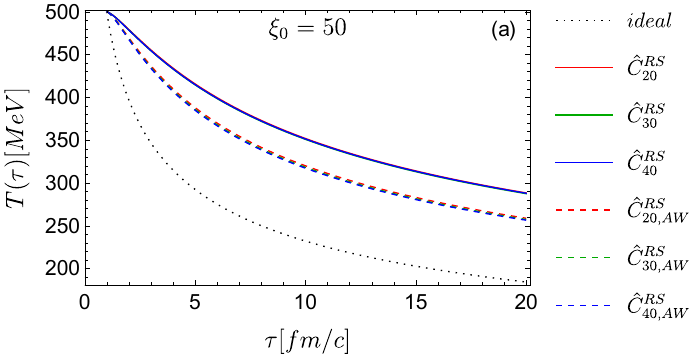} & %
\includegraphics[width=.49\linewidth]{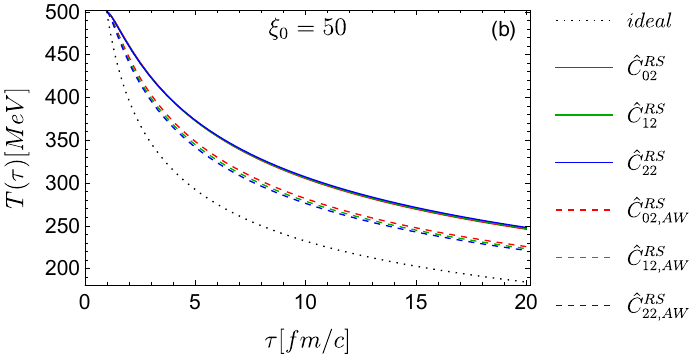} \\ 
\includegraphics[width=.49\linewidth]{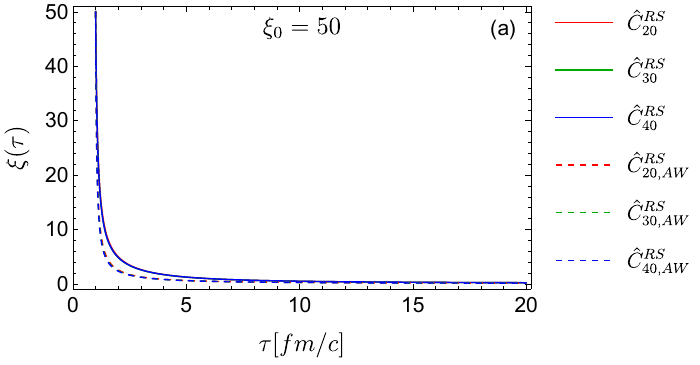} & %
\includegraphics[width=.49\linewidth]{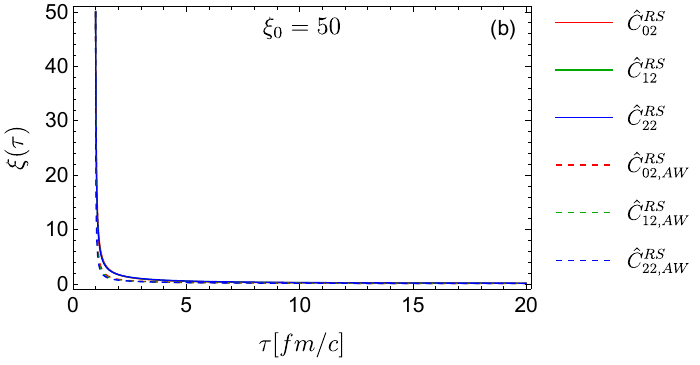} \\ 
\includegraphics[width=.49\linewidth]{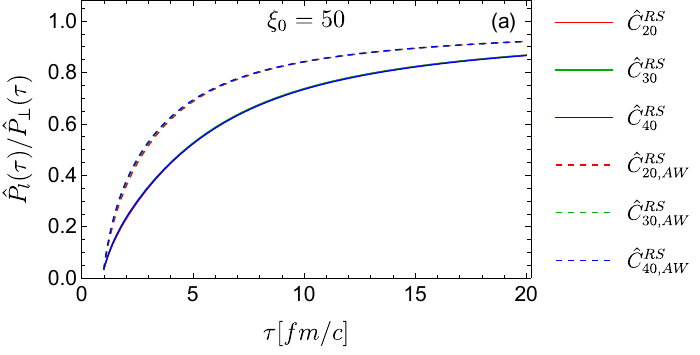} & %
\includegraphics[width=.49\linewidth]{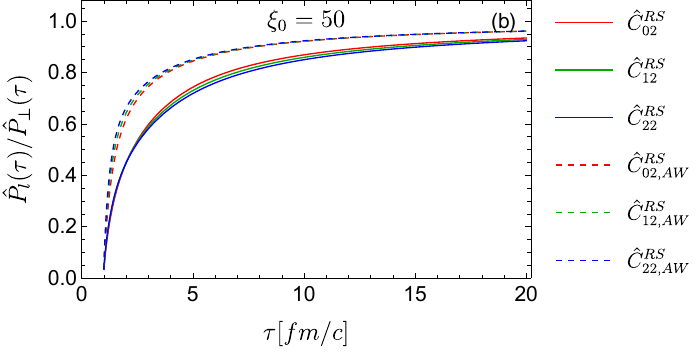}%
\end{tabular}%
\caption{The same as Fig.~\protect\ref{fig:set_xi_0=0} but for an initial
anisotropy of $\protect\xi_0 = 50$.  The black dotted lines represent 
the time evolution of the fugacity and temperature of an ideal fluid. 
The solid lines with red, with green, and with blue, represent 
the moments of the binary collision integral, the dashed lines 
of the same color represent the RTA of the collision integral.}
\label{fig:set_xi_0=50}
\end{figure*}

In all cases we have initialized the system at $\tau_0 = 1$ fm/c and
with a temperature of $T(\tau_0) \equiv T_0 = 0.5$ GeV and a chemical
potential of $\mu_0=0$ GeV; hence, the initial fugacity is $\lambda(\tau_0)
\equiv \lambda_0=1$. In all cases the relaxation time is fixed as constant, 
$\tau_R = 0.5$ fm/c. 
Furthermore, all plots in Fig.~\ref{fig:set_xi_0=0} correspond to an
initially isotropic distribution, i.e., the initial value of anisotropy is $%
\xi(\tau_0) \equiv \xi_0 = 0$, while all plots in Fig.~\ref{fig:set_xi_0=50}
correspond to an initially oblate spheroidal distribution, i.e., the initial
anisotropy is $\xi(\tau_0) \equiv \xi_0 = 50$.


\begin{figure*}[tbp]
	\begin{tabular}{cc}
		\includegraphics[width=.49\linewidth]{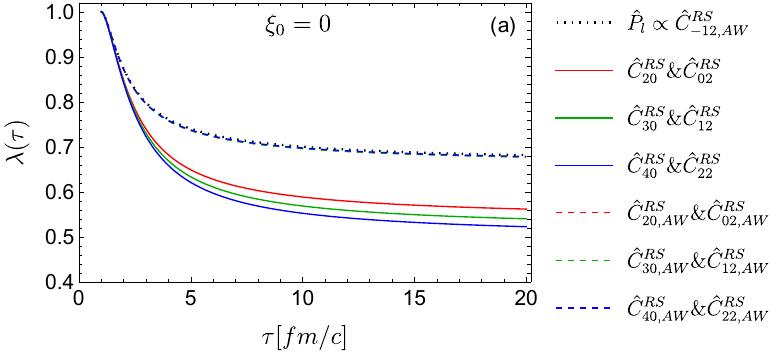} & %
		\includegraphics[width=.49\linewidth]{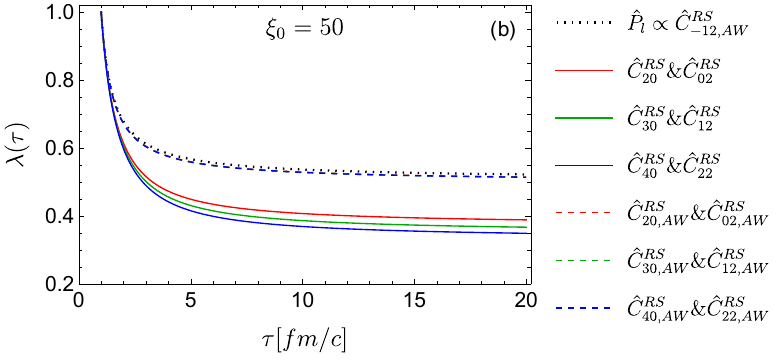} \\ 
		\includegraphics[width=.49\linewidth]{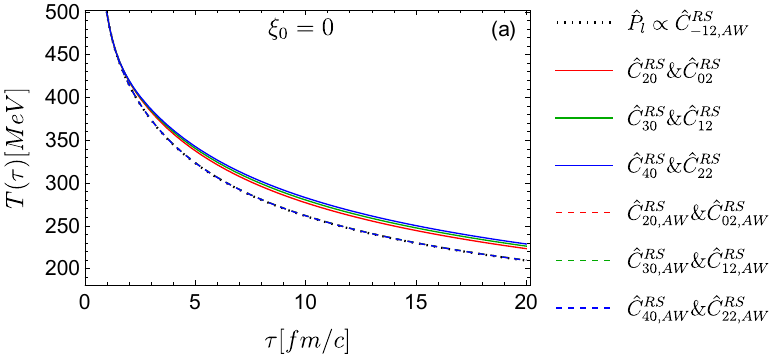} & %
		\includegraphics[width=.49\linewidth]{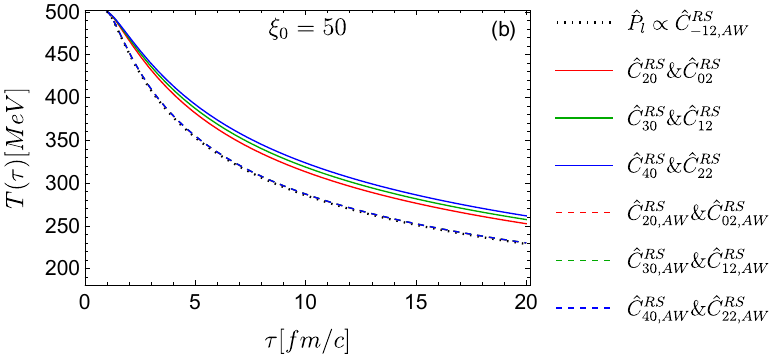} \\ 
		\includegraphics[width=.49\linewidth]{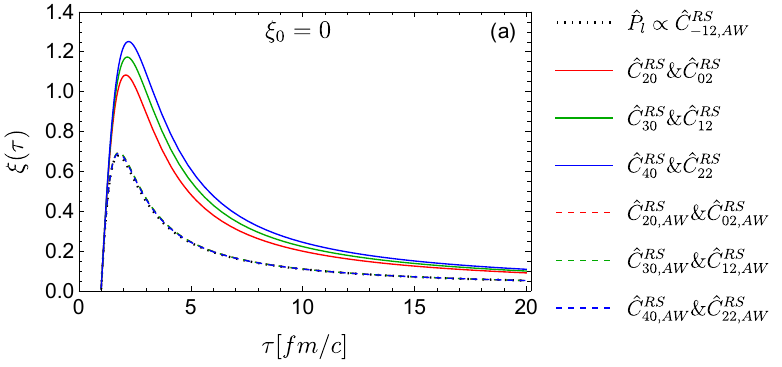} & %
		\includegraphics[width=.49\linewidth]{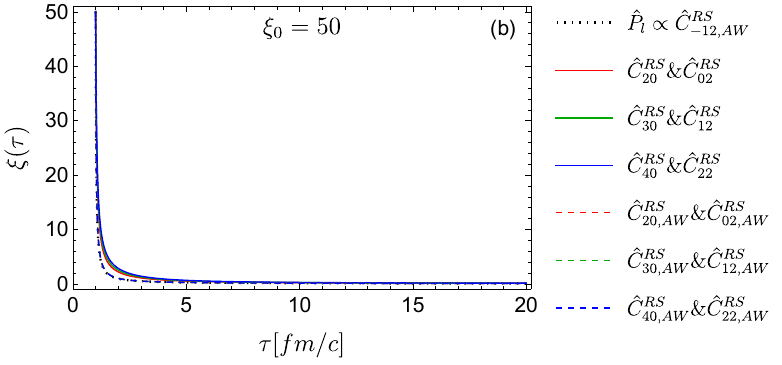} \\ 
		\includegraphics[width=.49\linewidth]{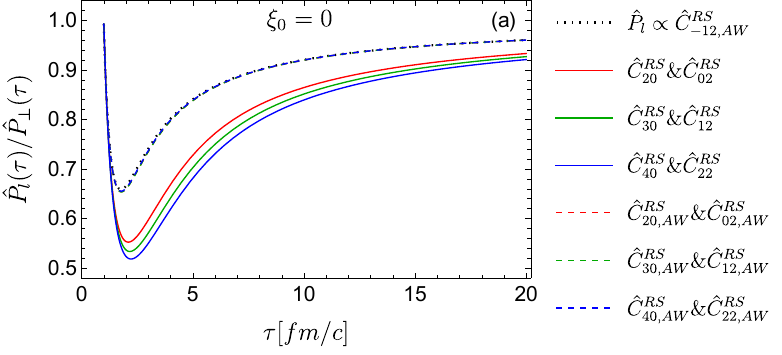} & %
		\includegraphics[width=.49\linewidth]{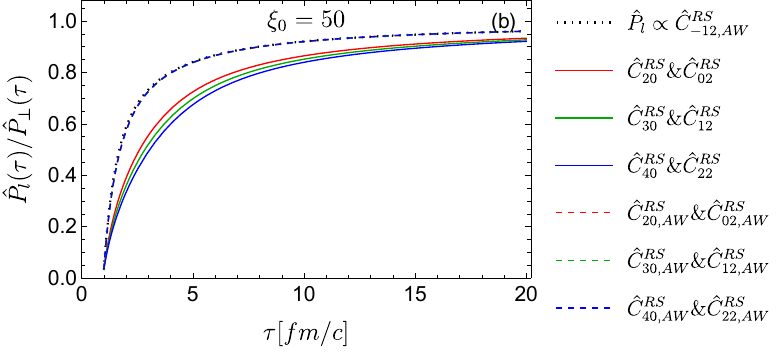}%
	\end{tabular}
	\caption{Similar to Fig.~\protect\ref{fig:set_xi_0=0} and Fig.~\protect\ref%
		{fig:set_xi_0=50}. The evolution of the fugacity $\protect\lambda( \protect%
		\tau)$, the temperature $T(\protect\tau)$, the anisotropy parameter $\protect%
		\xi(\protect\tau)$, and the ratio of the longitudinal and the transverse
		pressure components $\hat{P}_l(\protect\tau)/ \hat{P}_{\perp}( \protect\tau)$.
		Here the black dotted lines represent the solution of the conservation
		equations closed by the equation of $\hat{P}_l$ in the RTA from Eq.~(\protect
		\ref{BJ_Pl_relax}). The solid red, green, and blue lines are the solutions
		corresponding to the binary collision terms, while the dashed lines (with AW
		labels) of the same color coding represents the RTA. The left~(a) panels
		with $\protect\xi_0 = 0$ show the solutions of the conservation equations
		closed by the moment equations, (\protect\ref{BJ_I300-I320_relax}) with red,
		(\protect\ref{BJ_I400-I420_relax}) with green, and (\protect\ref%
		{BJ_I500-I520_relax}) with blue lines. Similarly the right~(b) panels show
		the solutions of the same dynamical equations but with an initial anisotropy
		of $\protect\xi_0 = 50$. Note that all solutions in the RTA presented in
		both~(a) and~(b) panels are very similar and less than the thickness of the
		line; hence, they overlap and cover each other.}
	\label{fig:set_xi_0=0_50}
\end{figure*}

Fig.~\ref{fig:set_xi_0=0} and Fig.~\ref{fig:set_xi_0=50} from top to bottom
show the evolution of the fugacity $\lambda(\tau)$, the temperature $T(\tau)$, 
the anisotropy parameter $\xi(\tau)$, and the ratio of the longitudinal pressure 
and the transverse pressure $\hat{P}_l(\tau)/\hat{P}_{\perp}(\tau)$ 
as a function of proper time. 
All figures in the left columns~(a), i.e., Fig.~\ref{fig:set_xi_0=0}(a) and 
Fig.~\ref{fig:set_xi_0=50}(a), correspond to the solution of the conservation 
laws closed by one of the moment equations: Eq.~(\ref{BJ_I300_relax}), 
Eq.~(\ref{BJ_I400_relax}), or Eq.~(\ref{BJ_I500_relax}), with red, green, 
and blue lines respectively. 
Similarly all figures in the right columns~(b), i.e., Fig.~\ref{fig:set_xi_0=0}(b)
and Fig.~\ref{fig:set_xi_0=50}(b), correspond to the solutions obtained by
one of the moment equations: Eq.~(\ref{BJ_I320_relax}), Eq.~(\ref{BJ_I420_relax}), 
or Eq.~(\ref{BJ_I520_relax}), with red, green, and blue lines, respectively. 
In all cases the solid lines always represent the solutions of anisotropic fluid dynamics
including the moments of the binary collision integral, while the dashed
lines represent the solutions of the fluid dynamical equations including the moments of the RTA.

The system expands and cools down; therefore, the fugacity and the temperature
of the system decrease as time passes (see the first and the second rows of
Fig.~\ref{fig:set_xi_0=0} and Fig.~\ref{fig:set_xi_0=50}). 
This change is faster for the fugacity and slower for the temperature for a nonzero initial
anisotropy. The reason is that any decrease in fugacity has to be
compensated by an increase in temperature; hence, a smaller $\lambda(\tau)$
requires a larger $T(\tau)$. An ideal fluid, without any dissipation, has the
largest pressure, $P_0=e_0/3 =n_0 T$; hence, it expands and cools fastest. 
Consequently, during the expansion the temperatures in anisotropic fluid dynamics are 
consistently higher than in ideal fluid dynamics, shown with a black dotted line in the figures.

The evolution of the anisotropy parameter $\xi$ and the ratio of pressures 
$\hat{P}_l/\hat{P}_\perp$ are shown in the third and fourth rows of
Fig.~\ref{fig:set_xi_0=0} and Fig.~\ref{fig:set_xi_0=50}. 
In case the system was initially in equilibrium, i.e., isotropic distribution 
with $\xi_0 = 0$, the left and right columns of Fig.~\ref{fig:set_xi_0=0}, 
the longitudinal expansion of the system drives the system out of equilibrium. 
This leads to an increase in the anisotropy parameter $\xi$ which in turn leads to a
decrease of the ratio of pressures $\hat{P}_l/\hat{P}_\perp$. 
This only lasts for about $2-3$ fm/c after which the longitudinal pressure 
builds up and the system starts to be driven toward a state of local equilibrium,
i.e., $\xi \rightarrow 0$ and $\hat{P}_l \rightarrow\hat{P}_\perp \simeq P_0$. 
For an initially anisotropic configuration when $\xi_0 = 50$, the
left and right columns of Fig.~\ref{fig:set_xi_0=50}, the
longitudinal pressure increases faster than for an initially isotropic configuration. 
Comparing both the left and the right columns of Fig.~\ref{fig:set_xi_0=0} 
to Fig.~\ref{fig:set_xi_0=50} reveals this distinction.
Nevertheless, the late-time behavior is quite similar in both cases 
as the system eventually approaches an isotropic state, $\xi \rightarrow 0$.

These conclusions are generally valid for both (binary and RTA) collision terms 
as well as for the six different choices of closure for the conservation equations.
As already discussed in detail in Ref.~\cite{Molnar:2016gwq}, the inclusion of 
some of these as well as other moments for closure (different from those considered here), 
also demonstrates a universal grouping of the corresponding results
based on the power of the longitudinal momentum $E_{\mathbf{k}l}^j$.
In the current cases of interest when $j=0$, hence corresponding to the anisotropic 
moments $\hat{I}^{RS}_{i00}$ with $i=3,4,5$, all the curves overlap entirely with 
differences that are less than the thickness of the line in the case of the 
binary collision integral. 
Similarly, in the RTA the dashed lines are also overlapping for the most part; 
see the left columns of Fig.~\ref{fig:set_xi_0=0} and Fig.~\ref{fig:set_xi_0=50}. 
The case when all the anisotropic moments include the second power, $j=2$, 
of the longitudinal momentum, i.e., $\hat{I}^{RS}_{i+2,2,0}$ with $i=1,2,3$, 
small differences, about the thickness of the line, between the solutions 
become visible, see the right columns of Fig.~\ref{fig:set_xi_0=0} and Fig.~\ref{fig:set_xi_0=50}. 
This can be observed about all relevant quantities irrespective of the initial 
anisotropy of the system; hence, the choice which closes the conservation equations strongly depends 
on the power of the longitudinal momentum and less on the power of energy. 
Here we showed that this universal grouping is present and persists regardless 
of the choice for the collision term.

The other important observation regarding these results are about 
the striking differences in the evolution of the system due to the
differences between the nonlinear binary collision integral and the RTA; 
compare the corresponding solid lines to the dashed lines in Fig.~\ref{fig:set_xi_0=0} and Fig.~\ref{fig:set_xi_0=50}. 
The relaxation-time approximation of the collision term drives the 
system toward equilibrium faster than the binary collision integral. 
This was expected based on the explicit formulas and discussions in Sec.~\ref{sec:RS_binary_vs_RTA}. 
We will return to this issue and explicitly study the behavior of the moments 
of the collision integral as a function of the anisotropy parameter 
in more detail in Sec.~\ref{sec:RTA_matching}.

\subsection{The importance of two dynamical moments}
\label{sec:Results_BJ_two}

The results of the previous section demonstrated the ambiguity in selecting a 
higher-order moment for closure, as different choices result in distinct groups 
of solutions based on the power of the longitudinal momentum.
In Refs.~\cite{Molnar:2016gwq,Niemi:2017stb} the so-called "judicious choice" of moment, 
$\hat{P}_l = \hat{I}_{220}$, i.e., the driving force in the energy conservation equation, 
offered the best overall agreement between anisotropic fluid dynamics and the solution of 
the Boltzmann equation in the RTA.
Furthermore, it was also shown that using other higher-order moments with the second power 
of the longitudinal momentum 
i.e., $\hat{I}_{i+2,2,0}$ with $i>0$, also leads to a very good agreement 
to the solution of the Boltzmann equation, similar to the equation for $\hat{P}_l$. 
However, when choosing other dynamical moments than $\hat{P}_l$ to close the conservation 
equations, the algebraically derived values for $\hat{P}_l$ deviate substantially from the
exact solutions as shown and discussed in Ref.~\cite{Molnar:2016gwq}.

These indicate that in a coupled hierarchy of moment equations including more than one dynamical 
moment to close the conservation equations would be beneficial. 
This is because the proper solution to the moment equations~(\ref{Main_eq_motion}) 
requires that all (infinitely many) higher-order moments should be included to obtain the
solution to the underlying microscopic equation, the Boltzmann equation.
The lower-order moments of the distribution function describe the lower frequency dynamics 
while moments of higher-order capture the higher frequency dynamics. 
Although an exact representation of the distribution function is practically unattainable 
through the moment method, a well-defined finite truncation of the coupled hierarchy is expected to 
yield a reasonably good approximation for the distribution function. 
Therefore including an increasing number of dynamical moments to close the conservation 
equations should also improve the solutions to the fluid dynamical equations 
(for the lower-order moments which are included)  and 
the approximation to the anisotropic distribution function.
A well-known example is transient fluid dynamics which is based on a series expansion around 
equilibrium, hence including additional higher-order moments dynamically, 
also improves the fluid dynamical solutions when compared to the Boltzmann equation~\cite{Denicol:2012vq}.

Inspecting the main equation of motion~(\ref{Main_eq_motion}) for any given set of indices 
$i,j>0$ we observe that the left hand side contains only two variables, $\hat{I}_{i+j,j,0}^{RS}$ and $\hat{I}_{i+j,j+2,0}^{RS}$.
Thus rewriting this equation for indices corresponding to $i\rightarrow i-2$ and 
$j \rightarrow j+2$ leads to a differential equation for $\hat{I}_{i+j,j+2,0}^{RS}$. 
From these two coupled differential equations we obtain the following 
equation of motion that governs the evolution of (a linear combination of) two dynamical moments
\begin{align}
&\frac{\partial \hat{I}_{i+j,j,0}^{RS}}{\partial \tau } -\frac{(i-1)}{(j+3)}%
\frac{\partial \hat{I}_{i+j,j+2,0}^{RS}}{\partial \tau } +\frac{1}{\tau }%
\left[ \left( j+1\right) \hat{I}_{i+j,j,0}^{RS} - \frac{(i-1)(i-3)}{(j+3)} 
\hat{I}_{i+j,j+4,0}^{RS}\right]  \notag \\
&= \hat{C}_{i-1,j}^{RS} - \frac{(i-1)}{(j+3)}\hat{C}_{i-3,j+2}^{RS} \approx 
\hat{C}_{i-1,j,AW}^{RS} - \frac{(i-1)}{(j+3)} \hat{C}_{i-3,j+2,AW}^{RS} \, .
\label{Main_eq_motion_2}
\end{align}%
This equation when contrasted to Eq.~(\ref{Main_eq_motion}) corresponds to a linear combination of 
two selected dynamical moments for the pair of indices $i$ and $j$.
This also means that the proper time evolution of the anisotropic distribution function will be 
represented by three new parameters, $\alpha^{RS}_{ij}(\tau)\rightarrow \alpha^{RS}_{i,j,j+2}(\tau)$, $\beta^{RS}_{ij}(\tau)\rightarrow \beta^{RS}_{i,j,j+2}(\tau)$, and 
$\xi_{ij}(\tau) \rightarrow \xi_{i,j,j+2}(\tau)$.

Now, using the previously given equations of motion from Sec.~\ref{sec:Results_BJ}, here we 
list the newly obtained moment equations.
Therefore, recalling Eq.~(\ref{BJ_I300_relax}) and Eq.~(\ref{BJ_I320_relax}), 
or equivalently writing Eq.~(\ref{Main_eq_motion_2}) for $i=3$ and $j=0$, leads to
the following equation of motion to close the conservation equations,
\begin{equation}
\frac{\partial \hat{I}_{300}^{RS}}{\partial \tau } - \frac{2}{3}\frac{%
\partial \hat{I}_{320}^{RS}}{\partial \tau } + \frac{1}{\tau } \hat{I}%
_{300}^{RS} =\hat{C}_{20}^{RS} - \frac{2}{3}\hat{C}_{02}^{RS}\approx \hat{C}%
_{20,AW}^{RS} - \frac{2}{3}\hat{C}_{02,AW}^{RS}\, .
\label{BJ_I300-I320_relax}
\end{equation}
Then similarly, using Eq.~(\ref{BJ_I400_relax}) and Eq.~(\ref{BJ_I420_relax}), or 
similarly writing Eq.~(\ref{Main_eq_motion_2}) for $i=4$ and $j=0$, we obtain 
\begin{equation}
\frac{\partial \hat{I}_{400}^{RS}}{\partial \tau } -\frac{\partial \hat{I}%
_{420}^{RS}}{\partial \tau } +\frac{1}{\tau }\left(\hat{I}_{400}^{RS}-\hat{I}%
_{440}^{RS}\right) =\hat{C}_{30}^{RS} - \hat{C}_{12}^{RS} \approx \hat{C}%
_{30,AW}^{RS} - \hat{C}_{12,AW}^{RS}\, ,  \label{BJ_I400-I420_relax}
\end{equation}
and finally, from Eq.~(\ref{BJ_I500_relax}) and Eq.~(\ref{BJ_I520_relax}), or from Eq.~(\ref{Main_eq_motion_2}) for $i=5$ and $j=0$, we get
\begin{equation}
\frac{\partial \hat{I}_{500}^{RS}}{\partial \tau } -\frac{4}{3}\frac{%
\partial \hat{I}_{520}^{RS}}{\partial \tau } +\frac{1}{\tau }\left(\hat{I}%
_{500}^{RS} - \frac{8}{3}\hat{I}_{540}^{RS}\right) =\hat{C}_{40}^{RS} -\frac{%
4}{3}\hat{C}_{22}^{RS} \approx \hat{C}_{40,AW}^{RS} -\frac{4}{3}\hat{C}%
_{22,AW}^{RS} \, .  \label{BJ_I500-I520_relax}
\end{equation}%
The solutions to the conservation equations~(\ref{BJ_n_cons}) and (\ref{BJ_e_cons}) 
closed by these new moment equations are shown in Fig.~\ref%
{fig:set_xi_0=0_50}. Similar to the previous results in Fig.~\ref%
{fig:set_xi_0=0} and Fig.~\ref{fig:set_xi_0=50} this new figure also shows
the evolution of the fugacity $\lambda(\tau)$, the temperature $T(\tau)$,
the anisotropy parameter $\xi(\tau)$ and the ratio of the longitudinal and
the transverse pressure components $\hat{P}_l(\tau)/\hat{P}_{\perp}(\tau)$ as
a function of proper time.

All figures in the left (a) and right (b) columns of Fig.~\ref%
{fig:set_xi_0=0_50} present the solutions of the conservation laws closed by
the moment equations corresponding to Eq.~(\ref{BJ_I300-I320_relax}), Eq.~(%
\ref{BJ_I400-I420_relax}), and Eq.~(\ref{BJ_I500-I520_relax}); red, green
and blue lines, respectively. These are denoted in the figure captions by $%
\hat{C}^{RS}_{20} \& \hat{C}^{RS}_{02}$, $\hat{C}^{RS}_{30} \& \hat{C}%
^{RS}_{12}$, and $\hat{C}^{RS}_{40} \& \hat{C}^{RS}_{22}$. The solid lines
always represent the solutions of anisotropic fluid dynamics including the
moments of the binary collision integral, while the dashed lines represent
the solutions of the fluid dynamical equations in the RTA and the captions
have the additional $AW$ label. The difference between the (a) and (b)
columns are due to the different initial values of the anisotropy parameter
that is $\xi_0 = 0$ and $\xi_0 = 50$ respectively. Furthermore, in Fig.~\ref%
{fig:set_xi_0=0_50} the black dotted lines represent the solution of the
conservation equations closed by the moment equation for $\hat{P}_l$ in the RTA, i.e., Eq.~(\ref{BJ_Pl_relax}).

As expected, now the solutions to the moment equations closed by
two dynamical moments~(\ref{BJ_I300-I320_relax})-(\ref{BJ_I500-I520_relax}) 
show a very good agreement with the "judicious choice" which represents the best 
match to the exact numerical solution of the Boltzmann equation in the 
RTA~\cite{Molnar:2016gwq,Niemi:2017stb,Tinti:2015xwa}.
These new solutions in the RTA presented in 
Fig.~\ref{fig:set_xi_0=0_50} differ 
from each other and from the best match by less than the thickness of the
line; hence, they overlap and cover each other.

Similarly, also the solutions with the nonlinear collision integral are 
in a better agreement with the solutions shown in the (b) panels of 
Fig.~\ref{fig:set_xi_0=0} and Fig.~\ref{fig:set_xi_0=50}; i.e., the solutions
for $\hat{I}_{i+2,2,0}$. 
To express it differently, the solutions corresponding to a
single dynamical moment $\hat{I}_{i,0,0}$  shown in the (a) panels of
Fig.~\ref{fig:set_xi_0=0} and Fig.~\ref{fig:set_xi_0=50}, improve
considerably by taking into account an additional dynamical moment of 
the second power of the longitudinal momentum, i.e.,
$\hat{I}_{i+2,2,0}$. 
Therefore, we also expect that using the general equation
of motion~(\ref{Main_eq_motion_2}) recursively to include additional
dynamical moments would also improve the solutions for the higher-order
moments when compared to the Boltzmann equation.

Even so important differences between the solutions corresponding to the
binary collision integral and the collision term in the RTA remain; this is
apparent when comparing the full lines to the dashed lines in all figures
presented so far. In the next section we aim to minimize these differences
through the rescaling of the relaxation-time parameter that optimally
matches the moments of the nonlinear binary collision integral to the
moments in the RTA.

\subsection{Matching to the RTA and the rescaling of the relaxation time}

\label{sec:RTA_matching}

\begin{figure*}[tbp]
\begin{tabular}{cc}
\includegraphics[width=.49\linewidth]{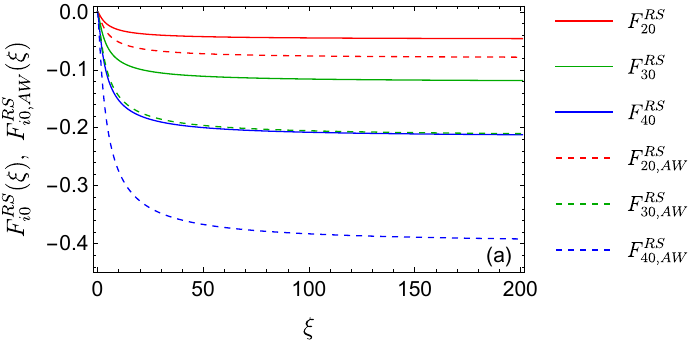} & %
\includegraphics[width=.49\linewidth]{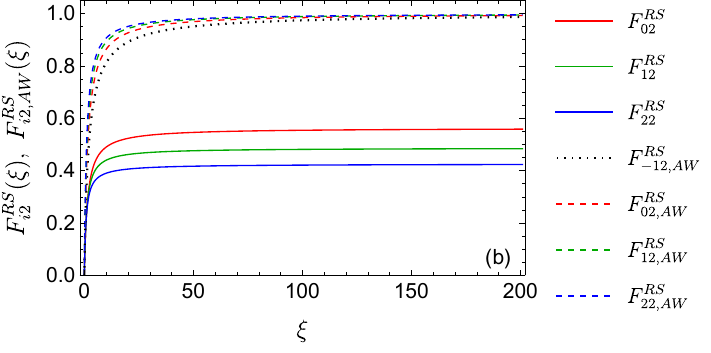}%
\end{tabular}%
\caption{The normalized and dimensionless collision terms $F_{ij}^{RS}$ with 
solid lines, $F_{ij,AW}^{RS}$ with dashed lines, and $F_{-12,AW}^{RS}$ with dotted black line. 
(a) Left: the red solid line is $F_{20}^{RS}$, the green solid line 
is $F_{30}^{RS}$, while the blue solid line is $F_{20}^{RS}$. The dashed lines
with the same color coding represent $F_{i0,AW}^{RS}$, i.e., the ratios in the RTA. 
(b) Right: $F_{02}^{RS}$, $F_{12}^{RS}$,
and $F_{22}^{RS}$, with red, green, and blue lines. 
The dashed lines with red, green and blue represent, $F_{02,AW}^{RS}$, $F_{12,AW}^{RS}$, and $F_{22,AW}^{RS}$.}
\label{fig:F_ij}
\end{figure*}

\begin{figure*}[tbp]
\begin{tabular}{cc}
\includegraphics[width=.49\linewidth]{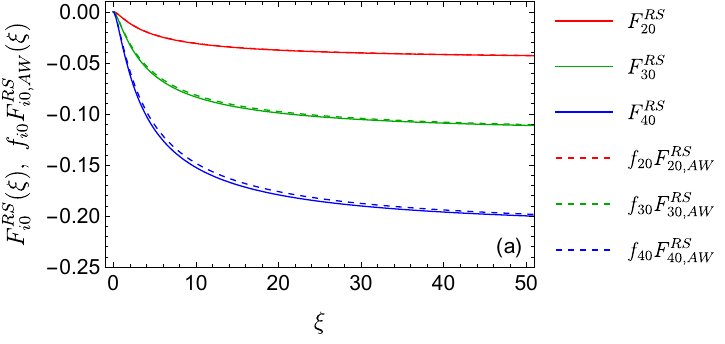} & %
\includegraphics[width=.49\linewidth]{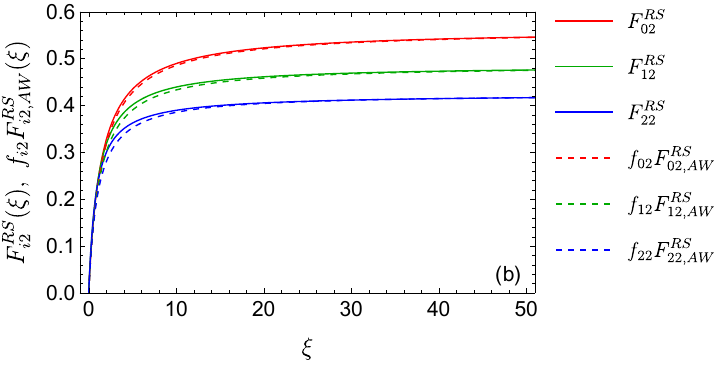}%
\end{tabular}%
\caption{The same as Fig.~\protect\ref{fig:F_ij}, but with scaled values
[see Eq.~(\ref{RTA_scale_factor})] of the collision terms in the RTA.
(a) Left:  the solid lines with red, green, and blue show $F_{20}^{RS}$, $F_{30}^{RS}$, and $F_{20}^{RS}$, while the dashed lines with the same color coding show $f_{i0} F_{i0,AW}^{RS}$.
(b) Right: with red, green, and blue solid lines show, $F_{02}^{RS}$, $F_{12}^{RS}$,
and $F_{22}^{RS}$, while the dashed lines
with the same color coding show $f_{i2} F_{i2,AW}^{RS}$.}
\label{fig:F_ij_scale}
\end{figure*}

\begin{figure*}[tbp]
\begin{tabular}{cc}
\includegraphics[width=.49\linewidth]{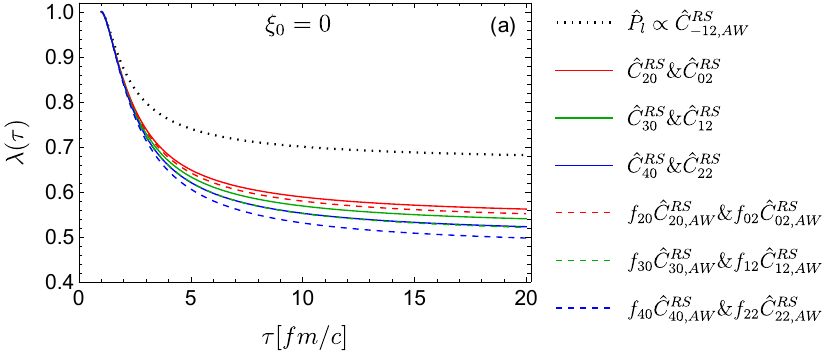} & %
\includegraphics[width=.49\linewidth]{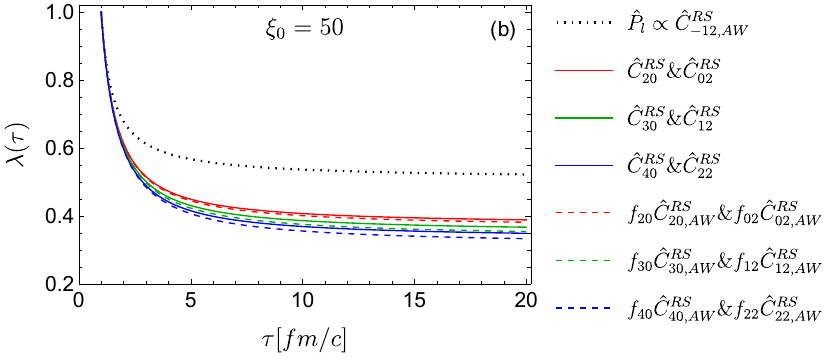} \\ 
\includegraphics[width=.49\linewidth]{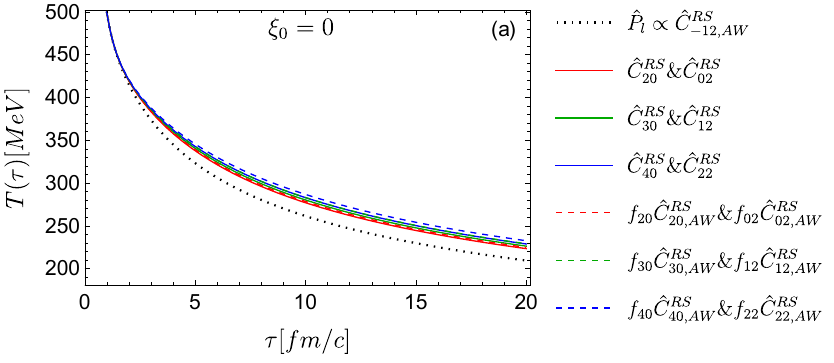} & %
\includegraphics[width=.49\linewidth]{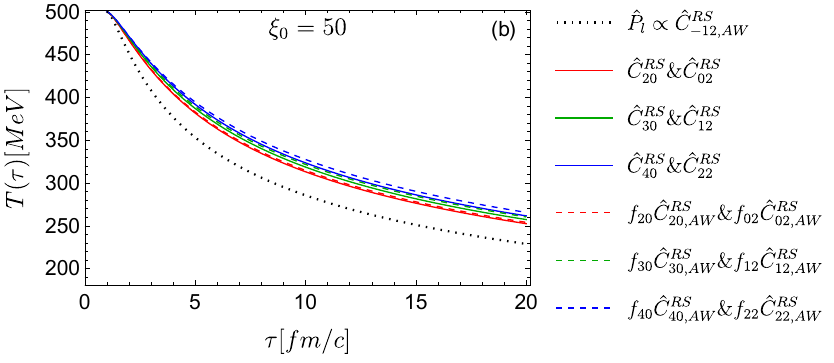} \\ 
\includegraphics[width=.49\linewidth]{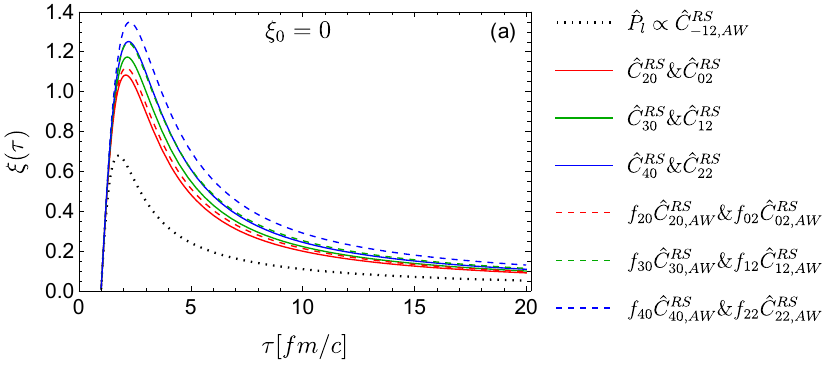} & %
\includegraphics[width=.49\linewidth]{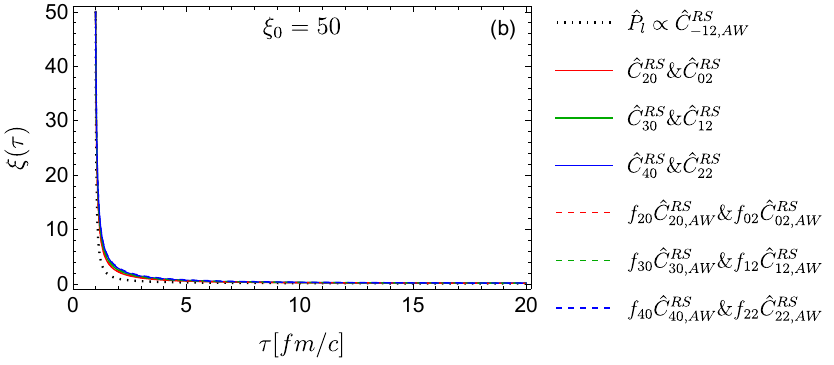} \\ 
\includegraphics[width=.49\linewidth]{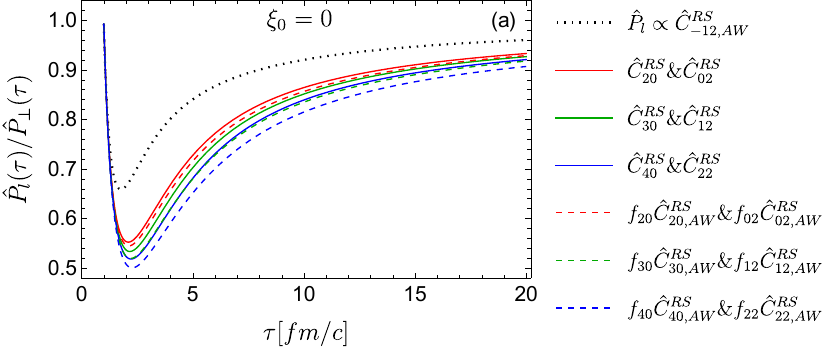} & %
\includegraphics[width=.49\linewidth]{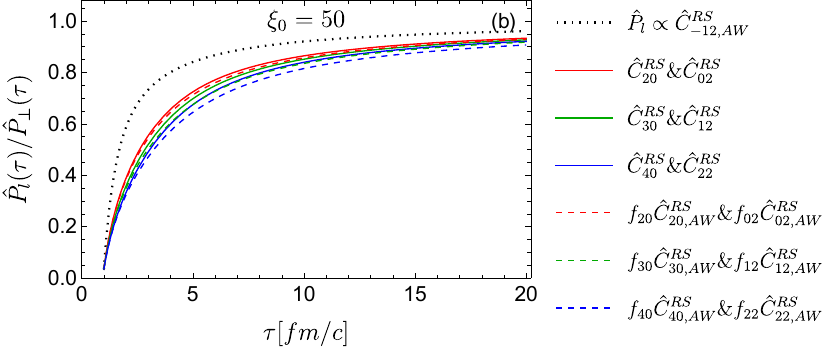}%
\end{tabular}
\caption{Similar to Fig.~\protect\ref{fig:set_xi_0=0_50}, showing the
evolution of the fugacity $\protect\lambda( \protect\tau)$, the temperature $%
T(\protect\tau)$, the anisotropy parameter $\protect\xi(\protect\tau)$ and
the ratio of the longitudinal and the transverse pressure components $\hat{P}_l%
(\protect\tau)/ \hat{P}_{\perp}( \protect\tau)$. The black dotted lines
represent the solution of the conservation equations closed by the equation
of $\hat{P}_l$ from Eq.~(\protect\ref{BJ_Pl_relax}). The solid lines with
red, green, and blue represent solutions of two dynamical moments and
corresponding binary collision terms, while the dashed lines of the same
color represent the solutions with two moments and rescaled RTA. All (a)
panels are for $\protect\xi_0 = 0$ while all (b) panels show the solutions
of the same equations but with an initial anisotropy of $\protect\xi_0 = 50$.}
\label{fig:set_xi_0=0_50_RTA}
\end{figure*}

To better understand the differences between the binary collision integral
and the RTA we will study the moments of the collision integral $\hat{C}_{ij}$, multiplied
by the relaxation time parameter $\tau_{R}$, and divided by the
corresponding equilibrium moment $I_{i+j+1,j,0}$. 
Thereby we define the following dimensionless ratios:
\begin{eqnarray}
F_{ij}^{RS}\left( \xi \right)  &\equiv &\tau _{R}\frac{\hat{C}_{ij}^{RS}(\xi
)}{I_{i+j+1,j,0}}\,, \\
F_{ij,AW}^{RS}\left( \xi \right)  &\equiv &\tau _{R}\frac{\hat{C}%
_{ij,AW}^{RS}(\xi )}{I_{i+j+1,j,0}}=1-\frac{\hat{I}_{i+j+1,j,0}^{RS}}{%
I_{i+j+1,j,0}}\,,
\end{eqnarray}%
where $F_{ij}^{RS}$ is defined using the moments of the binary collision integral,
while $F_{ij,AW}^{RS}$ makes use of Eq.~(\ref{Coll_Int_RTA}).
Note that these ratios are functions of the anisotropy parameter alone, and hence, 
we can study the behavior of the moments of the collision integral as
a function of $\xi$.

Now recalling the moments of the nonlinear binary collision integral from
Eqs.~(\ref{C_RS_20})-(\ref{C_RS_40}) and Eqs.~(\ref{C_RS_02})-(\ref{C_RS_22}) as well as 
the moments of the RTA approximation~(\ref{Coll_Int_RTA}), we present the corresponding 
dimensionless ratios in Fig.~\ref{fig:F_ij}. 
In Fig.~\ref{fig:F_ij}(a) on the left, the red solid line is $F_{20}^{RS}$,
the green solid line is $F_{30}^{RS}$, while the blue solid line is $%
F_{40}^{RS}$. The dashed lines with the same color coding show the
corresponding ratio in the RTA, i.e., $F_{20,AW}^{RS}$, $F_{30,AW}^{RS}$, and $%
F_{40,AW}^{RS}$. Similarly, in Fig.~\ref{fig:F_ij}(b) on the right, the red
solid line is $F_{02}^{RS}$, the green solid line is $F_{12}^{RS}$, while
the blue solid line is $F_{22}^{RS}$. The dashed lines
with the same color coding represent the ratios in the RTA, i.e., $F_{i2,AW}^{RS}$.

First and foremost, we observe that for all values of the anisotropy parameter the absolute
value of the moments of the normalized binary collision integral, $F_{ij}^{RS}
$, are always smaller than the absolute value corresponding to the same moments of the collision
integral in the RTA, $F_{ij,AW}^{RS}$. This also means that the lhs of the moment
equations is larger in the RTA which in turn drives the system faster to
equilibrium. As expected for $\xi=0$, the normalized moments of the
collision integrals are also zero; $\lim\limits_{\xi \rightarrow
0}F_{ij}^{RS}\left( \xi \right) =\lim\limits_{\xi \rightarrow
0}F_{ij,AW}^{RS}\left( \xi \right) =0$. Furthermore, by increasing $\xi>0$
these ratios also increase and approach their corresponding maximums (asymptote)
as a function of the anisotropy parameter. 

In the asymptotic limit when $\xi\rightarrow \infty $, 
the ratios $F_{i0,AW}^{RS}\left( \xi \right) $ and 
$F_{i0}^{RS}\left( \xi \right) $ lead to the following finite values
\begin{align}
\lim\limits_{\xi \rightarrow \infty }F_{20,AW}^{RS}\left( \xi \right) 
=1- \frac{32}{3\pi ^{2}}\, ,\quad & \lim\limits_{\xi \rightarrow\infty}
F_{20}^{RS}\left( \xi \right) =-\frac{512 - 45\pi^2}{144\pi ^{2}}\, ,
\\
\lim\limits_{\xi \rightarrow \infty }F_{30,AW}^{RS}\left( \xi \right) =1- 
\frac{12}{\pi ^{2}}\, ,\quad & \lim\limits_{\xi \rightarrow
\infty}F_{30}^{RS}\left( \xi \right) =-\frac{6}{5\pi ^{2}}\, , \\
\lim\limits_{\xi \rightarrow \infty }F_{40,AW}^{RS}\left( \xi \right) =1- 
\frac{2048}{15\pi ^{4}}\, ,\quad & \lim\limits_{\xi \rightarrow
\infty}F_{40}^{RS}\left( \xi \right) =-\frac{2816-189\pi ^{2}}{45\pi ^{4}}\, ,
\end{align}
and similarly the $F_{i2,AW}^{RS}\left( \xi \right) $ and $F_{i2}^{RS}\left(
\xi \right) $ lead to 
\begin{align}
\lim\limits_{\xi \rightarrow \infty }F_{02,AW}^{RS}\left( \xi \right) =1\, ,
\quad & \lim\limits_{\xi \rightarrow \infty }F_{02}^{RS}\left( \xi \right)
= \frac{9}{16}\, , \\
\lim\limits_{\xi \rightarrow \infty }F_{12,AW}^{RS}\left( \xi \right) =1\, ,
\quad & \lim\limits_{\xi \rightarrow \infty }F_{12}^{RS}\left( \xi \right)
= \frac{24}{5\pi ^{2}}\, , \\
\lim\limits_{\xi \rightarrow \infty }F_{22,AW}^{RS}\left( \xi \right) =1\, ,
\quad & \lim\limits_{\xi \rightarrow \infty }F_{22}^{RS}\left( \xi \right)
= \frac{6656+1215\pi ^{2}}{450\pi ^{4}}\, .
\end{align}%
The ratio of the (normalized) collision terms in the asymptotic limit,
\begin{equation}
f_{ij} \equiv \lim\limits_{\xi \rightarrow \infty }
\frac{F_{ij}^{RS} \left(\xi\right)}{F_{ij,AW}^{RS} \left(\xi\right)} 
= \lim\limits_{\xi \rightarrow \infty } \frac{\hat{C}_{ij}^{RS}}{\hat{C}_{ij,AW}^{RS}} \, ,
\label{RTA_scale_factor}
\end{equation}
defines the corresponding scaling factor of the relaxation time parameter $\tau_R$ of the RTA that optimally 
matches the asymptotic value of the binary collision integral as a function of the anisotropy parameter. 
Therefore, using the asymptotic scale factors we expect to improve the relaxation time approximation, similarly as in Eq.~(\ref{Coll_Int_RTA}), but now with properly scaled relaxation-time parameter,
$\tau_R \rightarrow \tau_{ij}$, as 
\begin{align}
\hat{C}_{ij}^{RS}(\xi) &\approx f_{ij} \hat{C}_{ij,AW}^{RS}(\xi) \equiv 
\frac{\tau_R}{\tau_{ij}}\hat{C}_{ij,AW}^{RS} (\xi)  \notag \\
&= -\frac{1}{\tau_{ij}}\hat{I}_{i+j+1,j,0}^{RS} +\frac{1}{\tau_{ij}}%
I_{i+j+1,j,0} \, , \label{RTA_ij}
\end{align}%
where the corresponding asymptotic relaxation times are 
\begin{equation}
\tau_{ij} \equiv \tau_R f_{ij}^{-1} = \tau_R \lim\limits_{\xi \rightarrow
\infty } \frac{\hat{C}_{ij,AW}^{RS}(\xi)}{\hat{C}_{ij}^{RS} (\xi)} \, .
\end{equation}
Henceforth, the inverse of the nonlinear binary collision term is proportional to
microscopic timescales and can be interpreted as an effective relaxation
time proportional to $\tau_R=1/(\sigma_T n_0)$.
However, the relaxation times corresponding to the binary collision integral have different 
relaxation rates or collision frequencies $f_{ij}$ for different moments.
For example, using the previously obtained asymptotic values, here we list the numerical results,
\begin{align}
& f^{-1}_{20} = \frac{1536 -144\pi^2}{512 - 45\pi^2} \approx 1.691 \, ,
\quad f^{-1}_{02} = \frac{16}{9} \approx 1.777 \, , \\
& f^{-1}_{30} = 10 -\frac{5\pi^2}{6} \approx 1.775 \, , \qquad \quad
f^{-1}_{12} = \frac{5 \pi^2}{24} \approx 2.056 \, , \\
& f^{-1}_{40} = \frac{6144 -45\pi^4}{2816 -189\pi^2} \approx 1.852 \, ,
\quad f^{-1}_{22} = \frac{450\pi^{4}}{6656+1215\pi ^{2}} \approx 2.350 \, .
\end{align}
These results show that the "proper" relaxation times of the binary collision integral 
$\tau_{ij}$ increase with the order of the moments, i.e., the power $i$ of energy 
$E_{\mathbf{k}u}$ and the power $j$ of momentum in the direction of the anisotropy $E_{\mathbf{k}l}$.
This is, in fact, also well-known from the linearized
binary collision integral in transient fluid dynamics, where various
physical phenomena, such as viscosity or diffusion processes are also
described by different moments which by their nature relax on different
timescales~\cite{Wagner:2023joq}.
The nonlinear collision dynamics and hence the differences in timescales cannot 
be captured by the simplistic RTA with a single relaxation-time parameter~\cite{Ambrus:2023ilm,Ambrus:2024qsa}.
However, inspecting the moments of the nonlinear binary collision term we find that
the time it takes to reach local equilibrium becomes longer for higher-order moments. 
This translates to slower relaxation timescales and hence the higher-order moments contributing 
to the higher-frequency dynamics should not be neglected from the fluid dynamical description.
This behavior arises from coupling effects characteristic of the nonlinear Boltzmann 
collision term which leads to a spectrum of microscopic relaxation timescales even 
in the isotropic limit~\cite{Bazow:2015dha,Bazow:2016oky}.

The outcome of these modifications to the RTA using Eq.~(\ref{RTA_ij}) 
is shown in Fig.~\ref{fig:F_ij_scale},
where we have plotted the moments of the binary collision integral $%
F_{i0}^{RS}(\xi)$ on the lhs (a) and $F_{i2}^{RS}(\xi)$ on rhs (b) with red,
green, and blue, lines. The corresponding scaled values $f_{i0}
F_{i0}^{RS}(\xi)$ are included on the lhs of Fig.~\ref{fig:F_ij_scale} panel (a), 
while $f_{i2} F_{i2}^{RS}(\xi)$ are on the rhs of 
Fig.~\ref{fig:F_ij_scale} panel (b) with dashed lines, red,
green, and blue. 
The overall agreement is excellent and only small differences are visible at small $\xi$, 
while the differences gradually diminish with the increase of the anisotropy parameter 
as we approach the asymptotic limit.

The evolution of the expanding system using the scaled relaxation
times $\tau_{ij}$ in the RTA according to Eq.~(\Ref{RTA_ij}) are shown 
in Fig.~\ref{fig:set_xi_0=0_50_RTA}. 
However, here for the sake of brevity we only present the fluid dynamical solutions 
in the case when the conservation laws were closed by two dynamical 
moments~(\ref{Main_eq_motion_2}) as in the previous section. 
Similar to Fig.~\ref{fig:set_xi_0=0_50}, we present the
evolution of the fugacity, the temperature, the anisotropy parameter, and
the ratio of the longitudinal and the transverse pressure components as
function of the proper time. The overall conclusions regarding the results
presented in Fig.~\ref{fig:set_xi_0=0_50_RTA} are that using the asymptotic
scaling for the RTA matches the solutions obtained with the nonlinear
binary collision integral reasonably well. The small differences 
between the binary collision integral and the scaled RTA presented in Fig.~\ref{fig:F_ij_scale}, 
drive the evolution of the system in slightly different ways.

\section{Conclusions and Outlook}
\label{sec:Conclusions}

In this paper we have computed the scalar moments of the nonlinear binary
collision integral in the ultrarelativistic hard-sphere approximation assuming
a constant and energy independent cross section. 
The moments of the collision integral are expressed in quadratic
products of anisotropic thermodynamic integrals and 
represent the nonlinear dependence of the binary collision integral 
on the distribution function.
These exact results are valid for arbitrary anisotropic distribution functions 
of the form $\hat{f}_{0\mathbf{k}}\left( \hat{\alpha},\hat{\beta}_{u}E_{\mathbf{k}u},
\hat{\beta}_{l}E_{\mathbf{k}l}\right)$, which are commonly interpreted having distinct
temperatures in the direction of the anisotropy $\hat{\beta}^{-1}_l$ and perpendicular to it $\hat{\beta}^{-1}_u$.

As a particular example we have chosen the well-known spheroidal
distribution function, also known as the Romatschke-Strickland distribution
function~\cite{Romatschke:2003ms}, to study the moments of the binary
collision integral. In this case the moments of the  binary
collision integral simplify considerably since the corresponding
anisotropic thermodynamical integrals vanish, i.e., $\hat{I}^{RS}_{nrq} =0$, 
for all odd powers of the momentum in the direction of the anisotropy $E^{r}_{\mathbf{k}l}$.  
Furthermore, the well-known and widely used relativistic relaxation-time
approximation of Anderson and Witting~\cite{Anderson:1974nyl} for the
collision integral was also included for comparison.

To better understand the differences between the nonlinear binary collision
integral and the RTA we have studied the time evolution of an
ultrarelativistic massless but particle conserving system in the well-known
longitudinal boost-invariant Bjorken expansion scenario. 
Then we discussed various possible closures for the conservation laws based on 
different higher-order moments of the Boltzmann equation.
Additionally, we demonstrated that the ambiguity in selecting a higher-order moment for closure 
can be effectively resolved by including not one but two corresponding dynamical moments from the hierarchy.
Although now the choice of moment(s) selected for the closure is no longer 
of primary concern, differences between the moments of the nonlinear binary collision 
integral and the corresponding moments computed in the relaxation-time 
approximation remain significant. 
These differences highlight the importance of nonlinear couplings between the moments 
and the limitations of the RTA.

Our main conclusion regarding the spheroidal distribution function is that
the collision term in the RTA drives the system toward equilibrium faster
than the binary collision integral irrespective of the choice of the
dynamical moment(s) used to close the conservation laws. To remedy such
inaccuracies inherent in the simplified relaxation-time approximation we
have rescaled the relaxation time of the RTA based on the asymptotic
values of the ratio of the binary collision integral and the RTA collision integral. 
This scaled RTA leads to the relaxation-time parameters 
of the binary collision integral of higher-order moments $\tau_{ij}(\tau_R)$ and 
reproduces reasonably well the properties of the nonlinear binary collision integral 
in the case of the RS distribution function.
Therefore, similar to earlier comparisons to second-order fluid dynamical theories~\cite{Bouras:2010hm,Gallmeister:2018mcn}, comparisons between 
leading-order anisotropic fluid dynamical models and relativistic kinetic models which solve 
the nonlinear Boltzmann collision term can also be made using our exact results.
Furthermore, the difference between the relaxation times should be observable in hydrodynamic simulations such as aHydroQP~\cite{Alqahtani:2016ayv,Nopoush:2016qas,Alqahtani:2017jwl,Alqahtani:2017tnq,
Alqahtani:2020paa,Strickland:2024oat}
since the pressure anisotropies driving the expansion of the QGP will relax on different 
timescales and hence lead to different lifetimes of the plasma. 

In closing we also mention that these exact analytic results are also
important in the case of a more complete anisotropic distribution function, 
$\hat{f}_{\mathbf{k}} = \hat{f}_{0 \mathbf{k}} + \delta \hat{f}_{\mathbf{k}}$, 
since the same steps and principles as presented here are required to
compute the moments of the more general collision integral, 
$C\!\left[ \hat{f}_{0} + \delta \hat{f}\right]$. 
Moreover, our exact results are also useful for investigating the fluid dynamical evolution 
of a wide range of anisotropic distribution functions utilized in plasma physics,
such as the anisotropic J\"uttner or the bi-Maxwellian distribution functions.

\begin{acknowledgments}
The author thanks P.~Huovinen and H.~Niemi for reading the manuscript and for constructive comments and useful discussions. 
The author also thanks S.~Horv\'at for his MaTeX package~\cite{MaTeX} which was used 
to create figure labels using Latex in Mathematica. 
E.M. was supported by the program Excellence Initiative--Research University of the 
University of Wroc{\l}aw of the Ministry of Education and Science. 
Furthermore, the author acknowledges support by the Deutsche Forschungsgemeinschaft 
(DFG, German Research Foundation) through the CRC-TR 211 ``Strong-interaction matter under extreme conditions'' 
-- Project No. 315477589 -- TRR 211.
\end{acknowledgments}

\appendix
\section{The general equations of motion of leading-order anisotropic fluid dynamics}
\label{app:equations_of_motion}

For reasons of completeness we recall here the general equations of
anisotropic fluid dynamics from Refs.~\cite{Molnar:2016gwq} see 
Eqs.~(24) and~(25) of Ref.~\cite{Molnar:2016gwq},
\begin{align}
\mathcal{C}_{i-1,j}& =D\hat{\mathcal{I}}_{ij}-D_{l}\hat{\mathcal{I}}_{i-1,j+1}
-\left[ i\hat{\mathcal{I}}_{i-1,j+1}+j\hat{\mathcal{I}}_{i+1,j-1}\right] 
l_{\lambda }Du^{\lambda } 
+\left[ \left( i-1\right) \hat{\mathcal{I}}_{i-2,j+2}+\left( j+1\right) 
\hat{\mathcal{I}}_{i,j}\right] l_{\lambda }D_{l}u^{\lambda }  \notag \\
& -\frac{1}{2}\tilde{\theta}\left[ m_{0}^{2}\left( i-1\right) 
\hat{\mathcal{I}}_{i-2,j}-\left( i+1\right) \hat{\mathcal{I}}_{i,j}+\left( i-1\right) 
\hat{\mathcal{I}}_{i-2,j+2}\right] 
+\frac{1}{2}\tilde{\theta}_{l}\left[ m_{0}^{2}j\hat{\mathcal{I}}_{i-1,j-1} 
-j\hat{\mathcal{I}}_{i+1,j-1}+\left( j+2\right) \hat{\mathcal{I}}_{i-1,j+1}\right] \, ,  
\label{DI_ij}
\end{align}%
and 
\begin{align}
\mathcal{C}_{i-1,j}^{\{\mu \}} & = \frac{1}{2}\tilde{\nabla}^{\mu}
\left( m_{0}^{2}\hat{\mathcal{I}}_{i-1,j}-\hat{\mathcal{I}}_{i+1,j}
+\hat{\mathcal{I}}_{i-1,j+2}\right)  \notag \\
&- \frac{\left( i-1\right) }{2}\left( m_{0}^{2}\hat{\mathcal{I}}_{i-2,j+1}
-\hat{\mathcal{I}}_{i,j+1}+\hat{\mathcal{I}}_{i-2,j+3}\right) l_{\lambda }%
\tilde{\nabla}^{\mu}u^{\lambda } 
- \frac{j}{2}\left( m_{0}^{2}\hat{\mathcal{I}}_{i,j-1}-\hat{\mathcal{I}}_{i+2,j-1}
+\hat{\mathcal{I}}_{i,j+1}\right) 
l_{\lambda }\tilde{\nabla}^{\mu}u^{\lambda }  \notag \\
&- \frac{1}{2}\left[ m_{0}^{2}i\hat{\mathcal{I}}_{i-1,j}-\left( i+2\right) 
\hat{\mathcal{I}}_{i+1,j}+i\hat{\mathcal{I}}_{i-1,j+2}\right] 
\Xi _{\lambda}^{\mu}Du^{\lambda } 
+ \frac{1}{2}\left[ m_{0}^{2}j\hat{\mathcal{I}}_{i,j-1}-j\hat{\mathcal{I}}_{i+2,j-1}
+\left( j+2\right) \hat{\mathcal{I}}_{i,j+1}\right] 
\Xi _{\lambda}^{\mu}Dl^{\lambda }  \notag \\
&+ \frac{1}{2}\left[ m_{0}^{2}\left( i-1\right) \hat{\mathcal{I}}_{i-2,j+1}
-\left( i+1\right) \hat{\mathcal{I}}_{i,j+1}+\left( i-1\right) 
\hat{\mathcal{I}}_{i-2,j+3}\right] \Xi _{\lambda }^{\mu}D_{l}u^{\lambda}  \notag \\
&- \frac{1}{2}\left[ m_{0}^{2}\left( j+1\right) \hat{\mathcal{I}}_{i-1,j}
-\left( j+1\right) \hat{\mathcal{I}}_{i+1,j}+\left( j+3\right) 
\hat{\mathcal{I}}_{i-1,j+2}\right] \Xi _{\lambda }^{\mu}D_{l}l^{\lambda }\, .
\label{DI_ij_mu1}
\end{align}%
Here, $\partial_\mu \equiv u_\mu D + l_\mu D_l + \tilde{\nabla}_{\mu }$,
such that $D \equiv u^{\mu }\partial _{\mu }$ denotes the comoving
derivative and $D_{l}\equiv -l^{\mu }\partial _{\mu }$ is the derivative in
the direction of the anisotropy, while the spatial gradient orthogonal to
both $u^{\mu }$ and $l^{\mu }$ is 
$\tilde{\nabla}_{\mu } \equiv \Xi _{\mu\nu }\partial ^{\nu }$. 
The expansion scalars are $\tilde{\theta}=\tilde{\nabla}_{\mu }u^{\mu }$ 
and $\tilde{\theta}_{l}=\tilde{\nabla}_{\mu }l^{\mu }$.

Now, noting that due to the conservation of the particle-number in binary
collisions $\hat{C}_{00}=0$, while $\hat{C}_{10}=0$ and $\hat{C}_{01}=0$
vanish due to energy conservation and momentum conservation in the direction
of the anisotropy. Thus the particle number conservation equation follows
from Eq.~(\ref{DI_ij}) by choosing $i=1$ and $j=0$, 
\begin{align}
& 0 =\partial _{\mu }\hat{N}^{\mu } \equiv D\hat{n}-D_{l}\hat{n}_{l} 
+\hat{n} \tilde{\theta} +\hat{n}_{l}\tilde{\theta}_{l} +\hat{n} l_{\mu }D_{l}u^{\mu }
- \hat{n}_{l} l_{\mu}Du^{\mu } \, ,  
\label{conservation_eq_particle}
\end{align}
while the energy-conservation equation is obtained by choosing $i=2$ and $j=0$, 
\begin{align}
& 0 =u_{\nu }\partial _{\mu }\hat{T}^{\mu \nu } \equiv D\hat{e}-D_{l}\hat{M}
+\left( \hat{e}+\hat{P}_{\perp }\right) \tilde{\theta}+\hat{M}\tilde{\theta}_{l} 
+\left( \hat{e}+\hat{P}_{l}\right) l_{\mu }D_{l}u^{\mu } - 2\hat{M}l_{\mu }Du^{\mu } \, ,  
\label{conservation_eq_energy}
\end{align}
and the conservation equation for the momentum in the direction of the
anisotropy follows for $i=1$ and $j=1$, 
\begin{align}
& 0 =l_{\nu }\partial _{\mu }\hat{T}^{\mu \nu } \equiv D\hat{M} - D_{l}\hat{P}_{l} 
+ \hat{M}\tilde{\theta} -\left( \hat{P}_{\perp }-\hat{P}_{l}\right) 
\tilde{\theta}_{l} + 2\hat{M} l_{\mu }D_{l}u^{\mu } 
- \left( \hat{e}+\hat{P}_{l}\right) l_{\mu }Du^{\mu }\, .  
\label{conservation_eq_momenta_l}
\end{align}
Finally, due to the conservation of transverse momenta in microscopic
collisions $\hat{C}_{00}^{\left\{ \mu \right\} }=0$, hence the conservation
equation for the momentum transverse to the direction of the anisotropy
follows from Eq.~(\ref{DI_ij_mu1}) for $i=1$ and $j=0$:
\begin{align}
0 =\Xi _{\nu }^{\alpha }\partial _{\mu }\hat{T}^{\mu \nu } 
&\equiv \left(\hat{e}+\hat{P}_{\perp }\right) \left( Du^{\alpha } 
+l^{\alpha }l_{\nu}Du^{\nu }\right) -\tilde{\nabla}^{\alpha }\hat{P}_{\perp } 
+\left(\hat{P}_{\perp }-\hat{P}_{l}\right) \left( D_{l}l^{\alpha } 
+u^{\alpha }l_{\nu }D_{l}u^{\nu}\right)  \notag \\
&+ \hat{M}\left( Dl^{\alpha }+u^{\alpha } l_{\nu }Du^{\nu }\right) 
-\hat{M}\left( D_{l}u^{\alpha }+l^{\alpha } l_{\nu}D_{l}u^{\nu }\right) \, .
\label{conservation_eq_momenta_t}
\end{align}

\section{The center of momentum frame}

\label{app:CM_frame}

It is useful to define the center-of-momentum (CM) frame where the total
momentum in binary collisions in Eq.~(\ref{P_T}) is 
$P_{T}^{\mu }\equiv \left(P_{T}^{0}, \mathbf{P}_{T} \right)\overset{\mathrm{CM}}{=}
\left( \sqrt{s},\mathbf{0}\right)$, such that 
\begin{align}
P_{T}^{0} &\overset{\mathrm{CM}}{=} k^{0}+k^{\prime 0} = p^{0}+p^{\prime 0} 
= \sqrt{s}\, , \\
\mathbf{P}_{T} &\overset{\mathrm{CM}}{=} \mathbf{k}+\mathbf{k}^{\prime } 
= \mathbf{p}+\mathbf{p}^{\prime }=\mathbf{0}\, .
\end{align}%
Furthermore, since $P_{T}^{\mu }P_{T,\mu }=s$, in the CM frame we also have, 
\begin{align}
P_{T}^{\mu }k_{\mu } &\equiv P_{T}^{\mu }k_{\mu }^{\prime
}=m_{0}^{2}+k^{\mu }k_{\mu }^{\prime }\overset{\mathrm{CM}}{=}\frac{s}{2}\, ,
\label{P_Tk_s2} \\
\Delta _{T}^{\mu \nu }k_{\mu }k_{\nu } &\equiv 
m_{0}^{2} -\left( P_{T}^{\mu}k_{\mu }\right)^{2}/s 
\overset{\mathrm{CM}}{=}m_{0}^{2}-\frac{s}{4}\, ,
\label{Delta_Tk_s4}
\end{align}%
and hence in the ultrarelativistic limit, when $k^{\mu }k_{\mu}=m_{0}^{2}\rightarrow 0$, 
and $k^{0}=|\mathbf{k}|\equiv k$, we also obtain,
\begin{equation}
s\equiv P_{T}^{\mu }P_{T,\mu }\underset{m_{0}\rightarrow 0}{=} 2k^{\mu}k_{\mu }^{\prime }\, .
\end{equation}

Recalling Eq.~(\ref{B_nq}) together with the definition of the transition
rate from Eq.~(\ref{W_kk_pp}) we obtain the $\mathcal{B}_{nq}$ coefficient
\begin{eqnarray}
\mathcal{B}_{nq} &\equiv &\frac{\left( -1\right) ^{q}}{\left( 2q+1\right) !!}%
\frac{(2\pi )^{6}}{2}\int_{P\!P^{\prime }\!}s\frac{d\sigma (s,\Omega )}{%
d\Omega }\delta (k^{\mu }+k^{\prime \mu }-p^{\mu }-p^{\prime \mu })\left( 
\frac{P_{T}^{\mu }p_{\mu }}{s}\right) ^{n-2q}\left( \Delta _{T}^{\mu \nu
}p_{\mu }p_{\nu }\right) ^{q}  \notag \\
&=&\frac{1}{\left( 2q+1\right) !!}\frac{\sigma _{T}(s)}{2^{n}}\int \frac{dp}{%
\left( p^{0}\right) ^{2}}p^{2}\delta (\sqrt{s}-2p^{0})s\left(
s-4m_{0}^{2}\right) ^{q}  \notag \\
&=&\frac{\sigma _{T}(s)}{2^{n+1}\left( 2q+1\right) !!}\sqrt{s}\left(
s-4m_{0}^{2}\right) ^{\left( 2q+1\right) /2}\, ,
\end{eqnarray}%
where we used that $p^{0}\equiv p^{\prime 0}=\sqrt{s}/2$ and $\int_{P\!P^{\prime }} \delta (k^{\mu }+k^{\prime \mu}-p^{\mu }-p^{\prime \mu }) = 1/(2\pi)^{5}$.
This is precisely the result obtained earlier; see, for example Eq.~(G6) in Ref.~\cite{Molnar:2013lta} 
and Eq.~(D3) in Ref.~\cite{Wagner:2023joq}. Note, however, that both references
contain the same typographical error in the definition of $\mathcal{B}_{nq}$
as an integral over $P$ and $P'$. 
This is easily remedied by replacing $\sqrt{s}\rightarrow s$ in the term 
$\left(P_{T}^{\mu }p_{\mu }/\sqrt{s}\right) \rightarrow \left(P_{T}^{\mu }p_{\mu }/s\right)$ 
in those references. 
Also note that this typographical error does not affect the results of 
Refs.~\cite{Molnar:2013lta,Wagner:2023joq} since they used the correct 
value of $\mathcal{B}_{nq}$ obtained after integration.

Using this result we obtain the 
$\Theta ^{\mu _{1}\cdots \mu _{n}}\equiv \mathcal{P}_{00}^{\mu _{1}\cdots \mu _{n}}$ tensors
defined in Eq.~(\ref{Theta_mu1_mun}); hence, for $n=0$ and $n=1$ we obtain
\begin{align}
\Theta &\equiv \mathcal{P}_{00}= \sigma _{T}\frac{s}{2} \, ,  \label{Theta} \\
\Theta^{\mu } &\equiv \mathcal{P}_{00}^{\mu }=\sigma _{T}\frac{s}{2^{2}}P_{T}^{\mu }\, ,  
\label{Theta_mu}
\end{align}%
where we also used that $b_{00}=b_{10}=1$. 
Similarly for $n=2$ and $n=3$, Eq.~(\ref{Theta_mu1_mun}) leads to
\begin{align}
\Theta^{\mu \nu } &\equiv \mathcal{P}_{00}^{\mu \nu }=\sigma _{T}\frac{s}{2^{3}}
\left( P_{T}^{\mu }P_{T}^{\nu }-\frac{1}{3}s\Delta _{T}^{\mu\nu}\right) \, ,  \label{Theta_mu_nu} \\
\Theta^{\mu \nu \alpha } &\equiv \mathcal{P}_{00}^{\mu \nu \alpha }=\sigma_{T} 
\frac{s}{2^{4}}\left( P_{T}^{\mu }P_{T}^{\nu }P_{T}^{\alpha } 
-s\Delta_{T}^{\left( \mu \nu \right. } P_{T}^{\left. \alpha \right) }\right) \, ,
\label{Theta_mu_nu_alpha}
\end{align}%
where $b_{20}=b_{21}=b_{30}=1$, and $b_{31}=3$ was used. Finally for $n=4$ with 
$b_{40} =1 $, $b_{41}=6$, and $b_{42}=3$ we get
\begin{equation}
\Theta^{\mu \nu \alpha \beta }\equiv \mathcal{P}_{00}^{\mu \nu \alpha \beta}
=\sigma _{T}\frac{s}{2^{5}}\left( P_{T}^{\mu }P_{T}^{\nu }P_{T}^{\alpha}P_{T}^{\beta }
-2s\Delta _{T}^{\left( \mu \nu \right. }P_{T}^{\alpha}P_{T}^{\left. \beta \right) } 
+\frac{1}{5}s^{2}\Delta_{T}^{\left( \mu \nu\right. }\Delta_{T}^{\left. \alpha \beta \right) }\right) \, .
\label{Theta_mu_nu_alpha_beta}
\end{equation}

For later use, here we also list the moments of the anisotropic distribution from
Eq.~(\ref{I_ij_tens}) in the following cases:
\begin{align}
\hat{\mathcal{I}}_{ij}^{\mu } &= \hat{I}_{i+j+1,j,0}u^{\mu }
+\hat{I}_{i+j+1,j+1,0}l^{\mu }\, ,  \label{I_mu_ij} \\
\hat{\mathcal{I}}_{ij}^{\mu \nu } &= \hat{I}_{i+j+2,j,0}u^{\mu }u^{\nu } 
+2\hat{I}_{i+j+2,j+1,0}u^{\left( \mu \right. }l^{\left. \nu \right) }
+\hat{I}_{i+j+2,j+2,0}l^{\mu }l^{\nu }-\hat{I}_{i+j+2,j,1}\Xi ^{\mu \nu }\, ,
\label{I_mu_nu_ij}
\end{align}%
and 
\begin{align}
\hat{\mathcal{I}}_{ij}^{\mu \nu \lambda }& =\hat{I}_{i+j+3,j,0}u^{\mu}u^{\nu }u^{\lambda } 
+3\hat{I}_{i+j+3,j+1,0}u^{\left( \mu \right. }u^{\nu}l^{\left. \lambda \right) }
+3\hat{I}_{i+j+3,j+2,0}u^{\left( \mu \right.}l^{\nu }l^{\left. \lambda \right) }  \notag \\
&+\hat{I}_{i+j+3,j+3,0}l^{\mu }l^{\nu }l^{\lambda } 
-3\hat{I}_{i+j+3,j,1}u^{\left( \mu \right. }\Xi ^{\left. \nu \lambda \right) }
-3\hat{I}_{i+j+3,j+1,1}l^{\left( \mu \right. }\Xi ^{\left. \nu \lambda \right) }\, .
\label{I_mu_nu_lambda_ij}
\end{align}


\section{The loss terms}
\label{app:loss_terms}

Substituting $n\rightarrow n+1$ into Eq.~(\ref{I_ij_tens}) the moments of
the anisotropic distribution function are generalized to $n+1$ tensor indices%
\begin{equation}
\hat{\mathcal{I}}_{ij}^{\mu _{1}\cdots \mu _{n+1}} 
\equiv \sum_{q=0}^{\lfloor\left( n+1\right) /2\rfloor }\sum_{r=0}^{n+1-2q} 
\left( -1\right)^{q}b_{n+1,r,q}\hat{I}_{i+j+n+1,j+r,q} 
\Xi ^{\left( \mu _{1}\mu _{2}\right.}\cdots \Xi ^{\mu _{2q-1}\mu _{2q}} 
l^{\mu _{2q+1}}\cdots l^{\mu_{2q+r}}u^{\mu _{2q+r+1}}\cdots u^{\left. \mu _{n+1}\right) }\, ,
\end{equation}%
where the symmetrized tensor product is 
\begin{align}
& \Xi ^{\left( \mu _{1}\mu _{2}\right. }\cdots \Xi ^{\mu _{2q-1}\mu_{2q}}
l^{\mu _{2q+1}}\cdots l^{\mu _{2q+r}}u^{\mu _{2q+r+1}}\cdots u^{\left.\mu _{n+1}\right) }  
\notag \\
&= \frac{1}{b_{n+1,r,q}}\sum_{\mathcal{P}_{\mu }^{n+1}}\Xi ^{\mu _{1}\mu_{2}}\cdots 
\Xi ^{\mu _{2q-1}\mu _{2q}} l^{\mu _{2q+1}}\cdots l^{\mu_{2q+r}}u^{\mu _{2q+r+1}}\cdots u^{\mu _{n+1}}\, .  
\label{Xi_n+1}
\end{align}%
The total number of permutations of $n+1$ tensor indices is $(n+1)!$. The number of
index permutations on symmetric rank 2 projections $\Xi ^{\mu _{i}\mu _{j}}$
equals $2^{q}q!$, while the four-vectors, $l^{\mu _{i}}$ and $u^{\mu _{j}}$, 
also have $r!$ and $\left( n+1-r-2q\right) !$ numbers of permutations, respectively. 
These index permutations will not lead to distinct terms, and hence one needs to divide 
the total number of index permutations by these factors. 
Therefore the number of distinct terms is set by the $b_{n+1,r,q}$ coefficient, which follows from  Eq.~(\ref{b_nrq}) substituting $n\rightarrow n+1$.

The symmetrized rank $n+1$ tensor product from Eq.~(\ref{Xi_n+1}) can be
separated into three distinct parts where each part will contain a well-defined 
symmetrized rank $n$ tensor product of $\Xi $'s, $u$'s, and $l$'s, 
\begin{align}
\hat{\mathcal{I}}_{ij}^{\mu _{1}\cdots \mu _{n+1}} 
& =\sum_{q=0}^{\lfloor\left( n+1\right) /2\rfloor }\sum_{r=0}^{n+1-2q}\left( -1\right)^{q}
b_{n+1,r,q}\frac{\left( n+1-r-2q\right) }{\left( n+1\right) }\hat{I}_{i+j+n+1,j+r,q}  \notag \\
& \times u^{\mu _{n+1}}\Xi ^{\left( \mu _{1}\mu _{2}\right. }\cdots \Xi^{\mu _{2q-1}\mu _{2q}} 
l^{\mu _{2q+1}}\cdots l^{\mu _{2q+r}}u^{\mu_{2q+r+1}}\cdots u^{\left. \mu _{n}\right) }  \notag \\
& +\sum_{q=0}^{\lfloor \left( n+1\right) /2\rfloor }\sum_{r=0}^{n-2q}\left(-1\right) ^{q}
b_{n+1,r+1,q}\frac{\left( r+1\right) }{\left( n+1\right) }\hat{I}_{i+j+n+1,j+r+1,q}  \notag \\
& \times l^{\mu _{n+1}}\Xi ^{\left( \mu _{1}\mu _{2}\right. }\cdots \Xi^{\mu _{2q-1}\mu _{2q}} 
l^{\mu _{2q+1}}\cdots l^{\mu _{2q+r}}u^{\mu_{2q+r+1}}\cdots u^{\left. \mu _{n}\right) }  \notag \\
& +\sum_{q=0}^{\lfloor \left( n+1\right) /2\rfloor}\sum_{r=0}^{n+1-2q}\left( -1\right) ^{q}
b_{n+1,r,q}\frac{\left( 2q\right) }{\left( n+1\right) }\hat{I}_{i+j+n+1,j+r,q}  \notag \\
& \times \Xi ^{\mu _{n+1}\left( \mu _{1}\right. }\cdots \Xi ^{\mu _{2q-1}\mu_{2q}} 
l^{\mu _{2q+1}}\cdots l^{\mu _{2q+r}}u^{\mu _{2q+r+1}}\cdots u^{\left. \mu _{n}\right) }\, .
\end{align}%
Here the first term removes $u^{\mu _{n+1}}$ from the tensor product~(\ref{Xi_n+1}), 
and therefore, the number of distinct terms remaining under symmetrization is
$b_{n+1,r,q}\left(n+1-r-2q\right) /\left( n+1\right) =b_{nrq}$. 
Similarly, the second term separates $l^{\mu _{n+1}}$ from a total of $r+1$ such 
space-like four-vectors; therefore, the number of remaining terms under symmetrization is 
$b_{n+1,r+1,q}\left( r+1\right) /\left( n+1\right) = b_{nrq}$. 
The last term isolates one tensor index from the rank 2 projection operator leading to terms of
type $\Xi ^{\mu _{n+1}\left( \mu _{1}\right. }\cdots u^{\left. \mu_{n}\right) }$, 
and thus the number of distinct terms remaining under symmetrization is 
$b_{n+1,r,q}\left( 2q\right) /\left(n+1\right)$. 
Furthermore using Eq.~(\ref{b_nrq}) one can show that
$b_{nrq}\equiv b_{n+1,r,q}\frac{\left( n+1-r-2q\right) }{\left( n+1\right) }
=b_{n+1,r+1,q}\frac{\left( r+1\right) }{\left( n+1\right) }$, and then
recalling the definition of thermodynamic integrals from 
Eq.~(\ref{I_ij_tens}) we obtain
\begin{align}
& \hat{\mathcal{I}}_{ij}^{\mu _{1}\cdots \mu _{n}\lambda }=u^{\lambda }\hat{%
\mathcal{I}}_{i+1,j}^{\mu _{1}\cdots \mu _{n}}+l^{\lambda }\hat{\mathcal{I}}%
_{i,j+1}^{\mu _{1}\cdots \mu _{n}}  \notag \\
& +\sum_{q=0}^{\lfloor \left( n+1\right) /2\rfloor
}\sum_{r=0}^{n+1-2q}\left( -1\right) ^{q}b_{n+1,r,q}\frac{\left( 2q\right) }{%
\left( n+1\right) }\hat{I}_{i+j+n+1,j+r,q}\Xi ^{\lambda \left( \mu
_{1}\right. }\cdots \Xi ^{\mu _{2q-1}\mu _{2q}}l^{\mu _{2q+1}}\cdots l^{\mu
_{2q+r}}u^{\mu _{2q+r+1}}\cdots u^{\left. \mu _{n}\right) }\, .
\label{I_ij_mu_lambda}
\end{align}

Now, using the decomposition of the particle four-current, 
$\hat{\mathcal{I}}_{00,\lambda }=\hat{I}_{100}u_{\lambda }+\hat{I}_{110}l_{\lambda }$, 
while also taking into account that all terms found in the second line of 
Eq.~(\ref{I_ij_mu_lambda}) are proportional to $\Xi ^{\lambda \mu }$. Since all such terms 
are by definition orthogonal to both $u_{\lambda }$ and $l_{\lambda }$, therefore all 
terms proportional to $\Xi ^{\lambda \mu }$ vanish from the tensor product 
$\hat{\mathcal{I}}_{ij}^{\mu _{1}\cdots \mu _{n}\lambda }\hat{\mathcal{I}}_{00,\lambda }$, 
and thus we obtain Eq.~(\ref{L_ij_final}).

\section{The gain terms}
\subsection{The $\hat{G}_{n0}$ gain terms}
\label{app:Gn0}

The simplest gain term $\hat{G}_{n0}$ is computed by inserting Eq.~(\ref{Theta_mu1_mun}) 
into Eq.~(\ref{hat_G_ij}), and hence, for $i=n$ and $j=0$ we have
\begin{equation}
\hat{G}_{n0}=\int_{K\!K^{\prime }\!}\hat{f}_{0\mathbf{k}}\hat{f}_{0\mathbf{k}^{\prime }}
\sum_{q=0}^{\lfloor n/2\rfloor }\left( -1\right)^{q}b_{nq}
\mathcal{B}_{nq}\Delta _{T}^{\left( \mu _{1}\mu _{2}\right. }\cdots 
\Delta_{T}^{\mu _{2q-1}\mu _{2q}}P_{T}^{\mu _{2q+1}}\cdots 
P_{T}^{\left. \mu_{n}\right) }u_{\mu _{1}}\cdots u_{\mu _{n}}\, .  
\label{app_Gn0}
\end{equation}%
Now note that the contraction of the symmetrized tensor product by
$u_{\mu _{1}}\cdots u_{\mu _{n}}$ leads to
\begin{align}
\Delta _{T}^{\left( \mu _{1}\mu _{2}\right. }\cdots \Delta _{T}^{\mu_{2q-1}\mu _{2q}} 
P_{T}^{\mu _{2q+1}}\cdots P_{T}^{\left. \mu _{n}\right)}u_{\mu _{1}}\cdots u_{\mu _{n}} 
&\equiv  u_{\mu _{1}}\cdots u_{\mu _{n}}%
\frac{1}{b_{nq}}\sum_{\mathcal{P}_{\mu }^{n}}\Delta _{T}^{\mu _{1}\mu_{2}} \cdots 
\Delta _{T}^{\mu _{2q-1}\mu _{2q}}P_{T}^{\mu _{2q+1}}\cdots
P_{T}^{\mu _{n}}  \notag \\
&= \left( P_{T}^{\mu }u_{\mu }\right)^{n-2q}\left( \Delta _{T}^{\mu \nu}u_{\mu }u_{\nu }\right) ^{q}\, ,
\label{tensor_prod_2}
\end{align}%
and hence, we obtain
\begin{equation}
\Theta ^{\mu _{1}\cdots \mu _{n}}u_{\mu _{1}}\cdots u_{\mu_{n}}
=\sum_{q=0}^{\lfloor n/2\rfloor }\left( -1\right)^{q}b_{nq}\mathcal{B}_{nq}
\left( P_{T}^{\mu }u_{\mu }\right) ^{n-2q}\left( \Delta _{T}^{\mu \nu}u_{\mu }u_{\nu }\right)^{q}\, ,
\end{equation}
which then proves the formula in Eq.~(\ref{Gn0}).

Now, recalling Eq.~(\ref{B_nq_massless}) in the massless limit when $s\equiv
2k^{\mu }k_{\mu }^{\prime }$, the gain term $\hat{G}_{n0}$ from Eq.~(\ref{Gn0}) for $n=0$ leads to 
\begin{align}
\hat{G}_{00} &\equiv \int_{K\!K^{\prime }\!}\hat{f}_{0\mathbf{k}} 
\hat{f}_{0\mathbf{k}^{\prime }}\mathcal{P}_{00} 
=\int_{K\!K^{\prime }\!}\hat{f}_{0\mathbf{k}}\hat{f}_{0\mathbf{k}^{\prime }}\Theta  \notag \\
&= \int_{K\!K^{\prime }\!}\hat{f}_{0\mathbf{k}}\hat{f}_{0\mathbf{k}^{\prime}}
\left[ b_{00}\mathcal{B}_{00}\right] =\sigma _{T}\int_{K\!K^{\prime }\!}
\hat{f}_{0\mathbf{k}}\hat{f}_{0\mathbf{k}^{\prime }}k^{\mu }k_{\mu }^{\prime}  \notag \\
&= \sigma _{T}\hat{\mathcal{I}}_{00}^{\mu }\hat{\mathcal{I}}_{00,\mu}
=\sigma _{T}\left( \hat{I}_{100}\hat{I}_{100}-\hat{I}_{110}\hat{I}_{110}\right) \, ,  
\label{app_G00}
\end{align}%
where $b_{00}=1$, $\mathcal{P}_{00}\equiv \mathcal{B}_{00}=\sigma _{T}s/2$,
and we used Eq.~(\ref{I_mu_ij}) for the decomposition of the tensor moments.

Similarly, Eq.~(\ref{Gn0}) for $n=1$ leads to
\begin{align}
\hat{G}_{10} &\equiv \int_{K\!K^{\prime }\!}\hat{f}_{0\mathbf{k}}
\hat{f}_{0\mathbf{k}^{\prime }}\mathcal{P}_{10} 
=\int_{K\!K^{\prime }\!}\hat{f}_{0\mathbf{k}}\hat{f}_{0\mathbf{k}^{\prime }}
\Theta ^{\mu _{1}}u_{\mu _{1}} \notag \\
&= \int_{K\!K^{\prime }\!}\hat{f}_{0\mathbf{k}}\hat{f}_{0\mathbf{k}^{\prime
}}\left[ b_{10}\mathcal{B}_{10}\left( P_{T}^{\mu }u_{\mu }\right) \right] =%
\frac{\sigma _{T}}{2}\int_{K\!K^{\prime }\!}\hat{f}_{0\mathbf{k}}\hat{f}_{0%
\mathbf{k}^{\prime }}\left( E_{\mathbf{k}u}+E_{\mathbf{k}^{\prime }u}\right)
k^{\mu }k_{\mu }^{\prime }  \notag \\
&= \sigma _{T}\hat{\mathcal{I}}_{10}^{\mu }\hat{\mathcal{I}}_{00,\mu}
=\sigma _{T}\left( \hat{I}_{200}\hat{I}_{100}-\hat{I}_{210}\hat{I}_{110}\right) \, ,  
\label{app_G10}
\end{align}%
where $b_{10}=1$, $\mathcal{B}_{10}=\sigma _{T}s/4$, and we used Eq.~(\ref{PT_u}) 
as well as Eq.~(\ref{I_mu_ij}). 

Likewise for $n=2$ we have 
\begin{align}
\hat{G}_{20} &\equiv \int_{K\!K^{\prime }\!}\hat{f}_{0\mathbf{k}}
\hat{f}_{0\mathbf{k}^{\prime }}\mathcal{P}_{20}=\int_{K\!K^{\prime }\!} 
\hat{f}_{0\mathbf{k}}\hat{f}_{0\mathbf{k}^{\prime }} 
\Theta ^{\mu _{1}\mu _{2}}u_{\mu_{1}}u_{\mu _{2}}  \notag \\
&= \int_{K\!K^{\prime }\!}\hat{f}_{0\mathbf{k}}\hat{f}_{0\mathbf{k}^{\prime}}
\left[ b_{20}\mathcal{B}_{20}\left( P_{T}^{\mu }u_{\mu }\right) ^{2}
-b_{21}\mathcal{B}_{21}\left( \Delta _{T}^{\mu \nu }u_{\mu }u_{\nu }\right) \right]
\notag \\
&= \frac{\sigma _{T}}{3}\int_{K\!K^{\prime }\!}\hat{f}_{0\mathbf{k}}
\hat{f}_{0\mathbf{k}^{\prime }}\left( E_{\mathbf{k}u}+E_{\mathbf{k}^{\prime}u}\right)^{2} 
k^{\mu }k_{\mu }^{\prime }-\frac{\sigma _{T}}{6}
\int_{K\!K^{\prime }\!}\hat{f}_{0\mathbf{k}}\hat{f}_{0\mathbf{k}^{\prime}}
\left( k^{\mu }k_{\mu }^{\prime }\right) ^{2}  \notag \\
&= \frac{2\sigma _{T}}{3}\left( \hat{\mathcal{I}}_{20}^{\mu }\hat{\mathcal{I}}_{00,\mu }
+\hat{\mathcal{I}}_{10}^{\mu }\hat{\mathcal{I}}_{10,\mu }\right) 
-\frac{\sigma _{T}}{6}\hat{\mathcal{I}}_{00}^{\mu \nu }\hat{\mathcal{I}}_{00,\mu \nu }  
\notag \\
&= \frac{\sigma _{T}}{6}\left( 4\hat{I}_{300}\hat{I}_{100}-4\hat{I}_{310}\hat{I}_{110}
+3\hat{I}_{200}\hat{I}_{200}-2\hat{I}_{210}\hat{I}_{210}-\hat{I}_{220}\hat{I}_{220}
-2\hat{I}_{201}\hat{I}_{201}\right) \, ,  
\label{app_G20}
\end{align}%
where $b_{20}=$ $b_{21}=1$, $\mathcal{B}_{20}=\sigma _{T}s/8$, and 
$\mathcal{B}_{21}=\sigma _{T}s^{2}/24$, while we also used 
Eqs.~(\ref{PT_u}) and (\ref{DeltaT_uu}) and Eqs.~(\ref{I_mu_ij}) and (\ref{I_mu_nu_ij}) 
for the tensor products. 

Similarly for $n=3$ we obtain
\begin{align}
\hat{G}_{30} &\equiv \int_{K\!K^{\prime }\!}\hat{f}_{0\mathbf{k}}
\hat{f}_{0\mathbf{k}^{\prime }}\mathcal{P}_{30} 
=\int_{K\!K^{\prime }\!}\hat{f}_{0\mathbf{k}}\hat{f}_{0\mathbf{k}^{\prime }}
\Theta ^{\mu _{1}\mu _{2}\mu_{3}}u_{\mu _{1}}u_{\mu _{2}}u_{\mu _{3}}  \notag \\
&= \int_{K\!K^{\prime }\!}\hat{f}_{0\mathbf{k}}\hat{f}_{0\mathbf{k}^{\prime}}
\left[ b_{30}\mathcal{B}_{30}\left( P_{T}^{\mu }u_{\mu }\right) ^{3} 
-b_{31}\mathcal{B}_{31}\left( \Delta _{T}^{\mu \nu }u_{\mu }u_{\nu }\right) 
\left(P_{T}^{\mu }u_{\mu }\right) \right]  \notag \\
&= \frac{\sigma _{T}}{4}\int_{K\!K^{\prime }\!}\hat{f}_{0\mathbf{k}} 
\hat{f}_{0\mathbf{k}^{\prime }}\left( E_{\mathbf{k}u} 
+E_{\mathbf{k}^{\prime}u}\right) ^{3}k^{\mu }k_{\mu }^{\prime }
-\frac{\sigma _{T}}{4}\int_{K\!K^{\prime }\!}\hat{f}_{0\mathbf{k}} 
\hat{f}_{0\mathbf{k}^{\prime}}\left( E_{\mathbf{k}u}+E_{\mathbf{k}^{\prime }u}\right) 
\left( k^{\mu}k_{\mu }^{\prime }\right) ^{2}  \notag \\
&= \frac{\sigma _{T}}{2}\left( \hat{\mathcal{I}}_{30}^{\mu }\hat{\mathcal{I}}_{00,\mu }
+3\hat{\mathcal{I}}_{20}^{\mu }\hat{\mathcal{I}}_{10,\mu }\right) 
-\frac{\sigma _{T}}{2}\hat{\mathcal{I}}_{10}^{\mu \nu }\hat{\mathcal{I}}_{10,\mu \nu }  \notag \\
&= \frac{\sigma _{T}}{2}\left( \hat{I}_{400}\hat{I}_{100}-\hat{I}_{410}\hat{I}_{110}
+2\hat{I}_{300}\hat{I}_{200}-\hat{I}_{310}\hat{I}_{210} 
-\hat{I}_{320}\hat{I}_{220}-2\hat{I}_{301}\hat{I}_{201}\right) \, ,
\label{app_G30}
\end{align}%
where $b_{30}=1$, $b_{31}=3$, $\mathcal{B}_{30}=\sigma_{T}s/16$, 
and $\mathcal{B}_{31}=\sigma_{T} s^{2}/48$. 

The last gain term follows from Eq.~(\ref{Gn0}) for $n=4$, 
\begin{align}
\hat{G}_{40}& \equiv \int_{K\!K^{\prime }\!}\hat{f}_{0\mathbf{k}}
\hat{f}_{0\mathbf{k}^{\prime }}\mathcal{P}_{40}=\int_{K\!K^{\prime }\!}
\hat{f}_{0\mathbf{k}}\hat{f}_{0\mathbf{k}^{\prime }}
\Theta ^{\mu _{1}\mu _{2}\mu_{3}\mu _{4}}u_{\mu _{1}}u_{\mu _{2}}u_{\mu _{3}}u_{\mu _{4}}  
\notag \\
&= \int_{K\!K^{\prime }\!}\hat{f}_{0\mathbf{k}}\hat{f}_{0\mathbf{k}^{\prime}}
\left[ b_{40}\mathcal{B}_{40}\left( P_{T}^{\mu }u_{\mu }\right) ^{4}
-b_{41}\mathcal{B}_{41}\left( \Delta _{T}^{\mu \nu }u_{\mu }u_{\nu }\right) 
\left(P_{T}^{\mu }u_{\mu }\right) ^{2}+b_{42}\mathcal{B}_{42} 
\left( \Delta_{T}^{\mu \nu }u_{\mu }u_{\nu }\right) ^{2}\right]  \notag \\
&= \frac{\sigma _{T}}{5}\int_{K\!K^{\prime }\!}\hat{f}_{0\mathbf{k}}
\hat{f}_{0\mathbf{k}^{\prime }}\left( E_{\mathbf{k}u}+E_{\mathbf{k}^{\prime}u}\right) ^{4} 
k^{\mu }k_{\mu }^{\prime }-\frac{3\sigma _{T}}{10}
\int_{K\!K^{\prime }\!}\hat{f}_{0\mathbf{k}}\hat{f}_{0\mathbf{k}^{\prime}}
\left( E_{\mathbf{k}u}+E_{\mathbf{k}^{\prime }u}\right) ^{2} 
\left( k^{\mu}k_{\mu }^{\prime }\right) ^{2} 
+\frac{\sigma _{T}}{20}\int_{K\!K^{\prime }\!}\hat{f}_{0\mathbf{k}}
\hat{f}_{0\mathbf{k}^{\prime }} \left( k^{\mu }k_{\mu }^{\prime }\right)^{3}  
\notag \\
&= \frac{2\sigma _{T}}{5}\left( \hat{\mathcal{I}}_{40}^{\mu }\hat{\mathcal{I}}_{00,\mu }
+4\hat{\mathcal{I}}_{30}^{\mu }\hat{\mathcal{I}}_{10,\mu }
+3\hat{\mathcal{I}}_{20}^{\mu }\hat{\mathcal{I}}_{20,\mu }\right) 
-\frac{3\sigma_{T}}{5}\left( \hat{\mathcal{I}}_{20}^{\mu \nu } \hat{\mathcal{I}}_{00,\mu\nu }
+\hat{\mathcal{I}}_{10}^{\mu \nu }\hat{\mathcal{I}}_{10,\mu \nu}\right) 
+\frac{\sigma _{T}}{20}\hat{\mathcal{I}}_{00}^{\mu \nu \alpha }
\hat{\mathcal{I}}_{00,\mu \nu \alpha }  \notag \\
&= \frac{\sigma _{T}}{5}\left( 2\hat{I}_{500}\hat{I}_{100}-2\hat{I}_{510}\hat{I}_{110}
+5\hat{I}_{400}\hat{I}_{200}-2\hat{I}_{410}\hat{I}_{210}-3\hat{I}_{420}\hat{I}_{220}
-6\hat{I}_{401}\hat{I}_{201}\right)  \notag \\
&+ \frac{\sigma _{T}}{20}\left( 13\hat{I}_{300}\hat{I}_{300}-3\hat{I}_{310}\hat{I}_{310}
-9\hat{I}_{320}\hat{I}_{320}-\hat{I}_{330}\hat{I}_{330}-18\hat{I}_{301}\hat{I}_{301}
-6\hat{I}_{311}\hat{I}_{311}\right) \, ,
\label{app_G40}
\end{align}%
where $b_{40}=1$, $b_{41}=6$, $b_{42}=3$, while 
$\mathcal{B}_{40}= \sigma _{T} s/32$, $\mathcal{B}_{41}= \sigma _{T}s^{2}/96$, 
and $\mathcal{B}_{42}= \sigma _{T}s^{3}/480$.

\subsection{The $\hat{G}_{n1}$ gain terms}
\label{app:Gn1}

The $\hat{G}_{n1}$ gain term differs from the $\hat{G}_{n0}$ gain term since it contains 
one projection in the direction of the anisotropy,
\begin{equation}
\hat{G}_{n1}=-\int_{K\!K^{\prime }\!}\hat{f}_{0\mathbf{k}}\hat{f}_{0\mathbf{k}^{\prime }}
\sum_{q=0}^{\lfloor \left( n+1\right) /2\rfloor }\left(-1\right) ^{q}
b_{n+1,q}\mathcal{B}_{n+1,q}\Delta _{T}^{\left( \mu _{1}\mu_{2}\right. }
\cdots \Delta _{T}^{\mu _{2q-1}\mu _{2q}}P_{T}^{\mu_{2q+1}}\cdots 
P_{T}^{\left. \mu _{n+1}\right) }u_{\mu _{1}}\cdots u_{\mu_{n}}l_{\mu _{n+1}}\, ,  
\label{app_Gn1}
\end{equation}%
where the coefficient of distinct terms in the symmetrized tensor product is 
$b_{n+1,q}\equiv \frac{\left( n+1\right) !}{2^{q}q!\left( n+1-2q\right) !}=%
\frac{\left( n+1\right) }{\left( n+1-2q\right) }b_{nq}$. 
The symmetrized tensor product 
$\Delta _{T}^{\left( \mu _{1}\mu _{2}\right. }\cdots P_{T}^{\left. \mu _{n+1}\right) }$ 
contains $\left( n+1\right) !$ index permutations; however, the symmetric rank 2 projection 
operators have $q!2^{q} $ possible permutations while the remaining four-vectors also have 
$\left( n+1-2q\right) !$ permutations which do not lead to distinct terms.

Here we are interested in the scalar product of the symmetrized $n+1$ rank tensor 
$\Delta _{T}^{\left( \mu _{1}\mu _{2}\right. }\cdots P_{T}^{\left. \mu_{n+1}\right) }$ 
by the product of four-vectors containing $n$ four-velocities $u_{\mu _{1}}\cdots u_{\mu _{n}}$ 
and only $1$ four-vector in the direction of the anisotropy $l_{\mu _{n+1}}$. 
Such scalar products can lead up to four distinct scalar terms. 
Here, similarly as in Eq.~(\ref{tensor_prod_2}), the contraction by the four-velocities 
results in $\left( P_{T}^{\mu}u_{\mu }\right) $ and $\left( \Delta _{T}^{\mu \nu }u_{\mu }u_{\nu }\right) $,
while the contractions by the anisotropy four-vector lead to two new scalar products,
$\left( P_{T}^{\mu }l_{\mu }\right) $ and $\left( \Delta _{T}^{\mu \nu }u_{\mu}l_{\nu }\right) $. 

To compute the complete scalar product we first separate the symmetrized tensor product 
$\Delta_{T}^{\left( \mu _{1}\mu _{2}\right. }\cdots P_{T}^{\left. \mu _{n+1}\right)}$ 
into two distinct parts as follows:
\begin{align}
& \Theta ^{\mu _{1}\cdots \mu _{n+1}}u_{\mu _{1}}\cdots u_{\mu _{n}}l_{\mu
_{n+1}}  \notag \\
& \equiv \sum_{q=0}^{\lfloor \left( n+1\right) /2\rfloor }\left( -1\right)
^{q}b_{n+1,q}\mathcal{B}_{n+1,q}\Delta _{T}^{\left( \mu _{1}\mu _{2}\right.
}\cdots \Delta _{T}^{\mu _{2q-1}\mu _{2q}}\,P_{T}^{\mu _{2q+1}}\cdots
P_{T}^{\left. \mu _{n+1}\right) }u_{\mu _{1}}\cdots u_{\mu _{n}}l_{\mu
_{n+1}}  \notag \\
& =\sum_{q=0}^{\lfloor \left( n+1\right) /2\rfloor }\left( -1\right)
^{q}b_{n+1,q}\mathcal{B}_{n+1,q}\frac{\left( n+1-2q\right) }{\left(
n+1\right) }l_{\mu _{n+1}}P_{T}^{\mu _{n+1}}\Delta _{T}^{\left( \mu _{1}\mu
_{2}\right. }\cdots \Delta _{T}^{\mu _{2q-1}\mu _{2q}}\,P_{T}^{\mu
_{2q+1}}\cdots P_{T}^{\left. \mu _{n}\right) }u_{\mu _{1}}\cdots u_{\mu _{n}}
\notag \\
& +\sum_{q=0}^{\lfloor \left( n+1\right) /2\rfloor }\left( -1\right)
^{q}b_{n+1,q}\mathcal{B}_{n+1,q}\frac{\left( 2q\right) }{\left( n+1\right) }%
l_{\mu _{n+1}}\Delta _{T}^{\mu _{n+1}\left( \mu _{1}\right. }\cdots \Delta
_{T}^{\mu _{2q-1}\mu _{2q}}\,P_{T}^{\mu _{2q+1}}\cdots P_{T}^{\left. \mu
_{n}\right) }u_{\mu _{1}}\cdots u_{\mu _{n}}  \notag \\
& =\left( l_{\mu _{n+1}}P_{T}^{\mu _{n+1}}\right) \sum_{q=0}^{\lfloor
n/2\rfloor }\left( -1\right) ^{q}b_{nq}\mathcal{B}_{n+1,q}\left( P_{T}^{\mu
}u_{\mu }\right) ^{n-2q}\left( \Delta _{T}^{\mu \nu }u_{\mu }u_{\nu }\right)
^{q}  \notag \\
& +\left( l_{\mu _{n+1}}\Delta _{T}^{\mu _{n+1}\mu _{n}}u_{\mu _{n}}\right)
\sum_{q=1}^{\lfloor \left( n+1\right) /2\rfloor }\left( -1\right) ^{q}\left(
b_{n+1,q}-b_{nq}\right) \mathcal{B}_{n+1,q}\left( P_{T}^{\mu }u_{\mu
}\right) ^{n+1-2q}\left( \Delta _{T}^{\mu \nu }u_{\mu }u_{\nu }\right)
^{q-1}\, ,  \label{app_Theta_n+1}
\end{align}%
where in the last line we used that $b_{n+1,q}-b_{nq}\equiv \frac{2q}{\left( n+1-2q\right) }%
b_{nq}=\frac{2q}{\left( n+1\right) }b_{n+1,q}$. 

The first term, i.e., $P_{T}^{\mu_{n+1}}\Delta _{T}^{\left( \mu _{1}\mu _{2}\right. }\cdots P_{T}^{\left. \mu
_{n}\right) }$ now contains only $n$ symmetrized indices; hence, by removing one
four-vector $P_{T}^{\mu _{n+1}}$ from the total of $n+1-2q$ different
four-vectors changes the number of distinct terms to 
$b_{n+1,q}\left( n+1-2q\right) /\left( n+1\right)=b_{nq}$, as it should be. 
The remaining terms can be collected into a symmetrized tensor product of the following type, 
$\Delta _{T}^{\mu _{n+1}\left( \mu _{1}\right.}\cdots \,P_{T}^{\left. \mu _{n}\right) }$.
This product is symmetric with respect to $n$ indices such that one index from a rank 2 
projection operator is not included.
Since there are only $2q$ different indices on elementary projection operators, 
then removing one tensor index changes the number of distinct terms under symmetrization 
and it is counted by the following formula:
$b_{n+1,q}\left(2q\right) /\left( n+1\right)$.

Here we will calculate the following $\hat{G}_{n1}$ gain terms of interest. The gain term
from Eq.~(\ref{Gn1}) for $n=0$ leads to
\begin{align}
\hat{G}_{01}& \equiv \int_{K\!K^{\prime }\!}\hat{f}_{0\mathbf{k}} 
\hat{f}_{0\mathbf{k}^{\prime }}\mathcal{P}_{01} 
=-\int_{K\!K^{\prime }\!}\hat{f}_{0\mathbf{k}}\hat{f}_{0\mathbf{k}^{\prime }}
\Theta ^{\mu _{1}}l_{\mu _{1}} \notag \\
&= -\int_{K\!K^{\prime }\!}\hat{f}_{0\mathbf{k}}\hat{f}_{0\mathbf{k}^{\prime}}
\left[ b_{00}\mathcal{B}_{10} \left( l_{\mu}P_{T}^{\mu}\right)  \right]
=\frac{\sigma _{T}}{2}\int_{K\!K^{\prime }\!}\hat{f}_{0\mathbf{k}}
\hat{f}_{0\mathbf{k}^{\prime }}\left( E_{\mathbf{k}l}+E_{\mathbf{k}^{\prime }l}\right) 
k^{\mu}k_{\mu }^{\prime }  \notag \\
&= \sigma _{T}\hat{\mathcal{I}}_{01}^{\mu }\hat{\mathcal{I}}_{00,\mu}
=\sigma _{T}\left( \hat{I}_{210}\hat{I}_{100}-\hat{I}_{220}\hat{I}_{110}\right) \, ,  
\label{app_G01}
\end{align}%
where $b_{00}=1$ and $\mathcal{B}_{10}=\sigma _{T}s/4$, and we used Eq.~(\ref{PT_l}) 
as well as Eq.~(\ref{I_mu_ij}).

Similarly for $n=1$ we obtain the following gain term, 
\begin{align}
\hat{G}_{11}& \equiv \int_{K\!K^{\prime }\!}\hat{f}_{0\mathbf{k}} 
\hat{f}_{0\mathbf{k}^{\prime }}\mathcal{P}_{11} 
=-\int_{K\!K^{\prime }\!}\hat{f}_{0\mathbf{k}}\hat{f}_{0\mathbf{k}^{\prime }}
\Theta ^{\mu _{1}\mu _{2}}u_{\mu_{1}}l_{\mu _{2}}  \notag \\
&= -\int_{K\!K^{\prime }\!}\hat{f}_{0\mathbf{k}}\hat{f}_{0\mathbf{k}^{\prime}}
\left[ \left( l_{\mu _{2}}P_{T}^{\mu _{2}}\right) b_{10}\mathcal{B}%
_{20}\left( P_{T}^{\mu _{1}}u_{\mu _{1}}\right) -\left( l_{\mu _{2}}\Delta
_{T}^{\mu _{2}\mu _{1}}u_{\mu _{1}}\right) \left( b_{21}-b_{11}\right) 
\mathcal{B}_{21}\right]  \notag \\
&= \frac{\sigma _{T}}{3}\int_{K\!K^{\prime }\!}\hat{f}_{0\mathbf{k}}
\hat{f}_{0\mathbf{k}^{\prime }}\left( E_{\mathbf{k}u}+E_{\mathbf{k}^{\prime}u}\right) 
\left( E_{\mathbf{k}l}+E_{\mathbf{k}^{\prime }l}\right) k^{\mu}k_{\mu }^{\prime }  \notag \\
&= \frac{2\sigma _{T}}{3}\left( \hat{\mathcal{I}}_{11}^{\mu } 
\hat{\mathcal{I}}_{00,\mu }+\hat{\mathcal{I}}_{10}^{\mu }\hat{\mathcal{I}}_{01,\mu }\right) 
= \frac{2\sigma _{T}}{3}\left( \hat{I}_{310}\hat{I}_{100}-\hat{I}_{320}\hat{I}_{110}
+\hat{I}_{200}\hat{I}_{210}-\hat{I}_{210}\hat{I}_{220}\right) \, ,
\label{app_G11}
\end{align}%
where $b_{10}=b_{21}=1$ and $b_{11}=0$, while $\mathcal{B}_{20}=\sigma _{T}s/8$
and $\mathcal{B}_{21}=\sigma_{T}s^{2}/24$.

Furthermore, from Eq.~(\ref{Gn1}) for $n=2$ we get
\begin{align}
\hat{G}_{21}& \equiv \int_{K\!K^{\prime }\!}\hat{f}_{0\mathbf{k}}
\hat{f}_{0\mathbf{k}^{\prime }}\mathcal{P}_{21}=-\int_{K\!K^{\prime }\!} 
\hat{f}_{0\mathbf{k}}\hat{f}_{0\mathbf{k}^{\prime }} 
\Theta ^{\mu _{1}\mu _{2}\mu_{3}}u_{\mu _{1}}u_{\mu _{2}}l_{\mu _{3}}  \notag \\
&= -\int_{K\!K^{\prime }\!}\hat{f}_{0\mathbf{k}}\hat{f}_{0\mathbf{k}^{\prime}}
\left( l_{\mu _{3}}P_{T}^{\mu _{3}}\right) \left[ b_{20}\mathcal{B}_{30}
\left( P_{T}^{\mu }u_{\mu }\right) ^{2}-b_{21}\mathcal{B}_{31}
\left(\Delta _{T}^{\mu \nu }u_{\mu }u_{\nu }\right) \right]  \notag \\
&+ \int_{K\!K^{\prime }\!}\hat{f}_{0\mathbf{k}}\hat{f}_{0\mathbf{k}^{\prime}}
\left( l_{\mu _{3}}\Delta _{T}^{\mu _{3}\mu _{2}}u_{\mu _{2}}\right)
\left( b_{31}-b_{21}\right) \mathcal{B}_{31}\left( P_{T}^{\mu }u_{\mu}\right)  \notag \\
&= \frac{\sigma _{T}}{4}\int_{K\!K^{\prime }\!}\hat{f}_{0\mathbf{k}} 
\hat{f}_{0\mathbf{k}^{\prime }}\left( E_{\mathbf{k}u}+E_{\mathbf{k}^{\prime}u}\right) ^{2}
\left( E_{\mathbf{k}l}+E_{\mathbf{k}^{\prime }l}\right)
k^{\mu }k_{\mu }^{\prime }-\frac{\sigma _{T}}{12}\int_{K\!K^{\prime }\!}
\hat{f}_{0\mathbf{k}}\hat{f}_{0\mathbf{k}^{\prime }}
\left( E_{\mathbf{k}l}+E_{\mathbf{k}^{\prime }l}\right) 
k^{\mu }k^{\nu }k_{\mu }^{\prime }k_{\nu}^{\prime }  \notag \\
&= \frac{\sigma _{T}}{2}\left( \hat{\mathcal{I}}_{21}^{\mu }\hat{\mathcal{I}}_{00,\mu }
+2\hat{\mathcal{I}}_{11}^{\mu }\hat{\mathcal{I}}_{10,\mu }
+\hat{\mathcal{I}}_{01}^{\mu }\hat{\mathcal{I}}_{20,\mu }\right) 
-\frac{\sigma _{T}}{6}\hat{\mathcal{I}}_{01}^{\mu \nu }\hat{\mathcal{I}}_{00,\mu \nu }  
\notag \\
&= \frac{\sigma _{T}}{6}\left( 3\hat{I}_{410}\hat{I}_{100}-3\hat{I}_{420}\hat{I}_{110}
+5\hat{I}_{310}\hat{I}_{200}-4\hat{I}_{320}\hat{I}_{210}\right)
+\frac{\sigma _{T}}{6}\left( 3\hat{I}_{300}\hat{I}_{210}-3\hat{I}_{310}\hat{I}_{220}
-\hat{I}_{330}\hat{I}_{220}-2\hat{I}_{311}\hat{I}_{201}\right) \, ,
\label{app_G21}
\end{align}%
where $b_{31}=3$ and $b_{21}=1$, and $\mathcal{B}_{30}=\sigma_Ts/16$ and 
$\mathcal{B}_{31}=\sigma_T s^{2}/48$. 

Finally the gain term for $n=3$ leads to 
\begin{align}
\hat{G}_{31}& \equiv \int_{K\!K^{\prime }\!}\hat{f}_{0\mathbf{k}}
\hat{f}_{0\mathbf{k}^{\prime }}\mathcal{P}_{31}=-\int_{K\!K^{\prime }\!}
\hat{f}_{0\mathbf{k}}\hat{f}_{0\mathbf{k}^{\prime }} 
\Theta ^{\mu _{1}\mu _{2}\mu_{3}\mu _{4}}u_{\mu _{1}}u_{\mu _{2}}u_{\mu _{3}}l_{\mu _{4}}  
\notag \\
&= -\int_{K\!K^{\prime }\!}\hat{f}_{0\mathbf{k}}\hat{f}_{0\mathbf{k}^{\prime}}
\left( l_{\mu _{4}}P_{T}^{\mu _{4}}\right) \left[ b_{30}\mathcal{B}_{40}
\left( P_{T}^{\mu }u_{\mu }\right) ^{3}-b_{31}\mathcal{B}_{41} 
\left(P_{T}^{\mu }u_{\mu }\right) \left( \Delta _{T}^{\mu \nu }u_{\mu }u_{\nu}\right) \right]  
\notag \\
&+ \int_{K\!K^{\prime }\!}\hat{f}_{0\mathbf{k}}\hat{f}_{0\mathbf{k}^{\prime}}
\left( l_{\mu _{4}}\Delta _{T}^{\mu _{4}\mu _{3}}u_{\mu _{3}}\right) 
\left[\left( b_{41}-b_{31}\right) \mathcal{B}_{41}\left( P_{T}^{\mu }u_{\mu}\right) ^{2}
-\left( b_{42}-b_{32}\right) \mathcal{B}_{42} 
\left( \Delta_{T}^{\mu \nu }u_{\mu }u_{\nu }\right) \right]  \notag \\
&= \frac{\sigma _{T}}{5}\int_{K\!K^{\prime }\!}\hat{f}_{0\mathbf{k}}
\hat{f}_{0\mathbf{k}^{\prime }}\left( E_{\mathbf{k}u}+E_{\mathbf{k}^{\prime}u}\right) ^{3} 
\left( E_{\mathbf{k}l}+E_{\mathbf{k}^{\prime }l}\right)k^{\mu }k_{\mu }^{\prime } 
-\frac{3\sigma _{T}}{20}\int_{K\!K^{\prime }\!}\hat{f}_{0\mathbf{k}} 
\hat{f}_{0\mathbf{k}^{\prime }}\left( E_{\mathbf{k}u}+E_{\mathbf{k}^{\prime }u}\right) 
\left( E_{\mathbf{k}l}+E_{\mathbf{k}^{\prime }l}\right) 
k^{\mu }k^{\nu }k_{\mu }^{\prime }k_{\nu }^{\prime } 
\notag \\
&= \frac{2\sigma _{T}}{5}\left( \hat{\mathcal{I}}_{31}^{\mu }\hat{\mathcal{I}}_{00,\mu }
+3\hat{\mathcal{I}}_{21}^{\mu }\hat{\mathcal{I}}_{10,\mu }
+3\hat{\mathcal{I}}_{11}^{\mu }\hat{\mathcal{I}}_{20,\mu }
+\hat{\mathcal{I}}_{30}^{\mu }\hat{\mathcal{I}}_{01,\mu }\right) 
-\frac{3\sigma _{T}}{10}\left( \hat{\mathcal{I}}_{11}^{\mu \nu }\hat{\mathcal{I}}_{00,\mu \nu }
+\hat{\mathcal{I}}_{10}^{\mu \nu }\hat{\mathcal{I}}_{01,\mu \nu }\right)  \notag \\
&= \frac{2\sigma _{T}}{5}\left( \hat{I}_{510}\hat{I}_{100}-\hat{I}_{520}\hat{I}_{110}
+\hat{I}_{400}\hat{I}_{210}-\hat{I}_{410}\hat{I}_{220}\right) 
+\frac{3\sigma _{T}}{10}\left( 3\hat{I}_{410}\hat{I}_{200}-2\hat{I}_{420}\hat{I}_{210}
-\hat{I}_{430}\hat{I}_{220}-2\hat{I}_{411}\hat{I}_{201}\right)  \notag\\
& +\frac{3\sigma _{T}}{10}\left( 3\hat{I}_{300}\hat{I}_{310}-2\hat{I}_{310}\hat{I}_{320}
-\hat{I}_{320}\hat{I}_{330}-2\hat{I}_{301}\hat{I}_{311}\right)\, ,  
\label{app_G31}
\end{align}%
where we used that $b_{30}=1$, $b_{31}=b_{42}=3$, $b_{32}=0$, $b_{41}=6$, $%
\mathcal{B}_{40}=\sigma_T s/32$, $\mathcal{B}_{41}=\sigma_T s^{2}/96$ and 
$\mathcal{B}_{42}=\sigma_T s^{3}/480$.

\subsection{The $\hat{G}_{n2}$ gain terms}
\label{app:Gn2}

The last gain term of interest is a generalization of our previous results,
now including two projections in the direction of the anisotropy, 
\begin{equation}
\hat{G}_{n2} =\int dKdK^{\prime }\hat{f}_{0\mathbf{k}}\hat{f}_{0\mathbf{k}^{\prime }}
\sum_{q=0}^{\lfloor \left( n+2\right) /2\rfloor }\left( -1\right)^{q}
b_{n+2,q}\mathcal{B}_{n+2,q}\Delta _{T}^{\left( \mu _{1}\mu _{2}\right.}
\cdots \Delta _{T}^{\mu _{2q-1}\mu _{2q}}\,P_{T}^{\mu _{2q+1}}\cdots
P_{T}^{\left. \mu _{n+2}\right) }u_{\mu _{1}}\cdots u_{\mu _{n}}l_{\mu_{n+1}}l_{\mu _{n+2}}\, ,
\label{app_Gn2}
\end{equation}%
where now the coefficient of the symmetrized tensor product is 
$b_{n+2,q}\equiv \frac{\left( n+2\right) }{\left( n+2-2q\right) }b_{n+1,q}
=\frac{\left(n+2\right) }{\left( n+2-2q\right) }
\frac{\left( n+1\right) }{\left(n+1-2q\right) }b_{nq}$. 
The scalar product of the symmetrized tensor 
$\Delta_{T}^{\left( \mu _{1}\mu _{2}\right. }\cdots P_{T}^{\left. \mu _{n+2}\right)}$ 
by the product of four-vectors containing $n$ four-velocities $u_{\mu
_{1}}\cdots u_{\mu _{n}}$ and 2 four-vectors in the direction of the
anisotropy $l_{\mu _{n+1}}l_{\mu _{n+2}}$ leads to only five different type
of scalars. 
The first four scalar products are the same as obtained before in Eq.~(\ref{app_Gn1}), namely
$\left( P_{T}^{\mu }u_{\mu}\right) $, $\left( \Delta_{T}^{\mu \nu }u_{\mu }u_{\nu }\right) $, 
$\left(P_{T}^{\mu }l_{\mu}\right) $, and 
$\left( \Delta _{T}^{\mu \nu }u_{\mu }l_{\nu}\right) $, 
while the fifth term is a new scalar product $\left( \Delta _{T}^{\mu\nu }l_{\mu }l_{\nu }\right)$ 
which follows from the projection by 2 four-vectors in the direction of anisotropy.

To calculate the complete scalar product we separate the symmetrized tensor
product $\Delta_{T}^{\left( \mu _{1}\mu _{2}\right. }\cdots P_{T}^{\left.
\mu _{n+2}\right) }$ iteratively. First similar to Eq. (\ref{app_Theta_n+1}%
) we construct symmetrized rank $n+1$ tensor products, 
\begin{align}
& \Theta ^{\mu _{1}\cdots \mu _{n+2}}u_{\mu _{1}}\cdots u_{\mu _{n}}l_{\mu
_{n+1}}l_{\mu _{n+2}}  \notag \\
& \equiv \sum_{q=0}^{\lfloor \left( n+2\right) /2\rfloor }\left( -1\right)
^{q}b_{n+2,q}\mathcal{B}_{n+2,q}\Delta _{T}^{\left( \mu _{1}\mu _{2}\right.
}\cdots \Delta _{T}^{\mu _{2q-1}\mu _{2q}}\,P_{T}^{\mu _{2q+1}}\cdots
P_{T}^{\left. \mu _{n+2}\right) }u_{\mu _{1}}\cdots u_{\mu _{n}}l_{\mu
_{n+1}}l_{\mu _{n+2}}  \notag \\
& =\sum_{q=0}^{\lfloor \left( n+2\right) /2\rfloor }\left( -1\right)
^{q}b_{n+2,q}\mathcal{B}_{n+2,q}\frac{\left( n+2-2q\right) }{\left(
n+2\right) }l_{\mu _{n+2}}P_{T}^{\mu _{n+2}}\Delta _{T}^{\left( \mu _{1}\mu
_{2}\right. }\cdots \Delta _{T}^{\mu _{2q-1}\mu _{2q}}\,P_{T}^{\mu
_{2q+1}}\cdots P_{T}^{\left. \mu _{n+1}\right) }u_{\mu _{1}}\cdots u_{\mu
_{n}}l_{\mu _{n+1}}  \notag \\
& +\sum_{q=0}^{\lfloor \left( n+2\right) /2\rfloor }\left( -1\right)
^{q}b_{n+2,q}\mathcal{B}_{n+2,q}\frac{\left( 2q\right) }{\left( n+2\right) }%
l_{\mu _{n+2}}\Delta _{T}^{\mu _{n+2}\left( \mu _{1}\right. }\Delta
_{T}^{\mu _{2}\mu _{3}}\cdots \Delta _{T}^{\mu _{2q-1}\mu _{2q}}\,P_{T}^{\mu
_{2q+1}}\cdots P_{T}^{\left. \mu _{n+1}\right) }u_{\mu _{1}}\cdots u_{\mu
_{n}}l_{\mu _{n+1}}\, ,
\end{align}%
where the coefficients of the symmetrized tensors are obtained by
substituting $n \rightarrow n+1$ in Eq.~(\ref{app_Theta_n+1}). Then, in the
subsequent step we once again separate both terms into symmetrized rank $n$
tensor products, and hence we obtain four different terms, 
\begin{align}
& \Theta ^{\mu _{1}\cdots \mu _{n+2}}u_{\mu _{1}}\cdots u_{\mu _{n}}l_{\mu
_{n+1}}l_{\mu _{n+2}}  \notag \\
& =\sum_{q=0}^{\lfloor \left( n+2\right) /2\rfloor }\left( -1\right) ^{q}%
\mathcal{B}_{n+2,q}b_{n+2,q}\frac{\left( n+2-2q\right) }{\left( n+2\right) }%
\times \frac{\left( n+1-2q\right) }{\left( n+1\right) }  \notag \\
& \times l_{\mu _{n+2}}P_{T}^{\mu _{n+2}}l_{\mu _{n+1}}P_{T}^{\mu
_{n+1}}\Delta _{T}^{\left( \mu _{1}\mu _{2}\right. }\cdots \Delta _{T}^{\mu
_{2q-1}\mu _{2q}}\,P_{T}^{\mu _{2q+1}}\cdots P_{T}^{\left. \mu _{n}\right)
}u_{\mu _{1}}\cdots u_{\mu _{n}}  \notag \\
& +\sum_{q=0}^{\lfloor \left( n+2\right) /2\rfloor }\left( -1\right) ^{q}%
\mathcal{B}_{n+2,q}b_{n+2,q}\frac{\left( n+2-2q\right) }{\left( n+2\right) }%
\times \frac{\left( 2q\right) }{\left( n+1\right) }\times 2  \notag \\
& \times l_{\mu _{n+2}}P_{T}^{\mu _{n+2}}l_{\mu _{n+1}}\Delta _{T}^{\mu
_{n+1}\left( \mu _{1}\right. }\Delta _{T}^{\mu _{2}\mu _{3}}\cdots \Delta
_{T}^{\mu _{2q-1}\mu _{2q}}\,P_{T}^{\mu _{2q+1}}\cdots P_{T}^{\left. \mu
_{n}\right) }u_{\mu _{1}}\cdots u_{\mu _{n}}  \notag \\
& +\sum_{q=0}^{\lfloor \left( n+2\right) /2\rfloor }\left( -1\right) ^{q}%
\mathcal{B}_{n+2,q}b_{n+2,q}\frac{\left( 2q\right) }{\left( n+2\right) }%
\times \frac{1}{\left( n+1\right) }  \notag \\
& \times l_{\mu _{n+2}}\Delta _{T}^{\mu _{n+2}\mu _{n+1}}l_{\mu
_{n+1}}\Delta _{T}^{\left( \mu _{1}\mu _{2}\right. }\cdots \Delta _{T}^{\mu
_{2q-1}\mu _{2q}}\,P_{T}^{\mu _{2q+1}}\cdots P_{T}^{\left. \mu _{n}\right)
}u_{\mu _{1}}\cdots u_{\mu _{n}}  \notag \\
& +\sum_{q=0}^{\lfloor \left( n+2\right) /2\rfloor }\left( -1\right) ^{q}%
\mathcal{B}_{n+2,q}b_{n+2,q}\frac{\left( 2q\right) }{\left( n+2\right) }%
\times \frac{2\left( q-1\right) }{\left( n+1\right) }  \notag \\
& \times l_{\mu _{n+2}}\Delta _{T}^{\mu _{n+2}\left( \mu _{1}\right. }\Delta
_{T}^{\mu _{2}\mu _{3}}\cdots \Delta _{T}^{\mu _{2q-1}\mu _{2q}}\,P_{T}^{\mu
_{2q+1}}\cdots P_{T}^{\mu _{n-1}}\Delta _{T}^{\left. \mu _{n}\right) \mu
_{n+1}}l_{\mu _{n+1}}u_{\mu _{1}}\cdots u_{\mu _{n}}\, .
\end{align}%
Now using the identities, $b_{n+1,q}-b_{nq}\equiv b_{n+2,q}\frac{\left(
n+2-2q\right) }{\left( n+2\right) }\frac{\left( 2q\right) }{\left(
n+1\right) }$, and $b_{n,q-1}\equiv b_{n+2,q}\frac{\left( 2q\right) }{\left(
n+2\right) }\frac{1}{\left( n+1\right) }$, we obtain 
\begin{align}
& \Theta ^{\mu _{1}\cdots \mu _{n+2}}u_{\mu _{1}}\cdots u_{\mu _{n}}l_{\mu
_{n+1}}l_{\mu _{n+2}}  \notag \\
& =\left( l_{\mu _{n+2}}P_{T}^{\mu _{n+2}}\right) \left( l_{\mu
_{n+1}}P_{T}^{\mu _{n+1}}\right) \sum_{q=0}^{\lfloor n/2\rfloor }\left(
-1\right) ^{q}b_{nq}\mathcal{B}_{n+2,q}\left( P_{T}^{\mu }u_{\mu }\right)
^{n-2q}\left( \Delta _{T}^{\mu \nu }u_{\mu }u_{\nu }\right) ^{q}  \notag \\
& +\left( l_{\mu _{n+2}}P_{T}^{\mu _{n+2}}\right) \left( l_{\mu
_{n+1}}\Delta _{T}^{\mu _{n+1}\mu _{n}}u_{\mu _{n}}\right)
\sum_{q=1}^{\lfloor \left( n+2\right) /2\rfloor }\left( -1\right)
^{q}2\left( b_{n+1,q}-b_{nq}\right) \mathcal{B}_{n+2,q}\left( P_{T}^{\mu
}u_{\mu }\right) ^{n+1-2q}\left( \Delta _{T}^{\mu \nu }u_{\mu }u_{\nu
}\right) ^{q-1}  \notag \\
& +\left( l_{\mu _{n+2}}\Delta _{T}^{\mu _{n+2}\mu _{n+1}}l_{\mu
_{n+1}}\right) \sum_{q=1}^{\lfloor \left( n+2\right) /2\rfloor }\left(
-1\right) ^{q}b_{n,q-1}\mathcal{B}_{n+2,q}\left( P_{T}^{\mu }u_{\mu }\right)
^{n+2-2q}\left( \Delta _{T}^{\mu \nu }u_{\mu }u_{\nu }\right) ^{q-1}  \notag
\\
& +\left( l_{\mu _{n+2}}\Delta _{T}^{\mu _{n+2}\mu _{n}}u_{\mu _{n}}\right)
\left( l_{\mu _{n+1}}\Delta _{T}^{\mu _{n+1}\mu _{n-1}}u_{\mu _{n-1}}\right)
\sum_{q=2}^{\lfloor \left( n+2\right) /2\rfloor }\left( -1\right)
^{q}2\left( q-1\right) b_{n,q-1}\mathcal{B}_{n+2,q}\left( P_{T}^{\mu }u_{\mu
}\right) ^{n+2-2q}\left( \Delta _{T}^{\mu \nu }u_{\mu }u_{\nu
}\right)^{q-2}\, .
\end{align}

The gain term from Eq.~(\ref{Gn2}) for $n=0$ leads to, 
\begin{align}
\hat{G}_{02} &\equiv \int_{K\!K^{\prime }\!}\hat{f}_{0\mathbf{k}} 
\hat{f}_{0\mathbf{k}^{\prime }}\mathcal{P}_{02}
=\int_{K\!K^{\prime }\!}\hat{f}_{0\mathbf{k}}\hat{f}_{0\mathbf{k}^{\prime }}
\Theta ^{\mu _{1}\mu _{2}}l_{\mu_{1}}l_{\mu _{2}}  \notag \\
&= \int_{K\!K^{\prime }\!}\hat{f}_{0\mathbf{k}}\hat{f}_{0\mathbf{k}^{\prime}}
\left[b_{00}\mathcal{B}_{20} \left( l_{\mu}P_{T}^{\mu}\right)^2 
- b_{00}\mathcal{B}_{21} \left( \Delta _{T}^{\mu\nu} l_{\mu} l_{\nu}\right)  \right]  \notag \\
&= \frac{\sigma _{T}}{3}\int_{K\!K^{\prime }\!}\hat{f}_{0\mathbf{k}} 
\hat{f}_{0\mathbf{k}^{\prime }}\left( E_{\mathbf{k}l}+E_{\mathbf{k}^{\prime}l}\right) ^{2} 
k^{\mu }k_{\mu }^{\prime }+\frac{\sigma _{T}}{6}
\int_{K\!K^{\prime }\!}\hat{f}_{0\mathbf{k}}\hat{f}_{0\mathbf{k}^{\prime}}
k^{\mu }k^{\nu }k_{\mu }^{\prime }k_{\nu }^{\prime }  \notag \\
&= \frac{2\sigma _{T}}{3}\left( \hat{\mathcal{I}}_{02}^{\mu }\hat{\mathcal{I}}_{00,\mu }
+\hat{\mathcal{I}}_{01}^{\mu }\hat{\mathcal{I}}_{01,\mu }\right) 
+\frac{\sigma _{T}}{6}\hat{\mathcal{I}}_{00}^{\mu \nu }\hat{\mathcal{I}}_{00,\mu \nu }  \notag \\
&= \frac{\sigma _{T}}{6}\left( 4\hat{I}_{320}\hat{I}_{100}-4\hat{I}_{330}\hat{I}_{110}
+\hat{I}_{200}\hat{I}_{200}+2\hat{I}_{210}\hat{I}_{210}-3\hat{I}_{220}\hat{I}_{220}
+2\hat{I}_{201}\hat{I}_{201}\right) \, .  
\label{app_G02}
\end{align}%
where $b_{00}=1$, $\mathcal{B}_{20}=\sigma_T s/8$, and $\mathcal{B}_{21}=\sigma_T s^{2}/24$.
Similarly, the next gain term for $n=1$ leads to
\begin{align}
\hat{G}_{12}& \equiv \int_{K\!K^{\prime }\!}\hat{f}_{0\mathbf{k}}
\hat{f}_{0\mathbf{k}^{\prime }}\mathcal{P}_{12}=\int_{K\!K^{\prime }\!}
\hat{f}_{0\mathbf{k}}\hat{f}_{0\mathbf{k}^{\prime }}\Theta ^{\mu _{1}\mu _{2}\mu_{3}}
u_{\mu _{1}}l_{\mu _{2}}l_{\mu _{3}}  \notag \\
&= \int_{K\!K^{\prime }\!}\hat{f}_{0\mathbf{k}}\hat{f}_{0\mathbf{k}^{\prime}}
\left( l_{\mu _{3}}P_{T}^{\mu _{3}}\right) \left( l_{\mu _{2}}P_{T}^{\mu_{2}}\right) 
b_{10}\mathcal{B}_{30}\left( P_{T}^{\mu }u_{\mu }\right)
-\int_{K\!K^{\prime }\!}\hat{f}_{0\mathbf{k}}\hat{f}_{0\mathbf{k}^{\prime}}
\left( l_{\mu _{3}}P_{T}^{\mu _{3}}\right) 
\left( l_{\mu _{2}}\Delta_{T}^{\mu _{2}\mu _{1}}u_{\mu _{1}}\right) 
2\left( b_{21}-b_{11}\right) \mathcal{B}_{31}  \notag \\
&-\int_{K\!K^{\prime }\!}\hat{f}_{0\mathbf{k}}\hat{f}_{0\mathbf{k}^{\prime}}
\left( l_{\mu _{3}}\Delta _{T}^{\mu _{3}\mu _{2}}l_{\mu _{2}}\right) 
b_{10}\mathcal{B}_{31}\left( P_{T}^{\mu }u_{\mu }\right)  \notag \\
&= \frac{\sigma _{T}}{4}\int_{K\!K^{\prime }\!}\hat{f}_{0\mathbf{k}}
\hat{f}_{0\mathbf{k}^{\prime }}\left( E_{\mathbf{k}l}+E_{\mathbf{k}^{\prime}l}\right) ^{2}
\left( E_{\mathbf{k}u}+E_{\mathbf{k}^{\prime }u}\right)
k^{\mu }k_{\mu }^{\prime }+\frac{\sigma _{T}}{12}\int_{K\!K^{\prime }\!}
\hat{f}_{0\mathbf{k}}\hat{f}_{0\mathbf{k}^{\prime }}
\left( E_{\mathbf{k}u}+E_{\mathbf{k}^{\prime }u}\right) 
k^{\mu }k^{\nu }k_{\mu }^{\prime }k_{\nu}^{\prime }  \notag \\
&= \frac{\sigma _{T}}{2}\left( \hat{\mathcal{I}}_{12}^{\mu }\hat{\mathcal{I}}_{00,\mu }
+2\hat{\mathcal{I}}_{11}^{\mu }\hat{\mathcal{I}}_{01,\mu }
+\hat{\mathcal{I}}_{02}^{\mu }\hat{\mathcal{I}}_{10,\mu }\right) 
+\frac{\sigma _{T}}{6}\hat{\mathcal{I}}_{10}^{\mu \nu }\hat{\mathcal{I}}_{00,\mu \nu }  \notag\\
&= \frac{\sigma _{T}}{6}\left( 3\hat{I}_{420}\hat{I}_{100}-3\hat{I}_{430}\hat{I}_{110}
+4\hat{I}_{310}\hat{I}_{210}-5\hat{I}_{320}\hat{I}_{220}\right)
+\frac{\sigma _{T}}{6}\left( 3\hat{I}_{320}\hat{I}_{200}-3\hat{I}_{330}\hat{I}_{210}
+\hat{I}_{300}\hat{I}_{200}+2\hat{I}_{301}\hat{I}_{201}\right) \, ,
\label{app_G12}
\end{align}%
where $b_{10}=b_{21}=1$, $b_{11}=0$, $\mathcal{B}_{30}=\sigma_Ts/16$
and $\mathcal{B}_{31}=\sigma_T s^{2}/48$. 
Finally the last gain term for $n=2$ gives
\begin{align}
\hat{G}_{22} &\equiv \int_{K\!K^{\prime }\!}\hat{f}_{0\mathbf{k}} 
\hat{f}_{0\mathbf{k}^{\prime }}\mathcal{P}_{22} 
=\int_{K\!K^{\prime }\!}\hat{f}_{0\mathbf{k}}\hat{f}_{0\mathbf{k}^{\prime }}
\Theta ^{\mu _{1}\mu _{2}\mu_{3}\mu _{4}} u_{\mu _{1}}u_{\mu _{2}}l_{\mu _{3}}l_{\mu _{4}}  
\notag \\
&= \int_{K\!K^{\prime }\!}\hat{f}_{0\mathbf{k}}\hat{f}_{0\mathbf{k}^{\prime}}
\left( l_{\mu _{4}}P_{T}^{\mu _{4}}\right) \left( l_{\mu _{3}}P_{T}^{\mu_{3}}\right) 
\left[ b_{20}\mathcal{B}_{40}\left( P_{T}^{\mu }u_{\mu }\right)^{2}
-b_{21}\mathcal{B}_{41}\left( \Delta _{T}^{\mu \nu }u_{\mu }u_{\nu}\right) \right]  \notag \\
&- \int_{K\!K^{\prime }\!}\hat{f}_{0\mathbf{k}}\hat{f}_{0\mathbf{k}^{\prime}}
\left( l_{\mu _{4}}P_{T}^{\mu _{4}}\right) 
\left( l_{\mu _{3}}\Delta_{T}^{\mu _{3}\mu _{2}}u_{\mu _{2}}\right) 
\left[ 2\left(b_{31}-b_{21}\right) \mathcal{B}_{41}\left( P_{T}^{\mu }u_{\mu }\right)\right]  
\notag \\
&- \int_{K\!K^{\prime }\!}\hat{f}_{0\mathbf{k}}\hat{f}_{0\mathbf{k}^{\prime}}
\left( l_{\mu _{4}}\Delta _{T}^{\mu _{4}\mu _{3}}l_{\mu _{3}}\right) 
\left[b_{20}\mathcal{B}_{41}\left( P_{T}^{\mu }u_{\mu }\right)^{2}
-b_{21}\mathcal{B}_{42}\left( \Delta _{T}^{\mu \nu }u_{\mu }u_{\nu }\right) \right]  
\notag \\
&+ \int_{K\!K^{\prime }\!}\hat{f}_{0\mathbf{k}}\hat{f}_{0\mathbf{k}^{\prime}}
\left( l_{\mu _{4}}\Delta _{T}^{\mu _{4}\mu _{2}}u_{\mu _{2}}\right)
\left( l_{\mu _{3}}\Delta _{T}^{\mu _{3}\mu _{1}}u_{\mu _{1}}\right) 
2b_{21}\mathcal{B}_{42}\, ,
\end{align}%
where replacing $b_{20}=1$, $b_{21}=1$, $b_{31}=3$, $\mathcal{B}_{40}= \sigma_T s/32$, 
$\mathcal{B}_{41}=\sigma_T s^{2}/96$, and $\mathcal{B}_{42}= \sigma_T s^{3}/480$ leads to 
\begin{align}
\hat{G}_{22}& =\frac{\sigma _{T}}{5}\int_{K\!K^{\prime }\!}\hat{f}_{0\mathbf{k}}
\hat{f}_{0\mathbf{k}^{\prime }}\left( E_{\mathbf{k}u}+E_{\mathbf{k}^{\prime }u}\right)^{2} 
\left( E_{\mathbf{k}l}+E_{\mathbf{k}^{\prime}l}\right) ^{2} 
k^{\mu }k_{\mu }^{\prime }-\frac{\sigma _{T}}{60}
\int_{K\!K^{\prime }\!}\hat{f}_{0\mathbf{k}}\hat{f}_{0\mathbf{k}^{\prime}}
k^{\mu }k^{\nu }k^{\alpha }k_{\mu }^{\prime }k_{\nu }^{\prime }k_{\alpha}^{\prime }  
\notag \\
&- \frac{\sigma _{T}}{20}\int_{K\!K^{\prime }\!}\hat{f}_{0\mathbf{k}}
\hat{f}_{0\mathbf{k}^{\prime }}\left( E_{\mathbf{k}l}+E_{\mathbf{k}^{\prime}l}\right) ^{2} 
k^{\mu }k^{\nu }k_{\mu }^{\prime }k_{\nu }^{\prime }
+\frac{\sigma _{T}}{20}\int_{K\!K^{\prime }\!}\hat{f}_{0\mathbf{k}}
\hat{f}_{0\mathbf{k}^{\prime }}\left( E_{\mathbf{k}u}+E_{\mathbf{k}^{\prime }u}\right)^{2}
k^{\mu }k^{\nu }k_{\mu }^{\prime }k_{\nu }^{\prime }  \notag \\
&= \frac{2\sigma _{T}}{5}\left( \hat{\mathcal{I}}_{22}^{\mu }\hat{\mathcal{I}}_{00,\mu }
+\hat{\mathcal{I}}_{02}^{\mu }\hat{\mathcal{I}}_{20,\mu }
+2\hat{\mathcal{I}}_{12}^{\mu }\hat{\mathcal{I}}_{10,\mu }
+2\hat{\mathcal{I}}_{21}^{\mu }\hat{\mathcal{I}}_{01,\mu }
+2\hat{\mathcal{I}}_{11}^{\mu }\hat{\mathcal{I}}_{11,\mu }\right)  \notag \\
&- \frac{\sigma _{T}}{10}\left( \hat{\mathcal{I}}_{02}^{\mu \nu }\hat{\mathcal{I}}_{00,\mu \nu }
+\hat{\mathcal{I}}_{01}^{\mu \nu }\hat{\mathcal{I}}_{01,\mu \nu }\right) 
+\frac{\sigma _{T}}{10}\left( \hat{\mathcal{I}}_{20}^{\mu \nu }\hat{\mathcal{I}}_{00,\mu \nu }
+\hat{\mathcal{I}}_{10}^{\mu\nu }\hat{\mathcal{I}}_{10,\mu \nu }\right) 
-\frac{\sigma _{T}}{60}\hat{\mathcal{I}}_{00}^{\mu \nu \alpha }\hat{\mathcal{I}}_{00,\mu \nu \alpha } 
\notag \\
&= \frac{\sigma _{T}}{10}\left( 4\hat{I}_{520}\hat{I}_{100}-4\hat{I}_{530}\hat{I}_{110}
+7\hat{I}_{420}\hat{I}_{200}-7\hat{I}_{420}\hat{I}_{220}+4\hat{I}_{320}\hat{I}_{300}
-4\hat{I}_{330}\hat{I}_{310}\right)  \notag \\
&+ \frac{\sigma _{T}}{10}\left( 6\hat{I}_{410}\hat{I}_{210}-6\hat{I}_{430}\hat{I}_{210}
+\hat{I}_{400}\hat{I}_{200}-\hat{I}_{440}\hat{I}_{220}+2\hat{I}_{401}\hat{I}_{201}
-2\hat{I}_{421}\hat{I}_{201}\right)  \notag \\
&+ \frac{\sigma _{T}}{60}\left( 5\hat{I}_{300}\hat{I}_{300}-5\hat{I}_{330}\hat{I}_{330}
+33\hat{I}_{310}\hat{I}_{310}-33\hat{I}_{320}\hat{I}_{320}+6\hat{I}_{301}\hat{I}_{301}
-6\hat{I}_{311}\hat{I}_{311}\right) \, .
\label{app_G22}
\end{align}

\section{The moments of the binary collision integrals}
\label{C_ij_binary}

Here we write down the full collision integrals of interest. 
Using Eqs.~(\ref{L20})-(\ref{L40}) and Eqs.~(\ref{G20})-(\ref{G40}) we obtain
\begin{align}
\hat{C}_{20} &= -\frac{\sigma _{T}}{6}\left( 2\hat{I}_{300}\hat{n} 
-2\hat{I}_{310}\hat{n}_{l}-3\hat{e}^{2}\right) -\frac{\sigma _{T}}{6}
\left( 2\hat{M}^{2}+\hat{P}_{l}^{2}+2\hat{P}_{\perp }^{2}\right) \, ,  
\label{C_20} \\
\hat{C}_{30} &= -\frac{\sigma _{T}}{2}\left( \hat{I}_{400}\hat{n} 
-\hat{I}_{410}\hat{n}_{l}-2\hat{I}_{300}\hat{e}\right) -\frac{\sigma _{T}}{2}
\left(\hat{I}_{310}\hat{M}+\hat{I}_{320}\hat{P}_{l}+2\hat{I}_{301}\hat{P}_{\perp}\right) \, ,  
\label{C_30}
\end{align}%
and
\begin{align}
\hat{C}_{40}& =-\frac{\sigma _{T}}{5}\left( 3\hat{I}_{500}\hat{n}
-3\hat{I}_{510}\hat{n}_{l}-5\hat{I}_{400}\hat{e}+2\hat{I}_{410}\hat{M}
+3\hat{I}_{420}\hat{P}_{l}+6\hat{I}_{401}\hat{P}_{\perp }\right)  \notag \\
&+ \frac{\sigma _{T}}{20}\left( 13\hat{I}_{300}^{2}-3\hat{I}_{310}^{2}
-9\hat{I}_{320}^{2}-\hat{I}_{330}^{2}-18\hat{I}_{301}^{2}-6\hat{I}_{311}^{2}\right)\, .  
\label{C_40}
\end{align}%
Similarly, using Eqs.~(\ref{L11}-\ref{L31}) and Eqs.~(\ref{G11}-\ref{G31})
we get 
\begin{align}
\hat{C}_{11}&= -\frac{\sigma _{T}}{3}\left( \hat{I}_{310}\hat{n}
-\hat{I}_{320}\hat{n}_{l}-2\hat{M}\hat{e}+2\hat{M}\hat{P}_{l}\right) \, ,
\label{C_11} \\
\hat{C}_{21}&= -\frac{\sigma _{T}}{6}\left( 3\hat{I}_{410}\hat{n}
-3\hat{I}_{420}\hat{n}_{l}-5\hat{I}_{310}\hat{e}+4\hat{I}_{320}\hat{M}\right) 
+\frac{\sigma _{T}}{6}\left( 3\hat{I}_{300}\hat{M}-3\hat{I}_{310}\hat{P}_{l}
-\hat{I}_{330}\hat{P}_{l}-2\hat{I}_{311}\hat{P}_{\perp }\right) \, ,  
\label{C_21}
\end{align}
and 
\begin{align}
\hat{C}_{31}& =-\frac{\sigma _{T}}{5}\left( 3\hat{I}_{510}\hat{n}
-3\hat{I}_{520}\hat{n}_{l}-2\hat{I}_{400}\hat{M}+2\hat{I}_{410}\hat{P}_{l}\right) 
+\frac{3\sigma _{T}}{10}\left( 3\hat{I}_{410}\hat{e}-2\hat{I}_{420}\hat{M}
-\hat{I}_{430}\hat{P}_{l}-2\hat{I}_{411}\hat{P}_{\perp }\right)  \notag \\
&+ \frac{3\sigma _{T}}{10}\left( 3\hat{I}_{300}\hat{I}_{310}-2\hat{I}_{310}\hat{I}_{320}
-\hat{I}_{320}\hat{I}_{330}-2\hat{I}_{301}\hat{I}_{311}\right)\, .  
\label{C_31}
\end{align}%
Finally, using Eqs.~(\ref{L02}-\ref{L22}) and Eqs.~(\ref{G02}-\ref{G22}) we obtain
\begin{align}
\hat{C}_{02}& =-\frac{\sigma _{T}}{6}\left( 2\hat{I}_{320}\hat{n}
-2\hat{I}_{330}\hat{n}_{l}-\hat{e}^{2}\right) 
+\frac{\sigma _{T}}{6}\left( 2\hat{M}^{2}-3\hat{P}_{l}^{2}+2\hat{P}_{\perp }^{2}\right) \, ,  
\label{C_02} \\
\hat{C}_{12}& =-\frac{\sigma _{T}}{6}\left( 3\hat{I}_{420}\hat{n}
-3\hat{I}_{430}\hat{n}_{l}-4\hat{I}_{310}\hat{M}+5\hat{I}_{320}\hat{P}_{l}\right) 
+\frac{\sigma _{T}}{6}\left( 3\hat{I}_{320}\hat{e}-3\hat{I}_{330}\hat{M}
+\hat{I}_{300}\hat{e}+2\hat{I}_{301}\hat{P}_{\perp }\right) \, ,  
\label{C_12}
\end{align}
and
\begin{align}
\hat{C}_{22}& =-\frac{\sigma _{T}}{10}\left( 6\hat{I}_{520}\hat{n}
-6\hat{I}_{530}\hat{n}_{l}-7\hat{I}_{420}\hat{e}+7\hat{I}_{420}\hat{P}_{l}
-4\hat{I}_{320}\hat{I}_{300}+4\hat{I}_{330}\hat{I}_{310}\right)  \notag \\
&+ \frac{\sigma _{T}}{10}\left( 6\hat{I}_{410}\hat{M}-6\hat{I}_{430}\hat{M}
+\hat{I}_{400}\hat{e}-\hat{I}_{440}\hat{P}_{l}+2\hat{I}_{401}\hat{P}_{\perp}
-2\hat{I}_{421}\hat{P}_{\perp }\right)  \notag \\
&+ \frac{\sigma _{T}}{60}\left( 5\hat{I}_{300}^{2}-5\hat{I}_{330}^{2}
+33\hat{I}_{310}^{2}-33\hat{I}_{320}^{2}+6\hat{I}_{301}^{2}-6\hat{I}_{311}^{2}\right) \, .  
\label{C_22}
\end{align}

\section{Thermodynamic integrals in the massless Boltzmann limit}
\label{appendix_aniso_integrals}

In the LR frame $u_{LR}^{\mu }=\left(1,0,0,0\right)$ and $l_{LR}^{\mu}=\left(0,0,0,1\right)$, 
and hence, $E_{\mathbf{k}u,LR}=k^{0}$ and $E_{\mathbf{k}l,LR}=k_{z}$. 
Assuming Boltzmann statistics, the corresponding Maxwell-J\"uttner 
distribution function from Eq.~(\ref{f_0k}) reads 
\begin{equation}
f_{0\mathbf{k}}=\lambda\exp \left( -\beta \sqrt{m_{0}^{2}+k^{2}}\right) \, ,
\end{equation}%
where $\lambda =\exp \alpha$ is the fugacity. Similarly, the RS distribution function 
from Eq.~(\ref{f_RS}) now reads 
\begin{equation}
\hat{f}_{RS}=\lambda_{RS}\exp \left( -\beta _{RS}\sqrt{m_{0}^{2}+k^{2}+\xi k_{z}^{2}}\right) \, ,
\end{equation}%
where $\lambda _{RS}=\exp \alpha _{RS}$ is corresponding to the fugacity.

In the massless limit of Eq.~(\ref{I_nq}), i.e., $\lim_{m_{0}\rightarrow 0}%
\sqrt{m_{0}^{2}+k^{2}}=|k|$, we obtain the following result for the thermodynamic integrals 
\begin{align}
\lim_{m_{0}\rightarrow 0} I_{nq} &\equiv \lambda 
\frac{\left( -1\right)^{2q}4\pi A_{0}}{\left( 2q+1\right) !!}
\int_{0}^{\infty }dkk^{n+1}\exp \left( -\beta k\right)  \notag \\
&= \lambda \frac{4\pi A_{0} \left( n+1\right) !}{\beta ^{n+2}\left(2q+1\right) !!}\, .  
\label{I_nq_massless}
\end{align}%
Here we list those thermodynamic integrals that are needed explicitly, 
\begin{align}
I_{10}\left( \alpha,\beta\right) &\equiv I_{100}=\lambda 
\frac{8\pi A_{0}}{\beta^{3}}=n_{0}\, ,  \label{app_I_00} \\
I_{20}\left( \alpha,\beta\right) &\equiv I_{200}=\lambda
\frac{24\pi A_{0}}{\beta^{4}}=e_{0}=3\frac{n_{0}}{\beta}\, ,
\label{app_I_20} \\
I_{21}\left( \alpha,\beta\right) &\equiv I_{201}=I_{220}=P_{0}
=\frac{1}{3}e_{0}=\frac{n_{0}}{\beta}\, ,  \label{app_I_21}
\end{align}%
and 
\begin{align}
I_{30}\left( \alpha,\beta\right) &\equiv I_{300}=\lambda
\frac{96\pi A_{0}}{\beta^{5}}=12\frac{n_{0}}{\beta^{2}}\, ,
\label{app_I_30} \\
I_{40}\left( \alpha,\beta\right) &\equiv I_{400}=\lambda
\frac{480\pi A_{0}}{\beta^{6}}=60\frac{n_{0}}{\beta^{3}}\, ,
\label{app_I_40} \\
I_{50}\left( \alpha,\beta\right) &\equiv I_{500}=\lambda
\frac{2880\pi A_{0}}{\beta^{7}}=360\frac{n_{0}}{\beta^{4}}\, ,
\label{app_I_50} \\
I_{31}\left( \alpha,\beta\right) &\equiv I_{301}=I_{320}
=\frac{1}{3}I_{30}\left( \alpha,\beta\right) \, ,  \label{app_I_31} \\
I_{41}\left( \alpha,\beta\right) &\equiv I_{401}=I_{420}
=\frac{1}{3}I_{40}\left( \alpha,\beta\right) \, ,  \label{app_I_41} \\
I_{51}\left( \alpha,\beta\right) &\equiv I_{501}=I_{520}
=\frac{1}{3}I_{50}\left( \alpha,\beta\right) \, .  \label{app_I_51}
\end{align}

Using the RS distribution function in the massless limit leads to the following anisotropic 
thermodynamical integrals
\begin{align}
\lim_{m_{0}\rightarrow 0}\hat{I}_{nrq}^{RS}& \equiv \lambda _{RS}
\frac{\left( -1\right) ^{2q}2\pi A_{0}}{\left( 2q\right) !!}\int_{0}^{\infty}
dkk^{n+1}\int_{-1}^{1}dxx^{r}\left( 1-x^{2}\right) ^{q} 
\exp \left( -\beta_{RS}k\sqrt{1+\xi x^{2}}\right)  \notag \\
& =\lambda _{RS}\frac{2\pi A_{0}\left( n+1\right) !}{\beta _{RS}^{n+2} 
\left(2q\right) !!}\int_{-1}^{1}dx \frac{x^{r}\left( 1-x^{2}\right) ^{q}}
{\left(1+\xi x^{2}\right) ^{\frac{n+2}{2}}}\, ,  \label{I_nrq_RS_massless}
\end{align}%
while the ratio between the anisotropic and equilibrium thermodynamical integrals
defined in Eq.~(\ref{R_nrq_RS}) reads [see Eq.~(A11) in Ref.~\cite{Molnar:2016gwq}]
\begin{equation}
R_{nrq}=\frac{\left( 2q+1\right) !!}{2\left( 2q\right) !!}\int_{0}^{\pi} d\theta 
\frac{\cos^{r}\theta \sin ^{2q+1}\theta }{\left( 1+\xi \cos^{2}\theta \right)^{\frac{n+2}{2}}} \, .
\end{equation}%
The values of $R_{nrq}(\xi )$ that correspond to Eqs.~(\ref{n_hat_RS})-(\ref{Pt_hat_RS}) and to Eq.~(\ref{BJ_Pl_relax}) are 
\begin{align}
R_{100}\left( \xi \right) &= \frac{1}{\sqrt{1+\xi }}\, ,  \label{R_100} \\
R_{200}\left( \xi \right) &= \frac{1}{2}\left( \frac{1}{1+\xi } 
+\frac{\arctan \sqrt{\xi }}{\sqrt{\xi }}\right) \, ,  \label{R_200} \\
R_{201}\left( \xi \right) &= \frac{3}{2\xi }\left( \frac{1}{1+\xi } 
-\left(1-\xi \right) R_{200}\left( \xi \right) \right) \, ,  \label{R_201} \\
R_{220}\left( \xi \right) &= -\frac{1}{\xi }\left( \frac{1}{1+\xi }
-R_{200}\left( \xi \right) \right) \, ,  \label{R_220} \\
R_{240}\left( \xi \right) &= \frac{1}{\xi ^{2}}\left( \frac{3+\xi }{1+\xi }
-3R_{200}\left( \xi \right) \right) \, .
\end{align}

Furthermore, for the moment equations listed in Eqs.~(\ref{BJ_I300_relax})-(\ref{BJ_I500_relax}), 
and Eqs.~(\ref{BJ_I320_relax})-(\ref{BJ_I520_relax}), we also need
to specify the following $R_{nrq}(\xi )$ ratios:
\begin{align}  
R_{300}\left( \xi \right) &= \frac{3+2\xi }{3\left( 1+\xi \right) ^{3/2}}\, ,
\label{R_300} \\
R_{301}\left( \xi \right) &= R_{100}\left( \xi \right) \, , \\
R_{320}\left( \xi \right) &= \frac{1}{3\left( 1+\xi \right) ^{3/2}} \, ,
\label{R_320}
\end{align}%
and 
\begin{align}
R_{400}\left( \xi \right) &= \frac{1}{8}\left( \frac{5+3\xi }{(1+\xi )^{2}}
+3\frac{\arctan \sqrt{\xi }}{\sqrt{\xi }}\right) \, , \\
R_{401}\left( \xi \right) &= \frac{3}{16\xi }\left( \frac{1+3\xi }{(1+\xi )}
-\left( 1-3\xi \right) \frac{\arctan \sqrt{\xi }}{\sqrt{\xi }}\right) \, , \\
R_{420}\left( \xi \right) &= -\frac{1}{8\xi }\left( \frac{1-\xi }{(1+\xi)^{2}} 
-\frac{\arctan \sqrt{\xi }}{\sqrt{\xi }}\right) \, , \\
R_{421}\left( \xi \right) &= \frac{3}{16\xi ^{2}}\left( \frac{3+\xi }{(1+\xi)}
-\left( 3-\xi \right) \frac{\arctan \sqrt{\xi }}{\sqrt{\xi }}\right) \, ,
\\
R_{440}\left( \xi \right) &= -\frac{1}{8\xi ^{2}}\left( \frac{3+5\xi }{(1+\xi )^{2}}
-3\frac{\arctan \sqrt{\xi }}{\sqrt{\xi }}\right) \, ,
\end{align}%
and finally,
\begin{align}
R_{500}\left( \xi \right) &= \frac{15+4\xi \left( 5+2\xi \right) }{15\left(1+\xi \right)^{5/2}}\, , \\
R_{520}\left( \xi \right) &= \frac{5+2\xi }{15\left( 1+\xi \right)^{5/2}}\, , \\
R_{540}\left( \xi \right) &= \frac{1}{5\left( 1+\xi \right)^{5/2}}\, .
\end{align}



\begin{thebibliography}{99}
\bibitem{Landau_hydro} 
L.~D.~Landau, 
``On the multiparticle production in high-energy collisions,'' 
Izv. Akad. Nauk SSSR \textbf{17}, 51 (1953). 

\bibitem{Landau_hydro56} 
Belen’kji, S.Z., Landau,
``Hydrodynamic theory of multiple production of particles,''
Nuovo Cim \textbf{3} (Suppl 1), 15–31 (1956)
https://doi.org/10.1007/BF02745507

\bibitem{Juttner} 
F.\ J\"uttner, 
``Das Maxwellsche Gesetz der Geschwindigkeitsverteilung in der Relativtheorie,''
Ann.\ Phys. \textbf{339}, 856 (1911)
https://doi.org/10.1002/andp.19113390503
 
 
\bibitem{Juttner_quantum} 
F.\ J\"uttner, 
``Die relativistische Quantentheorie des idealen Gases,'' 
Z.\ Phys.\ \textbf{47}, 542 (1928).
doi.org/10.1007/BF01340339

\bibitem{Rezzolla_book} 
L. Rezzolla and O. Zanotti, 
\textit{Relativistic Hydrodynamics} 
(Oxford University Press, Oxford, United Kingdom, 2013).

\bibitem{Denicol_Rischke_book} 
Gabriel S. Denicol and Dirk H. Rischke, 
\textit{Microscopic foundations of relativistic fluid dynamics}, 
(Springer International Publishing AG, 2022).


\bibitem{Chabanov:2021dee}
M.~Chabanov, L.~Rezzolla and D.~H.~Rischke,
``General-relativistic hydrodynamics of non-perfect fluids: 
3+1 conservative formulation and application to viscous black hole accretion,''
Mon. Not. Roy. Astron. Soc. \textbf{505}, no.4, 5910-5940 (2021)
doi:10.1093/mnras/stab1384
[arXiv:2102.10419 [gr-qc]].


\bibitem{Heinz:2024jwu}
U.~Heinz and B.~Schenke,
``Hydrodynamic Description of the Quark-Gluon Plasma,''
[arXiv:2412.19393 [nucl-th]].

\bibitem{Rocha:2023ilf} 
G.~S.~Rocha, D.~Wagner, G.~S.~Denicol, J.~Noronha and D.~H.~Rischke, 
``Theories of Relativistic Dissipative Fluid Dynamics,''
Entropy \textbf{26}, no.3, 189 (2024) 
doi:10.3390/e26030189
[arXiv:2311.15063 [nucl-th]].


\bibitem{Muller:1967zza} 
I.~Muller, ``Zum Paradoxon der Warmeleitungstheorie,'' 
Z. Phys. \textbf{198}, 329-344 (1967)
doi:10.1007/BF01326412


\bibitem{Israel:1979wp} 
W.~Israel and J.~M.~Stewart, 
``Transient relativistic thermodynamics and kinetic theory,'' 
Annals Phys. \textbf{118}, 341-372 (1979) 
doi:10.1016/0003-4916(79)90130-1

\bibitem{Hiscock:1983zz} 
W.~A.~Hiscock and L.~Lindblom, 
``Stability and causality in dissipative relativistic fluids,'' 
Annals Phys. \textbf{151}, 466-496 (1983) 
doi:10.1016/0003-4916(83)90288-9

\bibitem{Hiscock:1985zz} 
W.~A.~Hiscock and L.~Lindblom, 
``Generic instabilities in first-order dissipative relativistic fluid theories,''
Phys. Rev. D \textbf{31}, 725-733 (1985) 
doi:10.1103/PhysRevD.31.725

\bibitem{Hiscock:1987zz} 
W.~A.~Hiscock and L.~Lindblom, 
``Linear plane waves in dissipative relativistic fluids,'' 
Phys. Rev. D \textbf{35}, 3723-3732 (1987) 
doi:10.1103/PhysRevD.35.3723

\bibitem{Pu:2009fj} 
S.~Pu, T.~Koide and D.~H.~Rischke, 
``Does stability of relativistic dissipative fluid dynamics imply causality?,'' 
Phys. Rev. D \textbf{81}, 114039 (2010) 
doi:10.1103/PhysRevD.81.114039 
[arXiv:0907.3906[hep-ph]].

\bibitem{Barz:1987pq} 
H.~W.~Barz, B.~Kampfer, B.~Lukacs, K.~Martinas and G.~Wolf, 
``Deconfinement transition in anisotropic matter,'' 
Phys. Lett. B \textbf{194}, 15-19 (1987) 
doi:10.1016/0370-2693(87)90761-1

\bibitem{Kampfer:1990qg} 
B.~Kampfer, B.~Lukacs, G.~Wolf and H.~W.~Barz,
``Description of the nuclear stopping process within anisotropic thermohydrodynamics,'' 
Phys. Lett. B \textbf{240}, 297-300 (1990)
doi:10.1016/0370-2693(90)91101-G

\bibitem{Bowers:1974tgi} 
R.~L.~Bowers and E.~P.~T.~Liang, ``Anisotropic Spheres in General Relativity,'' 
Astrophys. J. \textbf{188}, 657-665 (1974)
doi:10.1086/152760

\bibitem{Anderlik:1998cb} 
C.~Anderlik, Z.~I.~Lazar, V.~K.~Magas, L.~P.~Csernai, H.~Stoecker and W.~Greiner, 
``Nonideal particle distributions from kinetic freezeout models,'' 
Phys. Rev. C \textbf{59}, 388-394 (1999)
doi:10.1103/PhysRevC.59.388 
[arXiv:nucl-th/9808024 [nucl-th]].

\bibitem{Anderlik:1998et} 
C.~Anderlik, L.~P.~Csernai, F.~Grassi, W.~Greiner, Y.~Hama, T.~Kodama, 
Z.~I.~Lazar, V.~K.~Magas and H.~Stoecker, 
``Freezeout in hydrodynamical models,'' 
Phys. Rev. C \textbf{59}, 3309-3316 (1999)
doi:10.1103/PhysRevC.59.3309 
[arXiv:nucl-th/9806004 [nucl-th]].

\bibitem{Florkowski:2008ag}
W.~Florkowski,
``Anisotropic fluid dynamics in the early stage of relativistic heavy-ion collisions,''
Phys. Lett. B \textbf{668}, 32-35 (2008)
doi:10.1016/j.physletb.2008.07.101
[arXiv:0806.2268 [nucl-th]].

\bibitem{Florkowski:2010cf}
W.~Florkowski and R.~Ryblewski,
``Highly-anisotropic and strongly-dissipative hydrodynamics for early stages 
of relativistic heavy-ion collisions,''
Phys. Rev. C \textbf{83}, 034907 (2011)
doi:10.1103/PhysRevC.83.034907
[arXiv:1007.0130 [nucl-th]].

\bibitem{Ryblewski:2010bs}
R.~Ryblewski and W.~Florkowski,
``Non-boost-invariant motion of dissipative and highly anisotropic fluid,''
J. Phys. G \textbf{38}, 015104 (2011)
doi:10.1088/0954-3899/38/1/015104
[arXiv:1007.4662 [nucl-th]].


\bibitem{Ryblewski:2011aq}
R.~Ryblewski and W.~Florkowski,
``Highly-anisotropic and strongly-dissipative hydrodynamics with transverse expansion,''
Eur. Phys. J. C \textbf{71}, 1761 (2011)
doi:10.1140/epjc/s10052-011-1761-8
[arXiv:1103.1260 [nucl-th]].

\bibitem{Ryblewski:2012rr}
R.~Ryblewski and W.~Florkowski,
``Highly-anisotropic hydrodynamics in 3+1 space-time dimensions,''
Phys. Rev. C \textbf{85}, 064901 (2012)
doi:10.1103/PhysRevC.85.064901
[arXiv:1204.2624 [nucl-th]].


\bibitem{Martinez:2009ry}
M.~Martinez and M.~Strickland,
``Matching pre-equilibrium dynamics and viscous hydrodynamics,''
Phys. Rev. C \textbf{81}, 024906 (2010)
doi:10.1103/PhysRevC.81.024906
[arXiv:0909.0264 [hep-ph]].

\bibitem{Martinez:2010sc}
M.~Martinez and M.~Strickland,
``Dissipative Dynamics of Highly Anisotropic Systems,''
Nucl. Phys. A \textbf{848}, 183-197 (2010)
doi:10.1016/j.nuclphysa.2010.08.011
[arXiv:1007.0889 [nucl-th]].

\bibitem{Martinez:2010sd}
M.~Martinez and M.~Strickland,
``Non-boost-invariant anisotropic dynamics,''
Nucl. Phys. A \textbf{856}, 68-87 (2011)
doi:10.1016/j.nuclphysa.2011.02.003
[arXiv:1011.3056 [nucl-th]].

\bibitem{Florkowski:2013lya}
W.~Florkowski, R.~Ryblewski and M.~Strickland,
``Testing viscous and anisotropic hydrodynamics in an exactly solvable case,''
Phys. Rev. C \textbf{88}, 024903 (2013)
doi:10.1103/PhysRevC.88.024903
[arXiv:1305.7234 [nucl-th]].

\bibitem{Tinti:2015xwa}
L.~Tinti,
``Anisotropic matching principle for the hydrodynamic expansion,''
Phys. Rev. C \textbf{94}, no.4, 044902 (2016)
doi:10.1103/PhysRevC.94.044902
[arXiv:1506.07164 [hep-ph]].


\bibitem{Alqahtani:2017mhy}
M.~Alqahtani, M.~Nopoush and M.~Strickland,
``Relativistic anisotropic hydrodynamics,''
Prog. Part. Nucl. Phys. \textbf{101}, 204-248 (2018)
doi:10.1016/j.ppnp.2018.05.004
[arXiv:1712.03282 [nucl-th]].


\bibitem{Bazow:2013ifa}
D.~Bazow, U.~W.~Heinz and M.~Strickland,
``Second-order (2+1)-dimensional anisotropic hydrodynamics,''
Phys. Rev. C \textbf{90}, no.5, 054910 (2014)
doi:10.1103/PhysRevC.90.054910
[arXiv:1311.6720 [nucl-th]].


\bibitem{Molnar:2016vvu}
E.~Molnar, H.~Niemi and D.~H.~Rischke,
``Derivation of anisotropic dissipative fluid dynamics from the Boltzmann equation,''
Phys. Rev. D \textbf{93}, no.11, 114025 (2016)
doi:10.1103/PhysRevD.93.114025
[arXiv:1602.00573 [nucl-th]].

\bibitem{Molnar:2024fgy}
E.~Moln\'ar and D.~H.~Rischke,
``Higher-order dissipative anisotropic magnetohydrodynamics from the Boltzmann-Vlasov equation,''
Phys. Rev. D \textbf{111}, no.3, 036035 (2025)
doi:10.1103/PhysRevD.111.036035


\bibitem{Bobylev:1976}
A. V. Bobylev, 
``Some properties of Boltzmann's equation for Maxwell molecules,''
Sov. Phys. Dokl. 20, 820 (1976).

\bibitem{Krook_Wu:1976}
M. Krook and T. T. Wu, 
``Formation of Maxwellian Tails,''
Phys. Rev. Lett. \textbf{36}, 1107 (1976).
https://doi.org/10.1103/PhysRevLett.36.1107


\bibitem{Krook_Wu:1977}
M. Krook and T. T. Wu, 
``Exact solutions of the Boltzmann equation,''
Physics of Fluids 20, 1589 (1977).
https://doi.org/10.1063/1.861780

\bibitem{Bazow:2015dha}
D.~Bazow, G.~S.~Denicol, U.~Heinz, M.~Martinez and J.~Noronha,
``Analytic solution of the Boltzmann equation in an expanding system,''
Phys. Rev. Lett. \textbf{116}, no.2, 022301 (2016)
doi:10.1103/PhysRevLett.116.022301
[arXiv:1507.07834 [hep-ph]].

\bibitem{Bazow:2016oky}
D.~Bazow, G.~S.~Denicol, U.~Heinz, M.~Martinez and J.~Noronha,
``Nonlinear dynamics from the relativistic Boltzmann equation in the Friedmann-Lema\^\i{}tre-Robertson-Walker spacetime,''
Phys. Rev. D \textbf{94}, no.12, 125006 (2016)
doi:10.1103/PhysRevD.94.125006
[arXiv:1607.05245 [hep-ph]].



\bibitem{Molnar:2016gwq}
E.~Moln\'ar, H.~Niemi and D.~H.~Rischke,
``Closing the equations of motion of anisotropic fluid dynamics by 
a judicious choice of a moment of the Boltzmann equation,''
Phys. Rev. D \textbf{94}, no.12, 125003 (2016)
doi:10.1103/PhysRevD.94.125003
[arXiv:1606.09019 [nucl-th]].

\bibitem{Niemi:2017stb}
H.~Niemi, E.~Moln\'ar and D.~H.~Rischke,
``The right choice of moment for anisotropic fluid dynamics,''
Nucl. Phys. A \textbf{967}, 409-412 (2017)
doi:10.1016/j.nuclphysa.2017.05.038
[arXiv:1705.01851 [nucl-th]].


\bibitem{Romatschke:2003ms}
P.~Romatschke and M.~Strickland,
``Collective modes of an anisotropic quark gluon plasma,''
Phys. Rev. D \textbf{68}, 036004 (2003)
doi:10.1103/PhysRevD.68.036004
[arXiv:hep-ph/0304092 [hep-ph]].


\bibitem{Anderson:1974nyl}
J.~L.~Anderson and H.~R.~Witting,
``A relativistic relaxation-time model for the Boltzmann equation,''
Physica \textbf{74}, no.3, 466-488 (1974)
doi:10.1016/0031-8914(74)90355-3


\bibitem{deGroot_book} 
S.R.\ de Groot, W.A.\ van Leeuwen and Ch.G.\ van Weert, 
\textit{Relativistic Kinetic Theory - Principles and applications}, 
(North Holland, Amsterdam, 1980).

\bibitem{Cercignani_book} 
C.\ Cercignani and G.M.\ Kremer, 
\textit{The Relativistic Boltzmann Equation: Theory and Applications}, 
(Birkh\"auser, Basel, 2002).

\bibitem{Gedalin_1991} 
M.\ Gedalin, 
``Relativistic hydrodynamics and thermodynamics of anisotropic plasmas,'' 
Phys.\ Fluids B \textbf{3}, 1871 (1991)
https://doi.org/10.1063/1.859656

\bibitem{Gedalin_1995} 
M.\ Gedalin and I.\ Oiberman, ``Generally covariant relativistic anisotropic magnetohydrodynamics,'' 
Phys.\ Rev.\ E \textbf{51}, 4901 (1995).
 https://doi.org/10.1103/PhysRevE.51.4901
 
\bibitem{Huang:2011dc}
X.~G.~Huang, A.~Sedrakian and D.~H.~Rischke,
``Kubo formulae for relativistic fluids in strong magnetic fields,''
Annals Phys. \textbf{326}, 3075-3094 (2011)
doi:10.1016/j.aop.2011.08.001
[arXiv:1108.0602 [astro-ph.HE]].

\bibitem{Van:2013sma}
P.~V\'an and T.~S.~Bir\'o,
``Dissipation flow-frames: particle, energy, thermometer,''
[arXiv:1305.3190 [gr-qc]].

\bibitem{Landau_book} 
L.D.\ Landau and E.M.\ Lifshitz, \textit{Fluid Dynamics}, 
Second Edition, (Butterworth-Heinemann, Oxford, 1987).

\bibitem{Molnar:2013lta}
E.~Moln\'ar, H.~Niemi, G.~S.~Denicol and D.~H.~Rischke,
``Relative importance of second-order terms in relativistic dissipative fluid dynamics,''
Phys. Rev. D \textbf{89}, no.7, 074010 (2014)
doi:10.1103/PhysRevD.89.074010
[arXiv:1308.0785 [nucl-th]].


\bibitem{Wagner:2023joq}
D.~Wagner, V.~E.~Ambrus and E.~Molnar,
``Analytical structure of the binary collision integral and the ultrarelativistic limit of transport coefficients of an ideal gas,''
Phys. Rev. D \textbf{109}, no.5, 056018 (2024)
doi:10.1103/PhysRevD.109.056018
[arXiv:2309.09335 [physics.flu-dyn]].


\bibitem{Eckart:1940te}
C.~Eckart,
``The Thermodynamics of irreversible processes. 3.. Relativistic theory of the simple fluid,''
Phys. Rev. \textbf{58}, 919-924 (1940)
doi:10.1103/PhysRev.58.919


\bibitem{Bhatnagar:1954} 
P.~L. Bhatnagar, E.~P. Gross, and M.~Krook. 
``A model for collision processes in gases. I. Small amplitude processes in
charged and neutral one-component systems,'' 
Phys. Rev. 94: \textbf{511} (1954) 
doi:10.1103/PhysRev.94.511


\bibitem{Ambrus:2022vif}
V.~E.~Ambrus, E.~Moln\'ar and D.~H.~Rischke,
``Transport coefficients of second-order relativistic fluid dynamics in the relaxation-time approximation,''
Phys. Rev. D \textbf{106}, no.7, 076005 (2022)
doi:10.1103/PhysRevD.106.076005
[arXiv:2207.05670 [nucl-th]].

\bibitem{Bjorken:1982qr}
J.~D.~Bjorken,
``Highly Relativistic Nucleus-Nucleus Collisions: The Central Rapidity Region,''
Phys. Rev. D \textbf{27}, 140-151 (1983)
doi:10.1103/PhysRevD.27.140

\bibitem{Ambrus:2023qcl}
V.~E.~Ambru\c{s}, E.~Moln\'ar and D.~H.~Rischke,
``Relativistic second-order dissipative and anisotropic fluid dynamics in the relaxation-time 
approximation for an ideal gas of massive particles,''
Phys. Rev. D \textbf{109}, no.7, 076001 (2024)
doi:10.1103/PhysRevD.109.076001
[arXiv:2311.00351 [nucl-th]].


\bibitem{Denicol:2012vq}
G.~S.~Denicol, H.~Niemi, I.~Bouras, E.~Molnar, Z.~Xu, D.~H.~Rischke and C.~Greiner,
``Solving the heat-flow problem with transient relativistic fluid dynamics,''
Phys. Rev. D \textbf{89}, no.7, 074005 (2014)
doi:10.1103/PhysRevD.89.074005
[arXiv:1207.6811 [nucl-th]].

\bibitem{Ambrus:2023ilm}
V.~E.~Ambru\c{s} and E.~Moln\'ar,
``Shakhov-type extension of the relaxation time approximation in relativistic kinetic theory and second-order fluid dynamics,''
Phys. Lett. B \textbf{855}, 138795 (2024)
doi:10.1016/j.physletb.2024.138795
[arXiv:2311.11603 [nucl-th]].

\bibitem{Ambrus:2024qsa}
V.~E.~Ambru\c{s} and D.~Wagner,
``High-order Shakhov-like extension of the relaxation time approximation in relativistic kinetic theory,''
Phys. Rev. D \textbf{110}, no.5, 056002 (2024)
doi:10.1103/PhysRevD.110.056002
[arXiv:2401.04017 [nucl-th]].


\bibitem{Bouras:2010hm}
I.~Bouras, E.~Molnar, H.~Niemi, Z.~Xu, A.~El, O.~Fochler, C.~Greiner and D.~H.~Rischke,
``Investigation of shock waves in the relativistic Riemann problem: A Comparison of viscous fluid dynamics to kinetic theory,''
Phys. Rev. C \textbf{82}, 024910 (2010)
doi:10.1103/PhysRevC.82.024910
[arXiv:1006.0387 [hep-ph]].

\bibitem{Gallmeister:2018mcn}
K.~Gallmeister, H.~Niemi, C.~Greiner and D.~H.~Rischke,
``Exploring the applicability of dissipative fluid dynamics to small systems by comparison to the Boltzmann equation,''
Phys. Rev. C \textbf{98}, no.2, 024912 (2018)
doi:10.1103/PhysRevC.98.024912
[arXiv:1804.09512 [nucl-th]].


\bibitem{Alqahtani:2016ayv}
M.~Alqahtani and M.~Strickland,
``Quasiparticle anisotropic hydrodynamics,''
J. Phys. Conf. Ser. \textbf{832}, no.1, 012051 (2017)
doi:10.1088/1742-6596/832/1/012051
[arXiv:1610.07643 [nucl-th]].

\bibitem{Nopoush:2016qas}
M.~Nopoush, M.~Strickland and R.~Ryblewski,
``Phenomenological predictions of 3+1d anisotropic hydrodynamics,''
J. Phys. Conf. Ser. \textbf{832}, no.1, 012054 (2017)
doi:10.1088/1742-6596/832/1/012054
[arXiv:1610.10055 [nucl-th]].

\bibitem{Alqahtani:2017jwl}
M.~Alqahtani, M.~Nopoush, R.~Ryblewski and M.~Strickland,
``(3+1)D Quasiparticle Anisotropic Hydrodynamics for Ultrarelativistic Heavy-Ion Collisions,''
Phys. Rev. Lett. \textbf{119}, no.4, 042301 (2017)
doi:10.1103/PhysRevLett.119.042301
[arXiv:1703.05808 [nucl-th]].

\bibitem{Alqahtani:2017tnq}
M.~Alqahtani, M.~Nopoush, R.~Ryblewski and M.~Strickland,
``Anisotropic hydrodynamic modeling of 2.76 TeV Pb-Pb collisions,''
Phys. Rev. C \textbf{96}, no.4, 044910 (2017)
doi:10.1103/PhysRevC.96.044910
[arXiv:1705.10191 [nucl-th]].

\bibitem{Alqahtani:2020paa}
M.~Alqahtani and M.~Strickland,
``Bulk observables at 5.02~TeV using quasiparticle anisotropic hydrodynamics,''
Eur. Phys. J. C \textbf{81}, no.11, 1022 (2021)
doi:10.1140/epjc/s10052-021-09832-z
[arXiv:2008.07657 [nucl-th]].

\bibitem{Strickland:2024oat}
M.~Strickland, S.~Thapa and R.~Vogt,
``Bottomonium suppression in 5.02 and 8.16~TeV p-Pb collisions,''
Phys. Rev. D \textbf{109}, no.9, 096016 (2024)
doi:10.1103/PhysRevD.109.096016
[arXiv:2401.16704 [nucl-th]].


\bibitem{MaTeX}
Szabolcs Horvát, 
``MaTeX - LaTeX typesetting in Mathematica,''
https://doi.org/10.5281/zenodo.10828124

\end{thebibliography}
\end{document}